\documentclass[12pt,english,floatfix,nofootinbib,superscriptaddress,aps,prd,preprint]{revtex4}
\usepackage[utf8]{inputenc}       
\usepackage[english]{babel}       
\usepackage[norefs,nocites]{refcheck}
\usepackage[utf8]{inputenc}
\usepackage{textcomp}
\usepackage[a4paper, margin=1.6cm]{geometry}
\usepackage{amsmath,amsthm,amssymb,amsfonts,mathrsfs,amsbsy} 
\usepackage{tensor}               
\usepackage{slashed}  
\usepackage{bbm}
\usepackage{esint}                
\usepackage{cancel}               
\usepackage{graphicx}             
\usepackage{natbib}
\usepackage{float}                
\usepackage{subfig}               
\usepackage[font=small,labelfont=bf]{caption} 
\usepackage{multirow}             
\usepackage{array}                
\usepackage{tikz}                 
\usetikzlibrary{quotes,angles,arrows,decorations.markings} 
\usepackage{makecell} 

\usepackage{lipsum}

\usepackage{xcolor}               
\usepackage{color}                
\usepackage{textcomp}             
\usepackage{units}                

\newcommand{\be}{\begin{equation}}
\newcommand{\ee}{\end{equation}}
\newcommand{\Be}{\begin{eqnarray}}
\newcommand{\Ee}{\end{eqnarray}}

\newcommand{\mincir}{\raise
-3.truept\hbox{\rlap{\hbox{$\sim$}}\raise4.truept\hbox{$<$}\ }}
\newcommand{\magcir}{\raise
-3.truept\hbox{\rlap{\hbox{$\sim$}}\raise4.truept\hbox{$>$}\ }}

\newcolumntype{Y}{>{\centering\arraybackslash}X}
\providecommand{\U}[1]

\usepackage[dvips]{epsfig}        
\usepackage[dvips]{graphicx}      
\usepackage{hyperref}             
\hypersetup{
    colorlinks=true,              
    breaklinks=true,              
    citecolor=blue,               
    linkcolor=[rgb]{0,0.5,0.9},   
    urlcolor=red,                 
    filecolor=green               
}

\newcommand{\ie}{\begin{equation}}
\newcommand{\fe}{\end{equation}}
\newcommand{\se}{\begin{eqnarray}}
\newcommand{\ff}{\end{eqnarray}}

\begin{document}

\title{Quantum particle production and radiative properties of a new bumblebee black hole}


\author{N. Heidari}
\email{heidari.n@gmail.com}

\affiliation{Center for Theoretical Physics, Khazar University, 41 Mehseti Street, Baku, AZ-1096, Azerbaijan.}
\affiliation{School of Physics, Damghan University, Damghan, 3671641167, Iran.}
\affiliation{Departamento de Física, Universidade Federal de Campina Grande Caixa Postal 10071, 58429-900 Campina Grande, Paraíba, Brazil.}


\author{A. A. Ara\'{u}jo Filho}
\email{dilto@fisica.ufc.br}
\affiliation{Departamento de Física, Universidade Federal da Paraíba, Caixa Postal 5008, 58051--970, João Pessoa, Paraíba,  Brazil.}
\affiliation{Departamento de Física, Universidade Federal de Campina Grande Caixa Postal 10071, 58429-900 Campina Grande, Paraíba, Brazil.}
\affiliation{Center for Theoretical Physics, Khazar University, 41 Mehseti Street, Baku, AZ-1096, Azerbaijan.}


\date{\today}

\begin{abstract}

In this work, we investigate the quantum and radiative properties of a recently proposed static bumblebee black hole arising from a general Lorentz--violating vacuum configuration. The analysis begins with the geometric structure of the solution and the thermodynamic temperature obtained from the surface--gravity prescription. The associated thermodynamic topological structure is also examined. Quantum particle production is then analyzed for bosonic and fermionic fields using the tunneling method. Analytic greybody bounds are derived for spin--0, spin--1, spin--2, and spin--1/2 fields. Furthermore, full greybody factors are computed with the sixth--order WKB method, together with the corresponding absorption cross sections and their characteristic spin--dependent peak patterns. These results support the evaluation of the evaporation lifetimes and the emission rates of energy and particle modes associated with each spin contribution, followed by a comparison of the high--frequency regime with other Lorentz--violating geometries, including the \textit{metric} bumblebee, \textit{metric--affine} bumblebee, Kalb--Ramond, and non--commutative Kalb--Ramond black holes. In addition, greybody factors are obtained using a quasinormal--mode--based prescription.

\end{abstract}


\maketitle

\tableofcontents


\section{Introduction }

Lorentz symmetry has long been treated as a foundational element of relativistic physics, yet several theoretical programs aimed at linking gravity with quantum phenomena have suggested that this symmetry might function only approximately. A recurring theme in these investigations is the possibility that new geometric features could arise at energy scales close to those accessible in current experiments \cite{kostelecky1989spontaneous,colladay1997cpt,kostelecky2004gravity,kostelecky1999constraints,kostelecky2011data}. One mechanism frequently invoked to account for such deviations relies on dynamical fields that settle into vacuum states with nonvanishing configurations. When this occurs, the vacuum itself selects a direction in spacetime, and Lorentz symmetry becomes spontaneously broken.
Within this broad class of proposals, bumblebee models emerged as a compact and prominent framework for representing Lorentz violation. Instead of enforcing symmetry breaking through external prescriptions, these constructions employ a vector field whose magnitude is fixed by a potential. The field reaches to a stable configuration with constant norm, and this background acts as an orientation that reshapes the underlying spacetime geometry. The resulting gravitational sector remains internally consistent and furnishes a structured setting in which modifications to the relativistic dynamics can be examined \cite{Bluhm:2019ato,Bluhm:2023kph,Maluf:2014dpa,Maluf:2013nva,bluhm2008spontaneous,bluhm2005spontaneous}.

Several theoretical frameworks that attempt to extend or reinterpret general relativity have pointed to the possibility that spacetime may host background vector configurations capable of reshaping its symmetry properties \cite{kostelecky1989spontaneous,jacobson2004einstein,kostelecky1991photon}. In many of these settings, the fields introduced in the effective action naturally evolve toward vacuum states that do not vanish. Once such a configuration is reached, the geometry ceases to respect exact Lorentz invariance, since the vacuum itself singles out a direction \cite{bluhm2005spontaneous,kostelecky2004gravity}.
A concise realization of this mechanism appears in the family of constructions known as bumblebee models. Instead of imposing symmetry breaking externally, these theories assign a special role to a vector field $B_{\mu}$ whose norm is not arbitrary but restricted by a potential $V(B_{\mu} B^{\mu} \mp b^{2})$ \cite{Liu:2022dcn}. The dynamics guided by this potential drive the system toward a stable configuration with fixed magnitude. When the field reaches that state, the chosen background defines an orientation in spacetime and, consequently, the spontaneous violation of Lorentz symmetry is achieved \cite{bluhm2008spontaneous,bluhm2005spontaneous}.
Small fluctuations around this vacuum separate into two characteristic types. Modes that oscillate without disturbing the fixed-norm requirement behave analogously to massless gauge excitations and share several features with photonlike fields \cite{bluhm2005spontaneous}. In contrast, perturbations that shift the magnitude away from the constrained value acquire mass through the same potential responsible for stabilizing the vacuum configuration \cite{bluhm2008spontaneous}.

Bringing the bumblebee mechanism into curved spacetime placed the vacuum configuration of the vector field in direct correlates with the gravitational degrees of freedom, and this step led to a wide range of applications across different sectors of gravitational physics \cite{Bertolami:2005bh}. Instead of following a single trajectory, the subsequent developments branched into several independent research programs. One of the earliest and most influential directions centered on compact objects. After the black hole geometry proposed in \cite{Casana:2017jkc} became available, it served as a reference point for investigations that probed how Lorentz--violating backgrounds reshape strong--field gravity. This metric supported analyses of horizon--scale processes, such as modifications in entanglement properties \cite{Liu:2024wpa} and changes in quantum particle emission resulting from deviations in the underlying geometry \cite{AraujoFilho:2025hkm}. Parallel studies extended the underlying symmetry--breaking mechanism to the antisymmetric sector through Kalb--Ramond fields, yielding additional classes of black hole solutions with Lorentz violation built into their structure \cite{AraujoFilho:2024ctw}. Another body of work focused on large--scale cosmological and astrophysical settings. Configurations that emulate anisotropic expansion reminiscent of Kasner--type cosmologies were formulated in \cite{Neves:2022qyb}, and the influence of the same vector background on anisotropic stellar models was explored in \cite{Neves:2024ggn}. The dynamics of gravitational waves also underwent revision in these scenarios, with results demonstrating departures from the predictions of general relativity \cite{Liang:2022hxd,amarilo2024gravitational}.
Further extensions considered modifications to the geometric sector itself. Among them were constructions that introduced a cosmological constant within the bumblebee framework, leading to alternative vacuum structures \cite{Maluf:2020kgf} and additional phenomenological consequences \cite{Uniyal:2022xnq}.

The landscape of bumblebee gravity has changed substantially since the early static solution of Ref.~\cite{Casana:2017jkc}. As different geometric formulations were explored, the framework evolved into a broad collection of models with distinct dynamical properties. One of the most dynamic research arenas arose in the \textit{metric--affine} formulation, where the connection is treated independently from the metric. In this context, a static geometry was obtained in \cite{Filho:2022yrk}, and this result later paved the way for an axially symmetric rotating configuration \cite{AraujoFilho:2024ykw}. These achievements also opened the possibility of incorporating non--commutativity into the theory \cite{AraujoFilho:2025rvn} and motivated parallel constructions in antisymmetric tensor sectors, particularly within Kalb--Ramond gravity \cite{AraujoFilho:2025jcu}. At the same time, the influence of a fixed--norm vector field has been studied in arenas that go well beyond black hole solutions. Several works demonstrated that this background can sustain wormhole geometries or modify the criteria associated with their traversability \cite{Ovgun:2018xys,AraujoFilho:2024iox,Magalhaes:2025lti,Magalhaes:2025nql}. Additional generalizations proposed black--bounce scenarios supported by $\kappa$--essence dynamics while still maintaining Lorentz--violating effects \cite{Pereira:2025xnw}. Propagation processes formed another active branch of the literature. Neutrino deflection and related phenomena were analyzed under multiple realizations of the theory, including purely metric constructions \cite{Shi:2025plr}, \textit{metric--affine} formulations \cite{Shi:2025ywa}, and tensorial versions extending the bumblebee mechanism \cite{Shi:2025rfq}. Other aspects of neutrino physics in Lorentz--breaking backgrounds—ranging from phenomenological constraints to additional propagation features—were also studied in \cite{Khodadi:2023yiw,Khodadi:2022mzt,Khodadi:2022dff,Khodadi:2021owg}.

The catalogue of Lorentz--violating black hole geometries has grown in the past few years, particularly with the appearance of solutions constructed explicitly from different vacuum configurations of bumblebee symmetry--breaking mechanism \cite{Liu:2025oho,Zhu:2025fiy}. After these new setups were proposed, a subsequent investigation examined the static case in detail, exploring both its gravitational behavior and the bounds that restrict its physical parameters \cite{AraujoFilho:2025zaj}. The same background later served as a platform for studying neutrino dynamics, where its influence on oscillation processes was evaluated \cite{Shi:2025tvu}.
Progress did not remain confined to nonrotating spacetimes. An axisymmetric counterpart was eventually generated through a refined Newman--Janis procedure, yielding a rotating solution built directly from the static seed \cite{Kumar:2025bim}. Additional developments have extended the analysis to astrophysical environments as well: the behavior of accreting matter around this new black hole has recently been investigated and presented in \cite{Shi:2025hfe}.

{Recent years have witnessed a clear broadening of the phenomenological applications of bumblebee gravity. In the strong--gravity regime, several recent works have explored black hole configurations and constraints from Event Horizon Telescope data \cite{Afrin:2024egb,Islam:2024eht,Singh:2024greybody}. At the same time, the cosmological sector has also advanced significantly, with bumblebee gravity being applied to anisotropic Bianchi--I and Kasner--type scenarios, as well as to observational tests of the cosmic expansion history, thereby showing that Lorentz--symmetry breaking can affect both the background evolution and anisotropic features of the Universe \cite{Sarmah:2025aniso,Neves:2023kasner,Zhu:2025cosmo}. More recently, the framework has even been extended to ultra--high--energy cosmic-ray physics, where bumblebee gravity has been used to study the propagation, flux suppression, and anisotropy of UHECRs, opening a new high--energy astrophysical window for testing Lorentz--violating gravity \cite{Sarmah:2026uhecr}. }

Beyond modifying the gravitational sector or introducing additional couplings, gravity can also shape cosmic evolution through quantum processes that arise solely from spacetime curvature. In a nonflat background, the very notion of a vacuum loses its universality: different observers identify distinct sets of modes, and a state that appears empty to one may contain excitations for another. This feature of quantum field theory in curved spacetime laid the foundation for what later became known as gravitationally induced particle production. Parker’s pioneering work in the late 1960s revealed that a time--dependent geometry does not preserve the particle content of the field, allowing quanta to emerge purely because the spacetime metric evolves in time \cite{Parker:1968mv,Parker:1969au}. The phenomenon manifests through Bogoliubov transformations that relate inequivalent vacuum states associated with different cosmological epochs. As a consequence, the background geometry can transfer energy into quantum fields, effectively creating matter or radiation \cite{dewitt1975quantum,wald1975particle,fulling1989aspects,lin2010quantum,calzetta1989dissipation}. This mechanism has played a central role in scenarios describing the early Universe, where rapid expansion naturally fosters particle generation. In several cosmological models, the same effect behaves as an additional contribution to the evolution equations, and under appropriate conditions, it can reproduce an accelerated expansion phase without invoking exotic fluids or modifying the fundamental gravitational action \cite{wald1994quantum}.

One of the most striking consequences of quantum fields evolving on curved backgrounds emerges not in cosmology but in the environment surrounding black holes. Hawking’s analysis in the 1970s revealed that horizons fundamentally modify the behavior of vacuum fluctuations \cite{Hawking:1974rv,Hawking:1975vcx}. When a field is quantized on a stationary spacetime containing an event horizon, observers at infinity and observers near the horizon no longer agree on what constitutes the vacuum. This mismatch produces a continuous outflow of particles detectable far from the black hole. The radiation associated with this mechanism carries a thermal spectrum whose temperature decreases as the black hole mass grows. Once this effect was established, it became clear that black holes cannot remain perfectly cold objects; instead, they behave as thermodynamic systems. The assignment of entropy proportional to the area of the event horizon and the existence of a nonzero temperature connected quantum theory, gravity, and statistical mechanics in an unexpected way. The framework that emerged from these results formed the basis of black hole thermodynamics and reshaped the conceptual picture of gravitational systems \cite{Gibbons:1977mu}.

The phenomena of cosmological particle production and Hawking radiation, though often discussed in separate contexts, trace back to a common principle: quantum fields respond directly to the structure of spacetime itself. Quantum field theory on curved backgrounds established that the geometry can influence the very notion of particles, leading to observable effects in situations where the metric evolves in time or possesses horizons \cite{Birrell:1982ix,Parker:2009uva}. In an expanding Universe, the absence of a single global vacuum allows time-dependent metrics to generate quanta, effectively channeling energy from the gravitational sector into matter. This behavior parallels, at a conceptual level, the appearance of effective interactions between curvature and the matter content in models with nonminimal couplings. In contrast, for black holes, it is the causal structure introduced by the event horizon that shapes the particle content seen by distant observers, giving rise to the thermal radiation identified by Hawking \cite{hawking1974black}.

This study addresses the semiclassical radiation and quantum processes associated with a new static black hole produced by a Lorentz--violating bumblebee background. The discussion first reconstructs the spacetime geometry and determines the thermal behavior of the solution through the surface--gravity approach, followed by an examination of its thermodynamic topological features. Subsequently, quantum creation of particles is explored for both bosonic and fermionic sectors by employing the tunneling framework. From the corresponding effective potentials, analytic bounds on the greybody factors are established for fields with spins 0, 1, 2, and 1/2. The full transmission spectra are then obtained via the sixth--order WKB method, which also yields the absorption cross sections and the characteristic spin--dependent structures that accompany them. These results allow the computation of emission rates and evaporation lifetimes for each spin contribution and enable a high--frequency comparison with several Lorentz--violating backgrounds, such as the \textit{metric} and \textit{metric--affine} bumblebee geometries, as well as Kalb--Ramond and non--commutative Kalb--Ramond black holes. Finally, an alternative estimation of greybody factors is presented through a prescription based on quasinormal modes.


\section{Overview of the black hole geometry}

{ We consider a vector--tensor gravitational theory in which Lorentz symmetry is not imposed at the level of the vacuum. This feature arises from the presence of a vector field $B_\mu$ whose ground state selects a preferred direction in spacetime. The dynamics of the system is governed by the action~\cite{Casana:2017jkc}
\begin{align}
S=&\int \mathrm{d}^{4}x\,\sqrt{-\mathrm{g}}\left[
\frac{1}{2\kappa}R
+\frac{\Tilde{\xi}}{2\kappa}B^{\mu}B^{\nu}R_{\mu\nu}
-\frac{1}{4}B^{\mu\nu}B_{\mu\nu}
-V(B^{\mu}B_{\mu}\pm b^{2})
\right] +S_{\mathrm{m}},
\label{action}
\end{align}
where $B_{\mu\nu}=\partial_{\mu}B_{\nu}-\partial_{\nu}B_{\mu}$, $\kappa=8\pi G$, and $\Tilde{\xi}$ controls the strength of the nonminimal interaction between the vector field and the geometry. The potential $V$ fixes the norm of the field through its extremum, enforcing the condition $B^{\mu}B_{\mu}=\mp b^{2}$,
with $b$ a positive constant. This constraint implies that the vacuum configuration is characterized by a nonvanishing expectation value $\langle B_{\mu}\rangle=b_{\mu},
\,\,
b_{\mu}b^{\mu}=\mp b^{2}$,
which explicitly selects a preferred direction and therefore breaks Lorentz invariance at the level of the background.

In what follows, we focus on vacuum configurations in which no additional matter sources are present, {\it i.e.} $(T_m)_{\mu\nu}=0$, and the vector field is frozen at its vacuum value. Under these conditions, the gravitational equations take the form
\begin{equation}
R_{\mu\nu}-\frac{1}{2}R\,\mathrm{g}_{\mu\nu}
=\kappa\,T^{B}_{\mu\nu},
\end{equation}
where the effective energy--momentum tensor generated by the vector background reads
\begin{equation}
\begin{aligned}
T_{\mu\nu}^{B}=&-B_{\mu\alpha}B_{\nu}^{\ \alpha}
-\frac{1}{4}B_{\alpha\beta}B^{\alpha\beta}\mathrm{g}_{\mu\nu} {- V \mathrm{g}_{\mu\nu} + 2 V^{\prime} B_{\mu} B_{\nu} }
\\
&+\frac{{\Tilde{\xi}}}{\kappa}\left[
\frac{1}{2}B^{\alpha}B^{\beta}R_{\alpha\beta}\mathrm{g}_{\mu\nu}
-B_{\mu}B^{\alpha}R_{\alpha\nu}
-B_{\nu}B^{\alpha}R_{\alpha\mu}
\right.
\\
& \left. +\frac{1}{2}\nabla_{\alpha}\nabla_{\mu}(B^{\alpha}B_{\nu})
+\frac{1}{2}\nabla_{\alpha}\nabla_{\nu}(B^{\alpha}B_{\mu}) \right.
\\
& \left. -\frac{1}{2}\nabla^{2}(B_{\mu}B_{\nu})
-\frac{1}{2}\mathrm{g}_{\mu\nu}\nabla_{\alpha}\nabla_{\beta}(B^{\alpha}B^{\beta}) \right].
\end{aligned}
\end{equation}
In addition, the vector field satisfies
\begin{equation}
\nabla^{\mu}B_{\mu\nu}+\frac{{\Tilde{\xi}}}{\kappa}B^{\mu}R_{\mu\nu}=0.
\end{equation}

To construct static and spherically symmetric configurations, one considers a background of the form
\begin{equation}
b_{\mu}=(b_{t}(r),\,b_{r}(r),\,0,\,0),
\end{equation}
{Substituting these expressions together with a general line element for the metric (see \cite{Zhu:2025fiy} for further details) into the field equations and imposing the fixed-norm condition for $b_{\mu}$ leads to a family of static solutions.}

}

{Moreover,} a new bumblebee black hole geometry presented in Refs.~\cite{Zhu:2025fiy} arises from a static solution whose line element differs from both the Schwarzschild metric and the earlier bumblebee configuration of Ref.~\cite{Casana:2017jkc}. The deviation is produced by the background vector field that triggers Lorentz--symmetry breaking, which in turn depends on the particular choice of vacuum expectation value $b_{\mu}$. This setup defines the spacetime adopted in the present analysis. Accordingly, {for the general case of $\Tilde{\xi}$ and $b$ (Case I of Ref. \cite{Zhu:2025fiy})}, the metric takes the form
\ie
\begin{split}
\label{maaaaianametric}
\mathrm{d}s^{2} & = {\mathrm{g}_{\mu\nu} \mathrm{d}x^{\mu} \mathrm{d}x^{\nu} = } { - A(r,\chi) \mathrm{d}t^{2} + \frac{1}{B(r,\chi)}\mathrm{d}r^{2} + r^{2}\mathrm{d}\Omega^{2}    } \\
& = - \frac{1}{1+\chi}\left(1 - \frac{2M}{r}         \right)\mathrm{d}t^{2} + \frac{1+\chi}{\left(1 - \frac{2M}{r} \right)} \mathrm{d}r^{2} + r^{2}\mathrm{d}\Omega^{2},
\end{split}
\fe
{where $\mathrm{d}\Omega \equiv  \mathrm{d}\theta^{2} + \sin \theta^{2} \mathrm{d} \varphi^{2}  $. As will be used in the next section, we define $A(r,\chi)\equiv \mathrm{g}_{tt}$ as the temporal component of the metric and $1/B(r,\chi)\equiv \mathrm{g}_{rr}$ as its radial component.}

In this spacetime, the parameter $\chi$ introduces the deviation from standard Lorentz symmetry and is defined through the combination $\chi=\alpha\,\ell$. The constant $\alpha$ arises from the integration of the field equations, while $\ell=\Tilde{\xi}\,b^{2}$ incorporates both the nonminimal coupling $\Tilde{\xi}$ and the fixed norm of the bumblebee field, $b^{2}=b_{\mu}b^{\mu}$.

At first glance, the constant factor modifying the temporal component of the metric might suggest that a rescaling of the time coordinate could absorb the term $1/(1+\chi)$ in $g_{tt}$, leaving the Lorentz–violating effects to appear only in $g_{rr}$, in analogy with the metric previously obtained in Ref.~\cite{Casana:2017jkc}. Such a procedure, however, does not hold once the structure of the theory is examined more carefully. As emphasized in Ref.~\cite{Shi:2025hfe}, the form of the metric is intertwined with the vacuum configuration of the bumblebee field. The background vector that triggers the symmetry breaking must satisfy a prescribed norm, and this requirement fixes the admissible forms of $b_{\mu}$ compatible with the black hole solution.

Because of this constraint, redefining the time coordinate would not merely shift a constant in $g_{tt}$; it would also alter the components of $b_{\mu}$, thereby changing the vacuum configuration on which the solution rests. Since the metric and the vector background must be solved simultaneously, any such modification would generate a different spacetime altogether. A schematic discussion of this point is presented in Sec.~II of Ref.~\cite{Shi:2025hfe}.

{In addition, the observational relevance of the present black hole solution is tied to the fact that the Lorentz--violating parameter $\chi$ directly modifies the semiclassical quantities that determine the radiative response of the geometry. More specifically, as will be shown in the following sections, $\chi$ affects the Hawking temperature, the particle densities obtained from the tunneling method, the greybody factors, the absorption cross sections, the evaporation lifetime, and the energy and particle emission rates. These quantities constitute the primary physical channels through which the geometry may be probed in future analyses of evaporating black holes.}


\section{Thermodynamics}

In this section, we turn to the thermal properties of the new bumblebee black hole introduced earlier. We derive the Hawking temperature—which sets the scale for the particle and energy emission rates and enters directly in the evaluation of the evaporation lifetime through the Stefan--Boltzmann law—as well as the topological temperature. These results will later be compared with those obtained from the analysis of quantum radiation in the subsequent section.


\subsection{Hawking temperature }

The spacetime in Eq.~(\ref{maaaaianametric}) admits a Killing symmetry along the temporal direction, encoded in the vector field $\xi^{\mu}=\partial_{t}$. The existence of this symmetry ensures the presence of a conserved quantity associated with the motion of test particles or fields. Making use of this Killing vector, one can introduce the corresponding invariant quantity through the relation:
\ie
\nabla^\nu(\xi^\mu\xi_\mu) = -2\kappa\xi^\nu.
\fe

In this case, $\nabla_{\nu}$ denotes the covariant derivative. The quantity $\kappa$ does not vary along the integral curves generated by $\xi^{\mu}$; in other words, it stays constant on the flow of the Killing field. This property is taken into account by the vanishing of its Lie derivative along $\xi^{\mu}$:
\ie
\mathcal{L}_\xi\kappa = 0.
\fe
The quantity $\kappa$ takes the same value at every point on the horizon and is identified with the surface gravity of the black hole. When written in the coordinate basis, the components of the timelike Killing field assume the form $\xi^{\mu}=(1,0,0,0)$. With this vector, the corresponding expression for the surface gravity can be written as:
\ie 
\kappa = {\left.\frac{f^{\prime}(r)}{2} \right|_{r = {r_{h}}}}.
\fe

In this expression, the function $f(r)$ stands for $\left(1-\frac{2M}{r}\right)/(1+\chi)$. Moreover, Hawking’s original analysis \cite{hawking1975particle} established that a black hole behaves as a thermal emitter, and the temperature associated with this phenomenon is determined by the relation $T_{H}=\kappa/(2\pi)$.
Alternatively, one may introduce the notation
$A(r,\chi)=\left(1-\frac{2M}{r}\right)/(1+\chi)$ and
$B(r,\chi)=\left(1-\frac{2M}{r}\right)/(1+\chi)$
for the metric functions.

\begin{figure}
    \centering
     \includegraphics[scale=0.55]{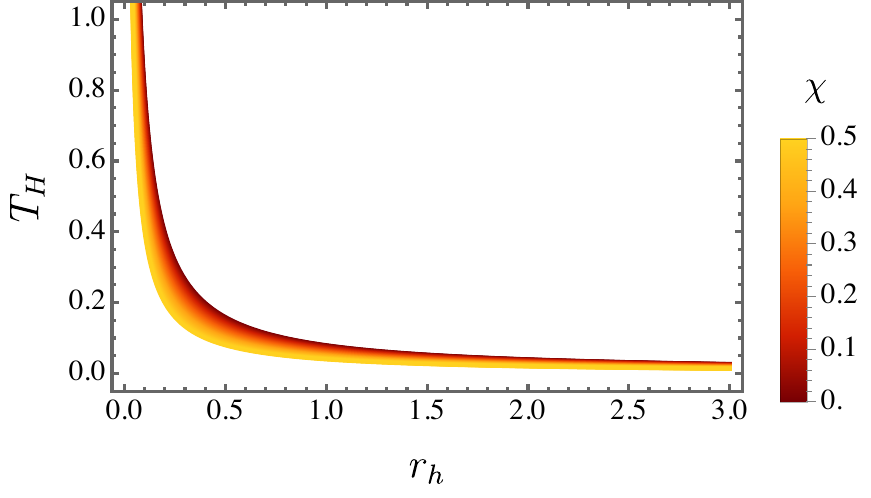}
     \includegraphics[scale=0.55]{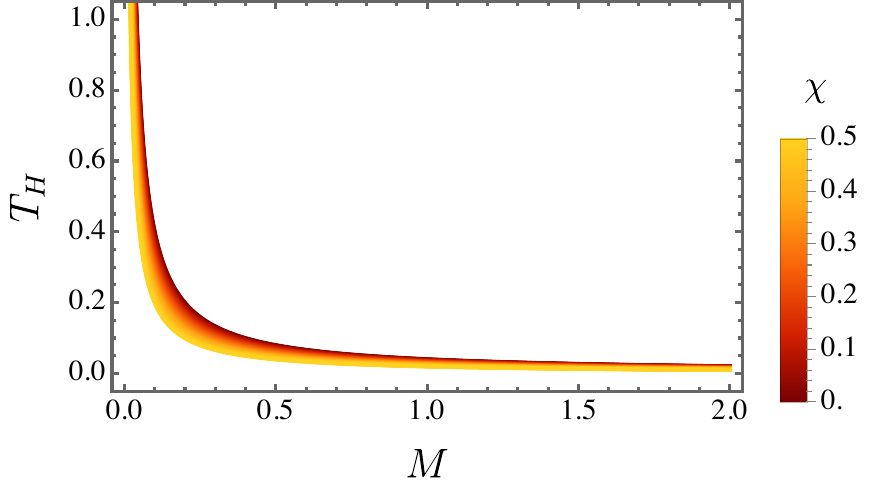}
    \caption{Hawking temperature $T_{H}$ as a function of $r_{h}$ (left panel) and $M$ (right panel), computed for several values of $\chi$.}
    \label{hawtemphorizon}
\end{figure}

When the surface--gravity prescription is applied to these components, the corresponding Hawking temperature takes the following form:
\ie
\label{hawviasgra}
T_{H} = \frac{1}{4\pi} \frac{1}{\sqrt{A(r,\chi),B^{-1}(r,\chi)}} \frac{\mathrm{d}}{\mathrm{d}r} \Big[ A(r,\chi)  \Big] \Bigg|_{r=r_{h}}  =  \frac{1}{4 \pi  r_{h} (\chi +1)} \approx \, \frac{1}{4 \pi  r_{h}}-\frac{\chi }{4 (\pi  r_{h})},
\fe
in which it is expanded only to first order in the Lorentz--violating parameter $\chi$, and $r_{h}$ denotes the event horizon. Moreover, rewriting Eq.~(\ref{hawviasgra}) in terms of the black hole mass is straightforward: inserting $r_{h}=2M$ into the result yields the Hawking temperature expressed as a function of $M$ in the form:
\ie
T_{H} =  \frac{1}{8 \pi  M (\chi +1)} \approx \, \frac{1}{8 \pi  M}-\frac{\chi }{8 (\pi  M)}.
\label{tmsss}
\fe
As will become clear in the discussion of the evaporation process, expressing $T_{H}$ in terms of the black hole mass is essential for determining the evaporation lifetime. Figure~\ref{hawtemphorizon} shows the Hawking temperature obtained from the surface gravity for several choices of $\chi$, displayed both as a function of the horizon radius $r_{h}$ (on the left panel) and of the mass $M$ (on the right panel). In each case, the parameter $\chi$ lowers $T_{H}$.

A natural question at this stage is whether the geometry in Eq.~(\ref{maaaaianametric}) admits a remnant mass. To test this possibility, we substitute Eq.~(\ref{tmsss}) into the condition $T_{H}=0$ and solve for $M$. The result is $M=0$, which indicates that this black hole does not develop a remnant in this formulation. It is also worth noting that neither the entropy nor the heat capacity will be examined here, as these quantities remain unchanged by the parameter $\chi$. For the new bumblebee black hole, both reduce to the standard Schwarzschild expressions.

As we shall confirm in the next section, the expression obtained in Eq.~(\ref{hawviasgra}) will be confronted with the temperature derived through an independent approach based on the quantum tunneling method.


\subsection{Topological Framework for Thermodynamic Criticality}

Topological methods have recently emerged as a powerful framework for characterizing phase transitions in black hole thermodynamics. Inspired by Duan’s $\phi$–mapping topological current theory \cite{Duane1984}, these approaches identify thermodynamic critical points as topological defects in a parameter space, with their nature determined by associated topological charges \cite{wei2022black,yerra2022topology,wu2023topological,wu2023topological1,zhang2023bulk,gogoi2023thermodynamic,fan2023topological}. Within this formulation, a scalar thermodynamic potential generates a two-dimensional vector field whose zeros encode potential phase transitions, while the associated winding numbers provide a topological classification of the critical behavior.

Using the Hawking temperature obtained in Eq.~\eqref{hawviasgra}, we introduce the thermodynamic potential
\begin{equation}
\Phi(r_h,\theta)=\frac{1}{\sin\theta}T_H
=\frac{\csc\theta}{4\pi r_h(1+\chi)}.
\end{equation}
The coordinates $(r_h,\theta)$ constitute a two--dimensional thermodynamic manifold on which the gradient of $\Phi$ defines the vector components
\begin{align}
\phi^{r_h} &= \frac{\partial \Phi}{\partial r_h}
= -\frac{\csc\theta}{4\pi r_h^{2}(1+\chi)},\\
\phi^{\theta} &= \frac{\partial \Phi}{\partial \theta}
= -\frac{\cot\theta\csc\theta}{4\pi r_h(1+\chi)}.
\end{align}
 To analyze the topology of this field, we extend the coordinates to the unit vector field as
\begin{equation}
n^{r_h}=\frac{\phi^{r_h}}{|\phi|},\qquad
n^{\theta}=\frac{\phi^{\theta}}{|\phi|},
\end{equation}
with $|\phi|=\sqrt{(\phi^{r_h})^{2}+(\phi^{\theta})^{2}}$, maps each point of the thermodynamic plane to the unit circle in the internal space.
The field topology is captured by the Duan topological current
\begin{equation}
j^{\mu}
=\frac{1}{2\pi}
\epsilon^{\mu\nu\lambda}\epsilon_{ab}
\partial_\nu n^{a}
\partial_\lambda n^{b},
\label{eq:jmu}
\end{equation}
where $\mu$, $\nu$, and $\lambda$ take the number: $0$, $1$, $2$ and $a$, $b$ are $r_h$, $\theta$. This current satisfies the conservation law
\begin{equation}
\partial_{\mu} j^{\mu}=0.
\end{equation}
The topological charge at a parameter region $\Sigma$ is computed as the spatial integral of the zeroth component $(j^0)$ of the associated topological current
\begin{equation}
Q=\frac{1}{2\pi}\int_{\Sigma} j^{0} \mathrm{d}^{2}x
=\sum_{i} \alpha_i\eta_i=\sum_{i} w_i .
\label{eq:Q}
\end{equation}
Here, $\alpha_i$ denotes the Hopf index and $\eta_i$ the Brouwer degree evaluated at the zero point. $\omega_i$ is the winding number associated with the $i$-th zero of the vector field. For $Q$, a value of $+1$ corresponds to a conventional (stable) critical point, $-1$ to an unstable or novel one, and $0$ indicates the absence of a thermodynamic phase transition.

\begin{figure}[t]
\centering
\includegraphics[scale=0.55]{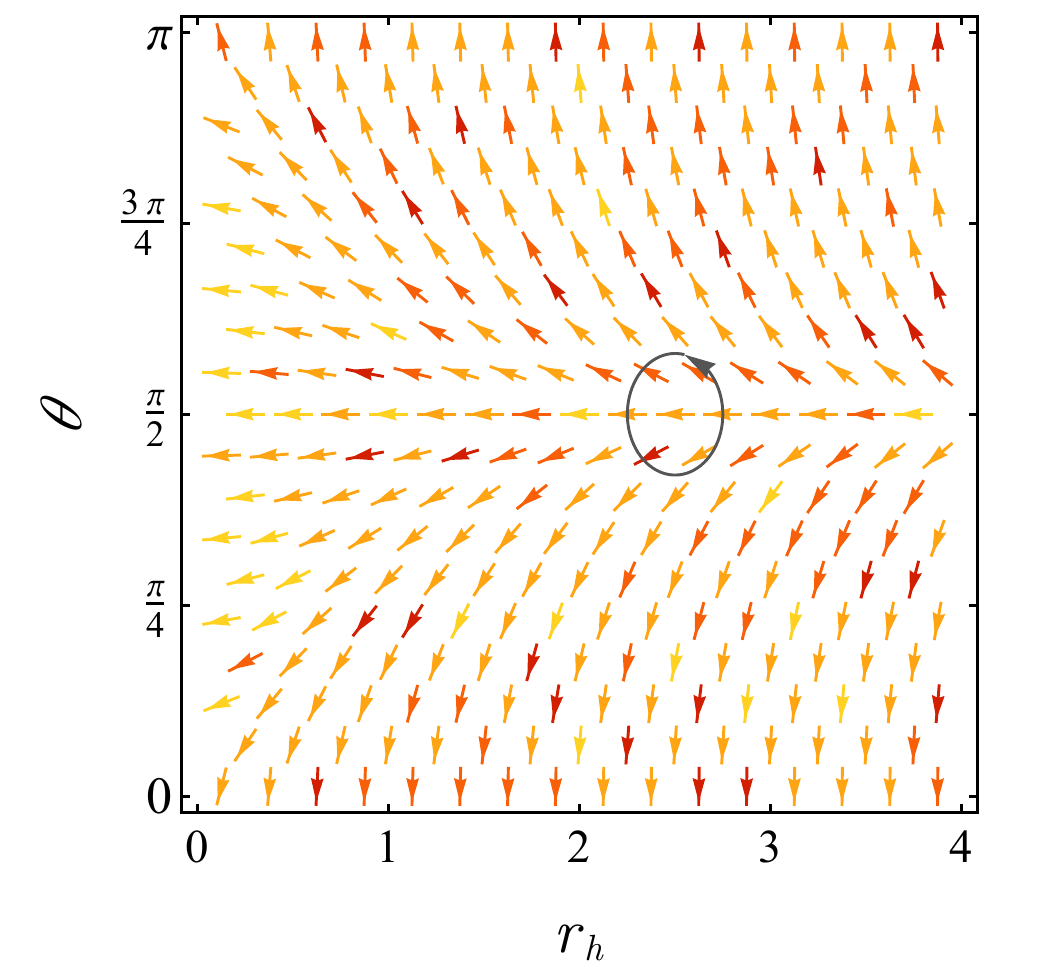}
\caption{
Normalized vector field derived from the thermodynamic potential in the $(r_h,\theta)$ plane for $M=1$ and $\chi=0.1$.
The field does not vanish anywhere, leading to a vanishing topological charge ($Q=0$) and confirming the absence of thermodynamic critical points.}
\label{TopologicalT}
\end{figure}

The normalized vector field for the present black hole is displayed in Fig.~\ref{TopologicalT}. The field is smooth across the entire $(r_h,\theta)$ domain and exhibits no zeroes. Consequently, any closed contour yields
$Q=0$, demonstrating that the system possesses no thermodynamic critical points. This result is consistent with the monotonic behavior of the Hawking temperature represented in Fig. \ref{hawtemphorizon}, and confirms that the black hole does not undergo a phase transition within this topological framework.

\section{Quantum particle production}


This part of the work addresses the mechanism of particle production in the recently obtained bumblebee black hole. The discussion starts with the bosonic sector, where the tunneling approach is applied. To handle the horizon behavior, the line element is first rewritten in Painlevé--Gullstrand coordinates, which eliminate the coordinate singularity at $r_{h}$. After this transformation, the relevant integrals—particularly the imaginary contribution to the classical action, $\text{Im}\,\mathcal{S}$—are evaluated through the residue technique, allowing one to extract the corresponding bosonic particle density $n$. The procedure follows the strategy outlined in Ref.~\cite{calmet2023quantum}.

The analysis proceeds by examining fermionic emission within the same tunneling framework. Here, the near--horizon expansion is adopted to streamline the computation and obtain the fermionic density $n_{\psi}$. The treatment of the spinor sector is based on the methods presented in Ref.~\cite{vanzo2011tunnelling}.

\subsection{Bosonic perturbations }


\subsubsection{Thermal radiation }

Hawking’s analysis in Ref.~\cite{hawking1975particle} focused on the behavior of a scalar field and introduced the following expression for its wave function, $\Psi$:
\ie
\frac{1}{\sqrt{-\mathrm{g}}}\partial_{\mu}(\mathrm{g}^{\mu\nu}\sqrt{-\mathrm{g}} \, \partial_{\nu}\Psi) = 0.
\fe
It is immediate to check that the metric tensor $\mathrm{g}$ used here corresponds to the newly obtained bumblebee black hole geometry. In that context, the associated field operator is written as:
\ie
\Psi = \sum_{i} \left (\mathrm{f}_{i}  a_i + \bar{\mathrm{f}}_{i} a^{\dagger}_{i} \right) = \sum_{i} \left( \mathrm{p}_{i} b_{i} + \bar{\mathrm{p}}_{i} b^{\dagger}_{i} + \mathrm{q}_{i}  c_{i} + \bar{\mathrm{q}}_{i}  c^{\dagger}_{i} \right ) .
\fe

Within this setting, the functions $\mathrm{f}_{i}$ and $\bar{\mathrm{f}}_{i}$ (the latter being their complex conjugates) correspond to modes that propagate exclusively toward the black hole. In contrast, $\mathrm{p}_{i}$ and $\bar{\mathrm{p}}_{i}$ describe modes that move purely outward, while $\mathrm{q}{i}$ and $\bar{\mathrm{q}}_{i}$ encode solutions without any outgoing component. The coefficients $a_{i}$, $b_{i}$, and $c_{i}$ act as annihilation operators, and $a_{i}^{\dagger}$, $b_{i}^{\dagger}$, and $c_{i}^{\dagger}$ serve as the associated creation operators.
The aim of this discussion is to show that all these mode functions—$\mathrm{f}_{i}$, $\bar{\mathrm{f}}_{i}$, $\mathrm{p}_{i}$, $\bar{\mathrm{p}}_{i}$, $\mathrm{q}_{i}$, and $\bar{\mathrm{q}}_{i}$—are altered when Lorentz violation is present. In other words, the analysis focuses on identifying how the Lorentz--violating parameter reshapes the structure of the modes originally introduced in Hawking’s treatment.

Because the new bumblebee black hole preserve spherical symmetry, the ingoing and outgoing field modes can be decomposed using spherical harmonics. In the exterior region of the black hole, this decomposition allows one to express the corresponding wave solutions in the form \cite{araujo2025particleasdasd,calmet2023quantum,araujo2025does,AraujoFilho:2025rwr}:
\ie
\begin{split}
f_{\omega^\prime l m} & =  \frac{1}{\sqrt{2 \pi \omega^\prime} r }  \mathcal{F}_{\omega^\prime}(r) e^{i \omega^\prime v} Y_{lm}(\theta,\phi)\ , \\ 
p_{\omega l m} & = \frac{1}{\sqrt{2 \pi \omega} r }  \mathcal{P}_\omega(r) e^{i \omega u} Y_{lm}(\theta,\phi). 
\end{split}
\fe
{Here, $\mathcal{F}_{\omega'}(r)$ and $\mathcal{P}_{\omega}(r)$ denote the radial mode functions associated with the ingoing and outgoing solutions of the scalar field, respectively.}  In this setting, it is convenient to introduce the advanced and retarded null coordinates, denoted by $v$ and $u$, respectively. For the spacetime under study, these coordinates take the form:
\ie
v = t + r^{*} = t  + r (1 + \chi) + 2 (1 + \chi) \, M \ln |r - 2M|,
\fe
and
\ie
u = t - r^{*} = t - r (1 + \chi) - 2 (1 + \chi) \, M \ln |r - 2M|,
\fe
{where $r^{*}$ denotes the tortoise coordinate, defined as
\ie
\mathrm{d}r^{*} = \frac{\mathrm{d}r}{\sqrt{A(r,\chi)B(r,\chi)}}.
\fe}

A practical route to identify how Lorentz--violating effects enter through these coordinate functions is to track the motion of a test particle following a geodesic of the background geometry. By introducing an affine parameter $\Tilde{\lambda}$ along the particle’s path, its momentum can be written as:
\ie
p_{\mu}=\mathrm{g}_{\mu\nu}\frac{\mathrm{d}x}{\mathrm{d}\Tilde{\lambda}}^\nu.
\fe
As expected, the momentum remains constant as the particle proceeds along its geodesic path. In this description, one also adopts the relation:
\ie
\mathcal{L} = \mathrm{g}_{\mu\nu} \frac{\mathrm{d}x^\mu}{\mathrm{d}\Tilde{\lambda}} \frac{\mathrm{d}x^\nu}{\mathrm{d}\Tilde{\lambda}}.
\fe
Notice that such a quantity remains fixed along any geodesic. For particles with mass, one sets $\mathcal{L}=-1$ and identifies $\tilde{\lambda}$ with the proper time $\tau$. In contrast, massless particles—the case of interest here—are described using an arbitrary affine parameter $\tilde{\lambda}$. Adopting a stationary and spherically symmetric background, and restricting the motion to radial null geodesics by imposing $p_{\varphi}=L=0$ and $\theta=\pi/2$, the corresponding relations take the form:
\ie
E =  A(r,\chi) \dot{t}.
\fe
In this setup, the quantity $E$ is introduced through the identification $p_{t}=-E$, while an overdot indicates differentiation with respect to the affine parameter $\tilde{\lambda}$. When these ingredients are combined with the previously stated geodesic relations, one arrives at the expression:
\ie
\label{fgfgfg}
    \left( \frac{\mathrm{d}r}{\mathrm{d}\Tilde{\lambda}} \right)^2 = \frac{E^2}{A(r,\chi)B(r,\chi)^{-1}}.
\fe

After carrying out the corresponding algebraic manipulations, the expression can be rewritten in the form
\ie
\frac{\mathrm{d}}{\mathrm{d}\Tilde{\lambda}}\left(t\mp r^{*}\right) = 0.
\fe
Rewriting the relation that defines the retarded coordinate yields
\ie
\label{nmnma}
\frac{\mathrm{d}u}{\mathrm{d}\Tilde{\lambda}}=\frac{2E}{A(r,\chi)}.
\fe

When analyzing an ingoing null geodesic labeled by the affine parameter $\tilde{\lambda}$, the retarded coordinate is regarded as a function of this parameter, $u(\tilde{\lambda})$. Rather than introducing this relation directly, one begins by determining how the radial position evolves along the geodesic; once $r(\tilde{\lambda})$ is known, the expression governing $u$ follows from the integral given in Eq.~\eqref{nmnma}. The structure of $u(\tilde{\lambda})$ is crucial, since it controls the form of the Bogoliubov coefficients that encode the quantum emission spectrum of the black hole.
The next step makes use of the metric functions $A(r,\chi)$ and $B(r,\chi)$. The integral involving the square root in Eq.~\eqref{fgfgfg} is then evaluated by integrating from the event horizon $r_{h}$ to a generic radius $r$, which corresponds to the interval $\tilde{\lambda}\in[0,\tilde{\lambda}]$ along the geodesic. Carrying out this procedure leads to the expression:
\ie
\label{rvv}
r = 2M - E\Tilde{\lambda}.
\fe
This expression follows from selecting the minus branch of the square root in Eq.~\eqref{fgfgfg}, which corresponds to a geodesic directed inward toward the horizon.

Substituting the radial trajectory $r(\tilde{\lambda})$ into the integral allows it to be evaluated, yielding
\ie
u(\Tilde{\lambda},\chi) = -4 (1 + \chi)\, M \ln \left( \frac{\Tilde{\lambda}}{C} \right).
\fe

The resulting expression contains an integration constant, here labeled $C$. To relate this solution to the null coordinates used for ingoing and outgoing rays, one invokes the geometric--optics correspondence between the affine parameter and the advanced coordinate. In this description, the parameter $\tilde{\lambda}$ is written as
$$
\tilde{\lambda}=\frac{v_{0}-v}{D},
$$
where $v_{0}$ marks the value of the advanced coordinate at the point where the ray meets the horizon (corresponding to $\tilde{\lambda}=0$), and $D$ is a positive constant setting the proportionality scale \cite{calmet2023quantum}.

With the preparatory relations in place, one can now turn to the modes that propagate outward. Solving the Klein--Gordon equation in the presence of the Lorentz--violating parameter $\ell$ yields outgoing solutions whose structure differs from the standard case. These modes can be written as follows:
\ie
p_{\omega} =\int_0^\infty \left ( \alpha_{\omega\omega^\prime} f_{\omega^\prime} + \beta_{\omega\omega^\prime} \bar{ f}_{\omega^\prime}  \right)\mathrm{d} \omega^\prime.
\fe
Here, the quantities $\alpha_{\omega\omega'}$ and $\beta_{\omega\omega'}$ are the Bogoliubov coefficients \cite{fulling1989aspects,hollands2015quantum,parker2009quantum,wald1994quantum}.
\begin{equation}
\begin{split}
    \alpha_{\omega\omega^\prime}=& -i K e^{i\omega^\prime v_0}e^{\pi \left[2 M (1+\chi) \right]\omega} \int_{-\infty}^{0} \,\mathrm{d}x\,\Big(\frac{\omega^\prime}{\omega}\Big)^{1/2}e^{\omega^\prime x}  \times e^{i\omega\left[4M(1+\chi)\right]\ln\left(\frac{|x|}{CD}\right)},
    \end{split}
\end{equation}
and
\begin{equation}
\begin{split}
    \beta_{\omega\omega'} &= i Ke^{-i\omega^\prime v_0}e^{-\pi \left[2 M (1+\chi) \right]\omega}
    \int_{-\infty}^{0} \,\mathrm{d}x\,\left(\frac{\omega^\prime}{\omega}\right)^{1/2}e^{\omega^\prime x} \times e^{i\omega\left[4M(1+\chi)\right]\ln\left(\frac{|x|}{CD}\right)}.
    \end{split}
\end{equation}

The appearance of $\chi$ inside the mode functions indicates that the Lorentz--violating sector modifies the amplitude associated with particle creation. In this picture, the altered spacetime structure allows for a channel through which quantum information can emerge from the black hole.
Even though the amplitude is affected by these corrections, the resulting radiation spectrum still exhibits a thermal character. To verify this point, one evaluates the quantity:
\ie
|\alpha_{\omega\omega'}|^2 = e^{\big[8\pi M (1+\chi) \big]\omega}|\beta_{\omega\omega'}|^2\,.
\fe
To isolate the portion of the radiation carried by modes of frequency near $\omega$, one inspects the flux contained in the infinitesimal interval $[\omega,\omega+\mathrm{d}\omega]$ \cite{o10}.
This calculation gives
\ie
\mathcal{P}(\omega, \chi)=\frac{\mathrm{d}\omega}{2\pi}\frac{1}{\left \lvert\frac{\alpha_{\omega\omega^\prime}}{\beta_{\omega\omega^\prime}}\right \rvert^2-1}\, ,
\fe
or, therefore, 
\ie
\mathcal{P}(\omega, \chi)=\frac{\mathrm{d}\omega}{2\pi}\frac{1}{e^{\left[8\pi M (1+\chi)\right]\omega}-1}\,.
\fe
A noteworthy feature becomes apparent upon confronting the obtained formula with the Planck distribution: it shows that
\ie
\mathcal{P}(\omega, \chi)=\frac{\mathrm{d}\omega}{2\pi}\frac{1}{e^{\frac{\omega}{T}}-1}.
\fe
From this standpoint, the resulting expression becomes
\ie
\label{hawtempmetricase}
    T = \frac{1}{8 \pi  (1+\chi) M}.
\fe 

Remarkably, notice that the temperature extracted from the particle--production analysis coincides exactly with the value obtained from the surface--gravity prescription in Eq.~(\ref{hawviasgra}), which confirms the consistency of both approaches.

This outcome indicates that black holes described by Lorentz--violating geometries radiate with an effective temperature $T$ given by Eq.~\eqref{hawtempmetricase}, resembling the behavior of a greybody spectrum. Up to this point, however, energy conservation during the emission process has not been incorporated. Each quantum of radiation reduces the black hole mass, altering its geometry over time. To include this backreaction, the following section adopts the tunneling picture developed by Parikh and Wilczek \cite{011}, which provides a dynamical framework for studying quantum tuneling process.


\subsubsection{Quantum tunneling method }

To incorporate energy conservation into the emission process, the analysis follows the tunneling framework developed in Refs.~\cite{011,vanzo2011tunnelling,parikh2004energy,calmet2023quantum}. The first step is to recast the geometry in Painlevé–Gullstrand form, for which the line element becomes
\ie
\mathrm{d}s^{2}=-A(r,\chi)\,\mathrm{d}t^{2}+2H(r,\chi)\,\mathrm{d}t\,\mathrm{d}r+\mathrm{d}r^{2}+r^{2}\mathrm{d}\Omega^{2},
\fe
with
\ie
H(r,\chi)=\sqrt{A(r,\chi)\left(B(r,\chi)^{-1}-1\right)}.
\fe
Within this coordinate system, the tunneling amplitude is governed by the imaginary component of the classical action, as emphasized in \cite{parikh2004energy,vanzo2011tunnelling,calmet2023quantum}.

The action for a particle traveling through a generic curved background is written as
\ie
\mathcal{S}=\int p_{\mu}\,\mathrm{d}x^{\mu}.
\fe
When isolating the imaginary part, only the radial contribution survives. The temporal term, $p_{t}\mathrm{d}t=-\omega\,\mathrm{d}t$, is purely real and therefore plays no role in $\mathrm{Im}\,\mathcal{S}$. Consequently, one is left with
\ie
\text{Im}\,\mathcal{S} = \text{Im}\,\int_{r_i}^{r_f} \,p_r\,\mathrm{d}r=\text{Im}\,\int_{r_i}^{r_f}\int_{0}^{p_r} \,\mathrm{d}p_r'\,\mathrm{d}r.
\fe

Starting from the Hamiltonian description in which the system evolves with
$ H = M - \omega'$,
the variation of the Hamiltonian follows directly from Hamilton’s equations. Since the emitted particle carries an energy $\omega'$ that ranges between $0$ and the total emission energy $\omega$, one finds $\mathrm{d}H = -\,\mathrm{d}\omega'$.
With this identification, the expression for the tunneling contribution becomes
\ie
\begin{split}
\text{Im}\, \mathcal{S} & = \text{Im}\,\int_{r_i}^{r_f}\int_{M}^{M-\omega} \,\frac{\mathrm{d}H}{\mathrm{d}r/\mathrm{d}t}\,\mathrm{d} r = \text{Im}\,\int_{r_i}^{r_f}\,\mathrm{d}r\int_{0}^{\omega} \,-\frac{\mathrm{d}\omega'}{\mathrm{d}r/\mathrm{d}t}\,.
\end{split}
\fe
After rearranging the integral and implementing the variable transformation
\ie
\frac{\mathrm{d}r}{\mathrm{d}t} = -H(r,\chi)+\sqrt{A(r,\chi) + H(r,\chi)^2} = 1 - \sqrt{\frac{\Delta(r,\omega^{\prime})}{r}}, 
\fe
and, for the sake of convenience, we define:
\ie
\Delta(r,\omega^{\prime})= \frac{\chi r + 2(M-\omega^{\prime})}{\chi+1}.
\fe 
In this manner, we obtain
\ie
\label{ims}
\text{Im}\, \mathcal{S} =\text{Im}\,\int_{0}^{\omega} -\mathrm{d}\omega'\int_{r_i}^{r_f}\,\frac{\mathrm{d}r}{\left(1-\sqrt{\frac{\Delta(r,\,\omega^\prime)}{r}}\right)}.
\fe

When the mass parameter in the geometry is shifted to $(M-\omega')$, the radial function $\Delta(r)$ acquires an explicit dependence on $\omega'$. This adjustment relocates the horizon and generates a pole at the corresponding radius. Evaluating the contribution from this pole by performing a counterclockwise contour integration yields
\begin{eqnarray}
    \text{Im}\, \mathcal{S}  = 4\pi (1 + \chi) \, \omega \left( M - \frac{\omega}{2} \right)  .
\end{eqnarray}
According to the treatment in Ref.~\cite{vanzo2011tunnelling}, in our case, the presence of Lorentz--violating terms alters the probability of Hawking emission. In that framework, the rate takes the form
\ie
\Gamma \sim e^{-2 \, \text{Im}\, \mathcal{S}}=e^{-8 (1 + \chi) \, \omega \left( M - \frac{\omega}{2} \right)} .
\fe
When the emitted energy approaches zero, $\omega\rightarrow 0$, the expression reduces to Hawking’s original thermal spectrum. In this regime, the distribution takes the form
\begin{equation}
    \mathcal{P}(\omega,\chi)=\frac{\mathrm{d}\omega}{2\pi}\frac{1}{e^{8 \pi(1 + \chi) \,\omega \left( M - \frac{\omega}{2} \right)
    }-1}.
\end{equation}

The frequency dependence of the tunneling probability leads to a radiation spectrum that no longer matches the usual blackbody profile; this deviation becomes apparent once the expression is examined in detail. In the low--energy regime, however, the spectrum approaches a Planck--type form, albeit characterized by a modified Hawking temperature. The corresponding particle occupation number, determined directly from the tunneling probability, is therefore given by:
\ie
n = \frac{\Gamma}{1 - \Gamma} = \frac{1}{e^{8 \pi  (1 + \chi) \, \omega  \left(M-\frac{\omega }{2}\right)} - 1}.
\fe

\begin{figure}
    \centering
      \includegraphics[scale=0.81]{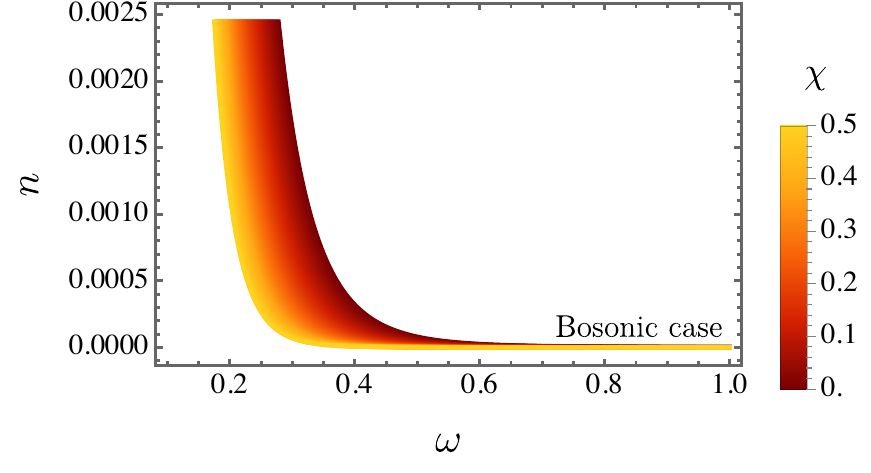}
    \caption{Particle density $n$ for bosons as a function of the frequency $\omega$ for several values of the Lorentz--violating parameter $\chi$.}
    \label{bosonparticledensity}
\end{figure}

Fig. \ref{bosonparticledensity} shows how the density of emitted bosonic quanta responds to variations of the Lorentz--violating parameter. The curves display a clear trend: larger values of $\chi$ suppress the number of particles produced. This behavior indicates that the spectrum carries information about the underlying geometry. In particular, the parameter $\chi$ reshapes the tunneling amplitudes, and once energy conservation is taken into account, the resulting power spectrum departs from the usual thermal distribution expected for a Schwarzschild black and the original bumblebee holes.

\begin{figure}
    \centering
      \includegraphics[scale=0.74]{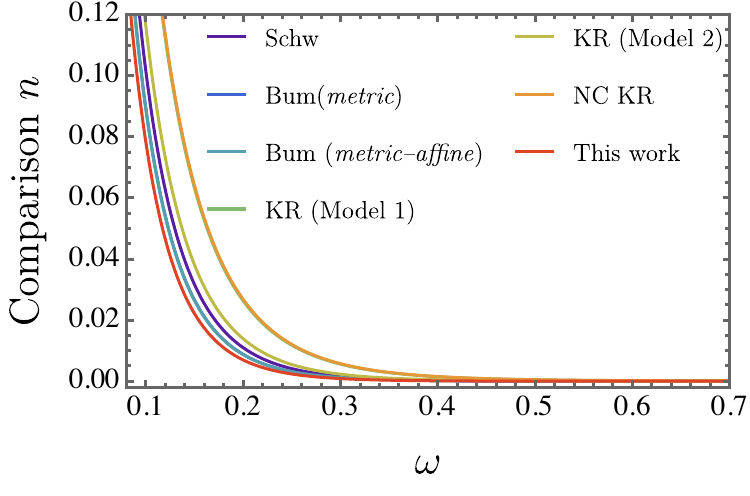}
       \includegraphics[scale=0.74]{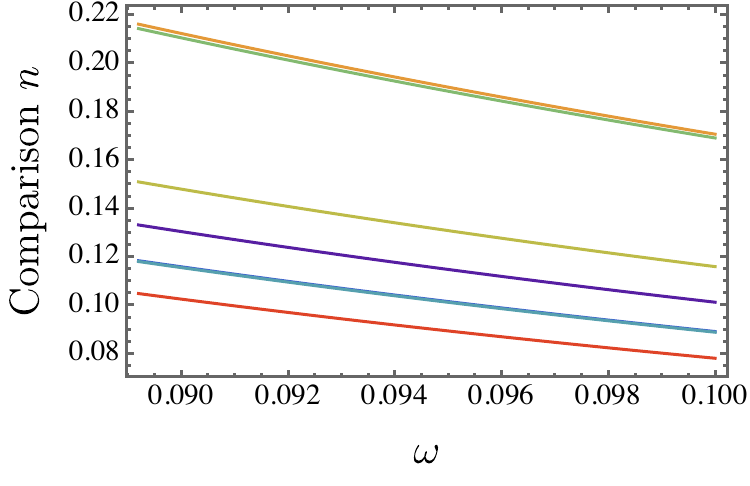}
    \caption{Comparison of bosonic particle creation for the present solution with the bumblebee (\textit{metric)} case, \textit{metric--affine} bumblebee, Schwarzschild, Kalb--Ramond (Models 1 and 2), and non--commutative Kalb--Ramond black holes.}
    \label{comparisonbosonparticledensity}
\end{figure}

A natural question emerges at this point: among the Lorentz--violating black holes considered here — the bumblebee solution (vector field) and the Kalb--Ramond solution (tensor field) — which one produces the largest bosonic particle density? To address this, Fig.~\ref{comparisonbosonparticledensity} presents a direct comparison. For simplicity, we set $\Theta = X = \ell = 0.1$ {(here, $\Theta$ and $X$ encode the non--commutative effects associated with the Kalb--Ramond sector and the Lorentz--violating parameter arising from the bumblebee field within the \textit{metric--affine} framework.)}.  {It should be emphasized that these values are not mandatory. They were selected only to provide a clearer visualization, following the same strategy adopted in Refs.~\cite{AraujoFilho:2025jcu,araujo2025particleasdasd,AraujoFilho:2025zaj}.}

Under this choice, one verifies the following hierarchy:
\[
n^{\text{this work}} < n^{\text{bum (\textit{metric})}} \approx n^{\text{bum (\textit{met--aff})}} < n^{\text{Schw}}  < n^{\text{KR (Model 2)}} < n^{\text{KR (Model 1)}} < n^{\text{NC KR}} . 
\]
In other words, within the set of Lorentz--violating black holes examined here, the new bumblebee solution evaporates the most slowly, whereas the non--commutative Kalb--Ramond black hole \cite{AraujoFilho:2025jcu} exhibits the fastest evaporation. {In addition, the remaining results displayed in Fig.~\ref{comparisonbosonparticledensity} are taken from \cite{AraujoFilho:2025hkm} for the \textit{metric} and \textit{metric--affine} bumblebee cases, \cite{AraujoFilho:2024ctw} for the Kalb--Ramond models, and \cite{AraujoFilho:2025jcu} for the non--commutative Kalb--Ramond black hole.}

\subsection{Fermionic modes }

Black holes behave as thermal objects and radiate with a characteristic temperature, though the observed spectrum is generally filtered by greybody effects. This radiation includes contributions from fields of different spins. Earlier analyses by Kerner and Mann \cite{o69}, together with subsequent developments \cite{o71,o75,o70,o74,o72,o73}, established that massless fermionic and bosonic modes originate at an identical temperature. Further investigations into spin--$1$ fields demonstrated that even when quantum corrections are incorporated, the Hawking temperature remains unchanged \cite{o77,o76}.

For fermions, the relevant action is typically connected to the phase of the spinor and is governed by a Hamilton--Jacobi--type equation. Alternative formulations have been discussed in \cite{o83,vanzo2011tunnelling,o84}. Modifications induced by the coupling between the spin and the spacetime connection do not produce divergences at the horizon; their effect is confined primarily to small corrections in spin precession. These contributions are negligible in the present context. In addition, emission of particles with opposite spin orientations tends to occur symmetrically, so nonrotating black holes with masses far above the Planck scale do not acquire angular momentum through this process \cite{vanzo2011tunnelling}.

Motivated by these considerations, we investigate the tunneling of fermionic modes across the horizon in the Lorentz--violating black hole background. The emission probability is computed within Schwarzschild--like coordinates, despite their well--known coordinate singularity at the horizon. Other coordinate choices—such as generalized Painlevé--Gullstrand or Kruskal--Szekeres charts—have been examined in earlier studies \cite{o69}. To set up the calculation, we begin with a general line element of the form
\ie
\mathrm{d}s^{2} = - A(r,\chi) \mathrm{d}t^{2} + [1/B(r,\chi)]\mathrm{d}r^{2} + C(r,\chi)[\mathrm{d}\theta^{2} + r^{2}\sin^{2}\theta ]\mathrm{d}\varphi^{2}.
\fe
In a curved background, the dynamics of a spin--$1/2$ field are governed by the Dirac equation, which takes the form
\ie
\left(\Tilde{\gamma}^\mu \Tilde{\nabla}_\mu + \frac{m}{\hbar}\right) \psi(t,r,\theta,\varphi) = 0
\fe
where, we also have
\ie
\Tilde{\nabla}_\mu = \partial_\mu + \frac{i}{2} {\Gamma^\alpha_{\;\mu}}^{\;\beta} \,\Tilde{\Sigma}_{\alpha\beta}
\fe and 
\ie
\Tilde{\Sigma}_{\alpha\beta} = \frac{i}{4} [\Tilde{\gamma}_\alpha,  \Tilde{\gamma}_\beta].
\fe
The generalized gamma matrices $\tilde{\gamma}^\mu$ are constrained by the Clifford algebra, which is imposed via
\ie
\{\Tilde{\gamma}_\alpha,\Tilde{\gamma}_\beta\} = 2 g_{\alpha\beta} \mathbb{1}.
\fe
The symbol $\mathbb{1}$ refers to the $4 \times 4$ identity operator. With this convention in place, the set of $\tilde{\gamma}$ matrices is taken to be
\begin{eqnarray*}
 \Tilde{\gamma} ^{t} &=&\frac{i}{\sqrt{A(r,\chi)}}\left( \begin{array}{cc}
\vec{1}& \vec{ 0} \\ 
\vec{ 0} & -\vec{ 1}%
\end{array}%
\right), \;\;
\Tilde{\gamma} ^{r} =\sqrt{B(r,\chi)}\left( 
\begin{array}{cc}
\vec{0} &  \vec{\sigma}_{3} \\ 
 \vec{\sigma}_{3} & \vec{0}%
\end{array}%
\right), \\
\Tilde{\gamma} ^{\theta } &=&\frac{1}{r}\left( 
\begin{array}{cc}
\vec{0} &  \vec{\sigma}_{1} \\ 
 \vec{\sigma}_{1} & \vec{0}%
\end{array}%
\right), \;\;
\Tilde{\gamma} ^{\varphi } =\frac{1}{r\sin \theta }\left( 
\begin{array}{cc}
\vec{0} &  \vec{\sigma}_{2} \\ 
 \vec{\sigma}_{2} & \vec{0}%
\end{array}%
\right).
\end{eqnarray*}%
Here, the vector $\vec{\sigma}$ denotes the Pauli matrices, whose algebra is fixed by the usual commutation rules:
\ie
 \sigma_i  \sigma_j = \vec{1} \delta_{ij} + i \varepsilon_{ijk} \sigma_k, \,\, \text{in which}\,\, i,j,k =1,2,3. 
\fe 
The corresponding $\tilde{\gamma}^{5}$ matrix is given by
\begin{equation*}
\Tilde{\gamma} ^{5}=i\Tilde{\gamma} ^{t}\Tilde{\gamma} ^{r}\Tilde{\gamma} ^{\theta }\Tilde{\gamma} ^{\varphi }=i\sqrt{\frac{B(r,\chi)}{A(r,\chi)}}\frac{1}{r^{2}\sin \theta }\left( 
\begin{array}{cc}
\vec{ 0} & - \vec{ 1} \\ 
\vec{ 1} & \vec{ 0}%
\end{array}%
\right)\:.
\end{equation*}
A Dirac field polarized along the $+r$ direction can be described through the following ansatz \cite{vanzo2011tunnelling}:
\begin{equation}
\psi^{+}(t,r,\theta ,\varphi ) = \left( \begin{array}{c}
\mathrm{H}(t,r,\theta ,\varphi ) \\ 
0 \\ 
\mathrm{Y}(t,r,\theta ,\varphi ) \\ 
0%
\end{array}%
\right) \exp \left[ \frac{i}{\hbar }\Tilde{\psi}^{+}(t,r,\theta ,\varphi )\right]\;.
\label{spinupbh} 
\end{equation} 
In what follows, we restrict attention to the spin--up configuration, noting that the opposite polarization—aligned with the negative radial direction—can be handled through the same steps. Substituting the ansatz in Eq.~(\ref{spinupbh}) into the curved space Dirac equation leads to the set of relations:
\ie
\begin{split}
-\left( \frac{i \,\mathrm{H}}{\sqrt{A(r,\chi)}}\,\partial _{t} \Tilde{\psi}_{+} + \mathrm{Y} \sqrt{B(r,\chi)} \,\partial_{r} \Tilde{\psi}_{+}\right) + \mathrm{H} m &=0, \\
-\frac{\mathrm{Y}}{r}\left( \partial _{\theta }\Tilde{\psi}_{+} +\frac{i}{\sin \theta } \, \partial _{\varphi }\Tilde{\psi}_{+}\right) &= 0, \\
\left( \frac{i \,\mathrm{Y}}{\sqrt{A(r,\chi)}}\,\partial _{t}\Tilde{\psi}_{+} - \mathrm{H} \sqrt{B(r,\chi)}\,\partial_{r}\Tilde{\psi}_{+}\right) + \mathrm{Y} m & = 0, \\
-\frac{\mathrm{H}}{r}\left(\partial _{\theta }\Tilde{\psi}_{+} + \frac{i}{\sin \theta }\,\partial _{\varphi }\Tilde{\psi}_{+}\right) &= 0,
\end{split}
\fe%

Focusing on the dominant contribution in the $\hbar$-expansion yields an action of the form
$
\Tilde{\psi}_{+}=- \omega\, t + \Xi(r) + L(\theta ,\varphi )  $
so that we have
\cite{vanzo2011tunnelling}
\begin{eqnarray}
\left( \frac{i\, \omega\, \mathrm{H}}{\sqrt{A(r,\chi)}} - \mathrm{Y} \sqrt{B(r,\chi)}\,  \Xi^{\prime }(r)\right) +m\, \mathrm{H} &=&0,
\label{bhspin5} \\
-\frac{\mathrm{H}}{r}\left( L_{\theta }+\frac{i}{\sin \theta }L_{\varphi }\right) &=&0,
\label{bhspin6} \\
-\left( \frac{i\,\omega\, \mathrm{Y}}{\sqrt{A(r,\chi)}} + \mathrm{H}\sqrt{B(r,\chi)}\,  \Xi^{\prime }(r)\right) +\mathrm{Y}\,m &=&0,
\label{bhspin7} \\
-\frac{\mathrm{H}}{r}\left( L_{\theta } + \frac{i}{\sin \theta }L_{\varphi }\right) &=& 0.
\label{bhspin8}
\end{eqnarray}
The explicit form of the angular functions ($\mathrm{H}$) and ($\mathrm{Y}$) plays no role in the restriction that follows from Eqs.~(\ref{bhspin6}) and (\ref{bhspin8}). These equations force the combination
$$
L_{\theta}+\, i\,(\sin\theta)^{-1} L_{\varphi}=0,
$$
which implies that the angular function ($L(\theta,\varphi)$) must be complex. This constraint arises for both ingoing and outgoing fermionic modes. Consequently, when computing the ratio between the corresponding tunneling probabilities, every factor involving ($L$) cancels, so the angular contribution does not influence the final result and can be omitted from the subsequent analysis.

For a massless spinor, Eqs.~(\ref{bhspin5}) and (\ref{bhspin7}) admit two distinct branches of solutions:
\ie
\mathrm{H} = -i \mathrm{Y}, \qquad \Xi^{\prime }(r) = \Xi_{\text{out}}' = \frac{\omega}{\sqrt{A(r,\chi)B(r,\chi)}},
\fe
\ie
\mathrm{H} = i \mathrm{Y}, \qquad \Xi^{\prime }(r) = \Xi_{\text{in}}' = - \frac{\omega}{\sqrt{A(r,\chi)B(r,\chi)}}.
\fe

\begin{figure}
    \centering
      \includegraphics[scale=0.7]{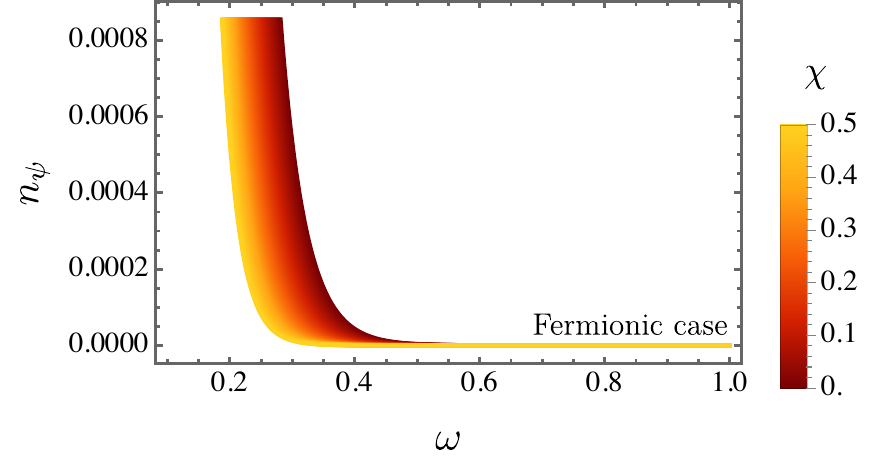}
    \caption{Particle creation density for fermions $n_{\psi}$ is shown as a function of the frequency $\omega$ for several values of the parameter $\chi$.}
    \label{partifermmns}
\end{figure}

Here, $\Xi_{\text{out}}$ and $\Xi_{\text{in}}$ correspond to the fermionic modes propagating outward and inward, respectively \cite{vanzo2011tunnelling}. The tunneling probability is governed by the difference between the imaginary parts of these two branches,
$$
\Gamma_{\psi} \propto \exp\!\left[-2\,\mathrm{Im}\!\left(\Xi_{\text{out}} - \Xi_{\text{in}}\right)\right].
$$
From this expression, one finds:
\ie
 \Xi_{\text{out}}(r)= -  \Xi_{ \text{in}} (r) = \int \mathrm{d} r \,\frac{\omega}{\sqrt{A(r,\chi)B(r,\chi)}}\,.
\fe
It should be emphasized that the dominant energy condition, together with the Einstein equations, implies that the functions $A(r,\chi)$ and $B(r,\chi)$ vanish at the same radial position. Consequently, in the vicinity of $r=r_h$, both functions may be expanded linearly as
\ie
A(r,\chi)B(r,\chi) = A'(r_{h},\chi) B'(r_{h}, \chi)(r - r_{h})^2 + \dots \, .
\fe
The expansions make clear that a simple pole emerges, carrying a definite coefficient. By invoking Feynman's rule for handling such singularities, we find:
\ie
2\mbox{ Im}\;\left(  \Xi_{ \text{out}} -  \Xi_{ \text{in}} \right) = \mbox{Im}\int \mathrm{d} r \,\frac{4\omega}{\sqrt{A(r,\chi)B(r,\chi)}}=\frac{2\pi\omega}{\kappa},
\fe
where the quantity $\kappa$ denotes the surface gravity, defined through
\ie
\kappa = \frac{1}{2} \sqrt{A'(r_{h},\chi) B'(r_{h},\chi)} .
\fe 
Knowing that $\Gamma_{\psi} \sim e^{-\frac{2 \pi \omega}{\kappa}}$, we can obtain therefore
\ie
n_{\psi} = \frac{\Gamma_{\psi}}{1+\Gamma_{\psi}}  = \frac{1}{e^{8 \pi  (1+\chi) \, M \omega }+1}.
\fe
Figure~\ref{partifermmns} illustrates how the fermionic density $n_{\psi}$ responds to changes in the parameter $\chi$. The curves show a clear trend: larger values of $\chi$ suppress the emission of fermions, mirroring the pattern previously identified in the bosonic sector.

\section{Greybody factors}\label{sec:GBF}

In this section, we investigate the scattering process using the WKB method. Another important aspect of gravitational perturbations around a black hole spacetime is the graybody factor. The probability for an outgoing wave to reach infinity, or equivalently the probability for an incoming wave to be absorbed by the black hole, is characterized by the greybody factor \cite{konoplya2019higher,cardoso2001quasinormal}. This quantity plays a key role in studying the tunneling probability of the field through the effective potential associated with a given black hole spacetime. In particular, we are interested in analyzing the influence of the parameter $\chi$ on the greybody factor.

Scattering via the WKB method requires imposing appropriate boundary conditions, as the fields near the horizon and at spatial infinity are expected to take the asymptotic forms \cite{konoplya2011quasinormal}

\begin{equation}
		R_{\omega l}=
		\begin{cases}
			{{e^{ - i\omega {r^*}}} + \mathcal{A}_R{e^{i\omega {r^*}}}}, \quad \text{if} \quad \ {r^*} \to -\infty ~ (r \to {r_h})\\
			{\mathcal{A}_T{e^{ - i\omega {r^*}}}}, \quad \quad \quad \quad \, \text{if} \,\, \quad {r^*} \to +\infty ~ (r \to \infty )\\
		\end{cases}
	\end{equation}
where $\mathcal{A}_R$ and $\mathcal{A}_T$ are the reflection and transmission coefficients, respectively. The reflection coefficient can be expressed as
\begin{equation}
\mathcal{A}_R = \left(\frac{1}{1 + e^{-2 i \pi \mathcal{K}}} \right)^{-\frac{1}2{}}.
\end{equation}

The parameter $\mathcal{K}$ is determined by the WKB expansion and is given by \cite{konoplya2003quasinormal,konoplya2011quasinormal}
\begin{equation}
\mathcal{K}
= \frac{i(\omega^{2} - \mathrm{V
}_{0})}{\sqrt{-2 \mathrm{V}''_{0}}}
- \sum_{j=2}^{6} \Lambda_{j},
\end{equation}
where $\mathrm{V}$ is the effective potential in different fields as $\mathrm{V}_{s,v,t,\psi}$ for scalar, vector, tensor, and spinoral fields, respectively.
$\mathrm{V}_{0}$ is the maximum of the effective potential, $\mathrm{V}''_{0}$ is the second derivative of the potential at this maximum with respect to the tortoise coordinate $r^{*}$, and the terms $\Lambda_{j}$ denote the higher--order WKB corrections, which depend on higher derivatives of the potential evaluated at the peak.

The effective potential for a bosonic field can be expressed in a generalized form as \cite{medved2004dirty,nomura2005continuous}
\begin{equation}
\mathrm{V}
_{s,v,t} = f(r) \left[ \frac{l(l+1)}{r^2} + s(s-1) \frac{1-f(r)}{r^2} + (1-s) \frac{f'(r)}{r} \right],
\label{eq:Veffg}
\end{equation}
where $s$ equals $0,1,2$ corresponding to scalar, vector, and tensor fields, respectively. Furthermore, the effective potential for the Dirac perturbation has the following form 
\begin{equation}
\mathrm{V}_{\psi} = \frac{(l + 1/2)^2}{r^2} f(r) + (l + 1/2) f(r) \frac{\mathrm{d}}{\mathrm{d}r}\!\left(\frac{\sqrt{f(r)}}{r}\right).
\label{eq:Veffpsi}
\end{equation}

On the other hand, the transmission coefficient can be computed via 
\begin{equation}\label{TrasRef1}
|\mathcal{A}_T|^2+|\mathcal{A}_R|^2 = 1.
\end{equation}
In addition, the greybody factor $\mathrm{T}$ is defined with transmission coeffieict as \cite{iyer1987black2,konoplya2011quasinormal}
\begin{equation}\label{TransRef2}
\mathrm{T}=|\mathcal{A}_T|^2=1-|\mathcal{A}_R|^2 = \frac{1}{1 + e^{+2 i \pi \mathcal{K}}}.
\end{equation}

In the following section, we explore the effect of both spin and Lorentz--violating parameter $\chi$ on the greybody factors.


\subsection{Spin 0}
\begin{figure}
    \centering
     \includegraphics[width=90mm]{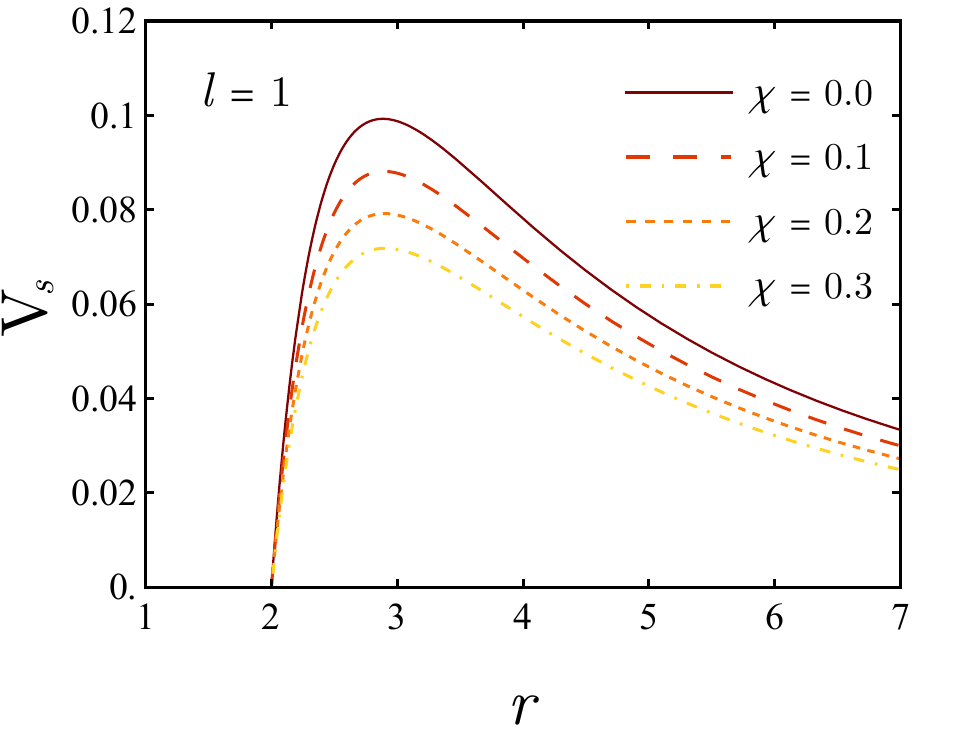}
    \caption{The effective potential for scalar perturbation $\mathrm{V}_s$ for $M = 1$, $l = 1$, and $\chi = 0$ - $0.3$ .}
    \label{fig:veffs}
\end{figure}

To calculate the greybody factor for the scalar field, we first investigate the effective potential in Eq. \eqref{eq:Veffg} by considering $(s = 0)$. The scalar effective potential $\text{V}_s$ for various values of the Lorentz–violating parameter is shown in Fig. \ref{fig:veffs}.

\begin{figure}
    \centering
     \includegraphics[width=90mm]{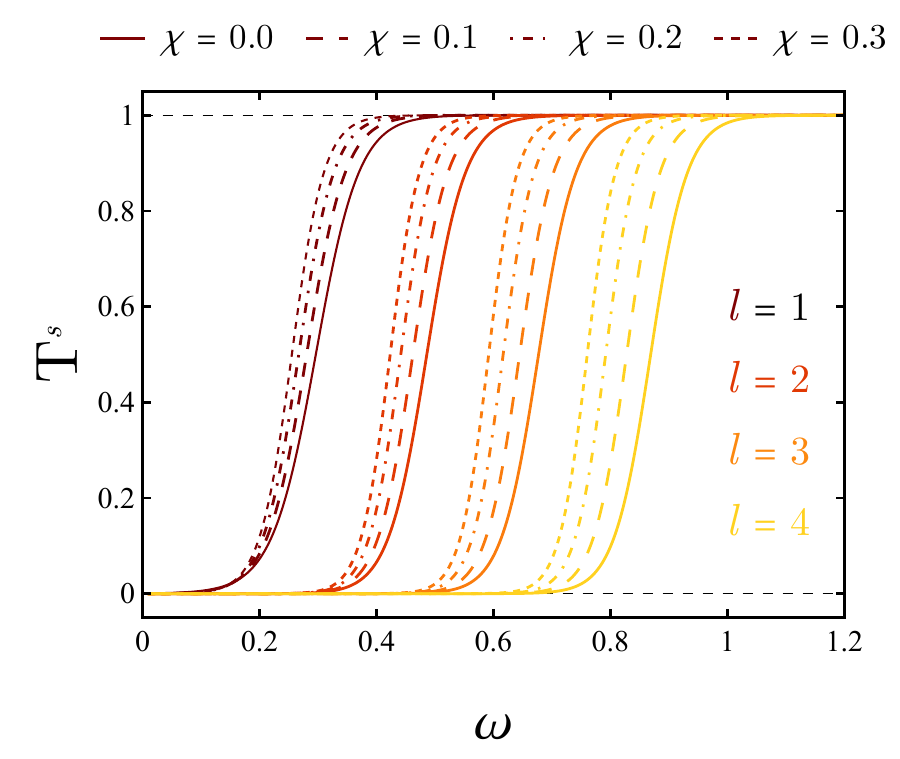}
    \caption{The greybody factors of the scalar field $\text{T}^s$, are obtained using the sixth--order WKB approximation for $M=1$, several multipole numbers $l$, and different values of the bumblebee parameter $\chi$.}
    \label{fig:Ts}
\end{figure}

Based on Fig. \ref{fig:veffs}, increasing the parameter $\chi$, suppresses the peak of the effective potential. Thus, we expect that the possibility of transmission increases with the Lorentz--violating parameter. In Fig. \ref{fig:Ts}, the greybody factor for different multipole numbers and bumblebee parameter is represented. For all cases of multipole number, the behavior of the greybody factor is similar. When the parameter $\chi$ goes higher, the probability of the transmission is increased, consistent with the behavior of the effective potential. Moreover, the impact of the Lorentz--violating parameter is stronger for higher multipole numbers.

\subsection{Spin 1}
\begin{figure}
    \centering
     \includegraphics[width=90mm]{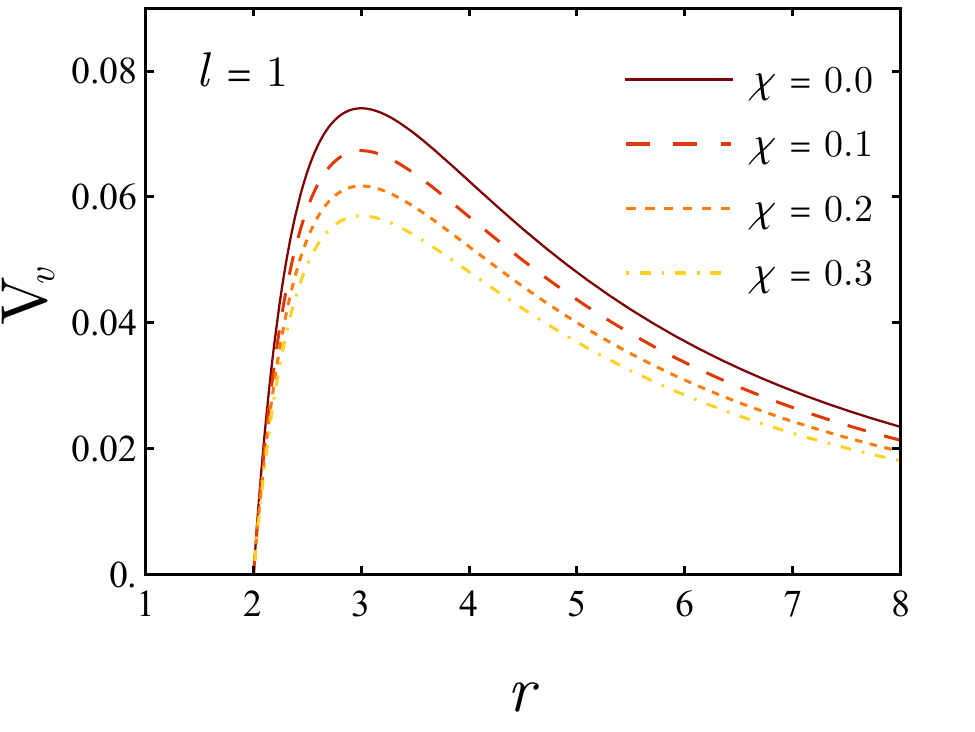}
    \caption{Effective potential $\mathrm{V}_v$ for vector perturbations with $M = 1$, $l = 1$, and varying Lorentz--violating parameter $\chi = 0.0$–$0.3$.}
    \label{fig:veffv}
\end{figure}

To compute the greybody factor for the vector field, we begin by examining the effective potential given in Eq.~\eqref{eq:Veffg} for the case of spin $s = 1$. The effective potential for vector perturbations, denoted $\mathrm{V}_v$, is illustrated in Fig.~\ref{fig:veffv} for different values of the Lorentz--violating parameter $\chi$.

\begin{figure}
    \centering
     \includegraphics[width=90mm]{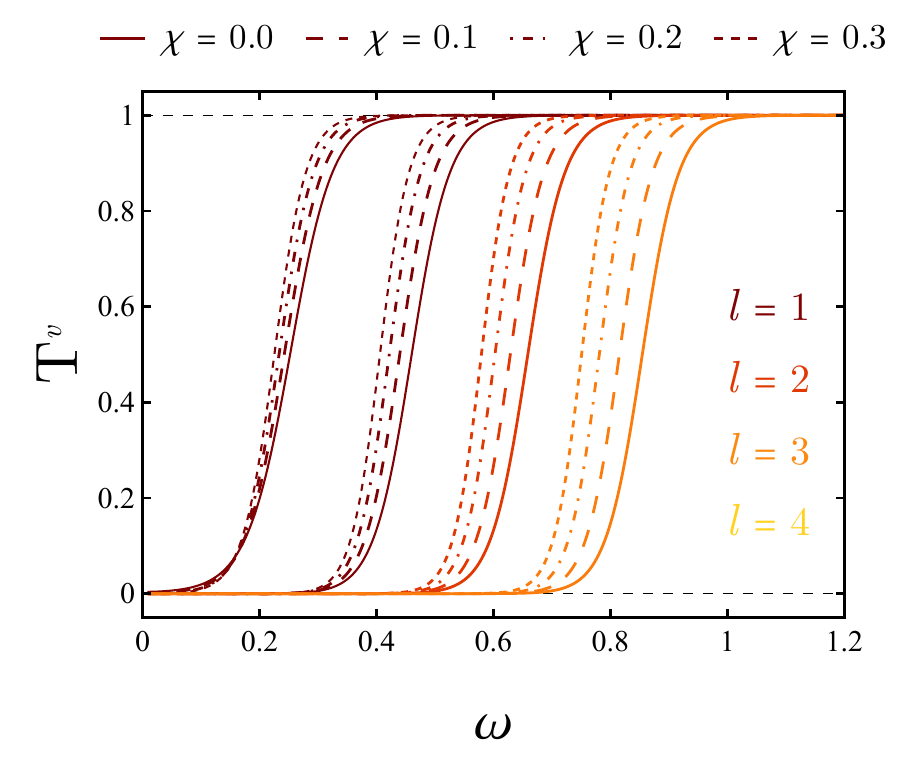}
    \caption{The greybody factors associated with vector perturbations $\text{T}^v$, are evaluated via the sixth--order WKB method for $M=1$, considering multiple angular modes $l$ and a range of the bumblebee parameter $\chi$.}
    \label{fig:Tv}
\end{figure}

As shown in Fig.~\ref{fig:veffv}, increasing the bumblebee parameter 
$\chi$ lowers the height of the effective potential barrier, similar to the scalar case. Consequently, we expect the transmission probability to grow with larger values of the Lorentz--violating parameter. This behavior is confirmed in Fig.~\ref{fig:Tv}, where the greybody factor for various multipole numbers 
$l$ and values of $\chi$ is presented. Additionally, the influence of Lorentz violation becomes more pronounced for larger $l$.

\subsection{Spin 2}
\begin{figure}
    \centering
     \includegraphics[width=90mm]{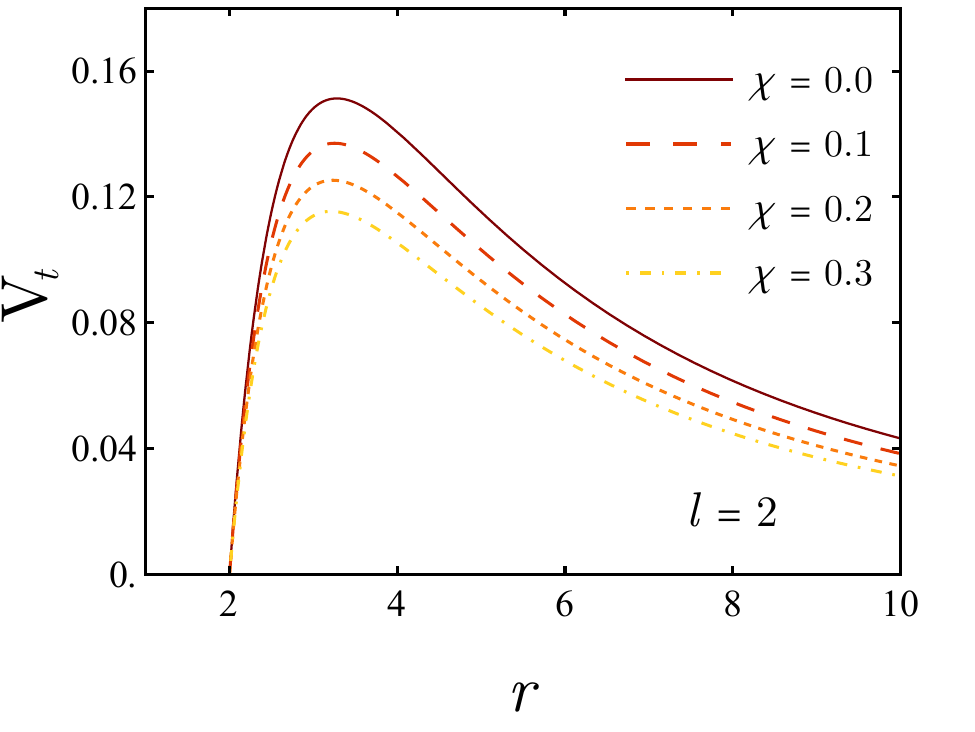}
    \caption{Effective potential for tensor perturbations $V_{t}$ with $M = 1$, $l = 2$, and $\chi = 0\text{–}0.3$.}
    \label{fig:vefft}
\end{figure}

We now turn to the tensor perturbations, corresponding to the choice $s = 2$ in Eq.~\eqref{eq:Veffg}. The resulting effective potential, denoted by $\text{V}_{t}$, is illustrated in Fig.~\ref{fig:vefft} for several representative values of the Lorentz–violating parameter $\chi$.

\begin{figure}
    \centering
     \includegraphics[width=90mm]{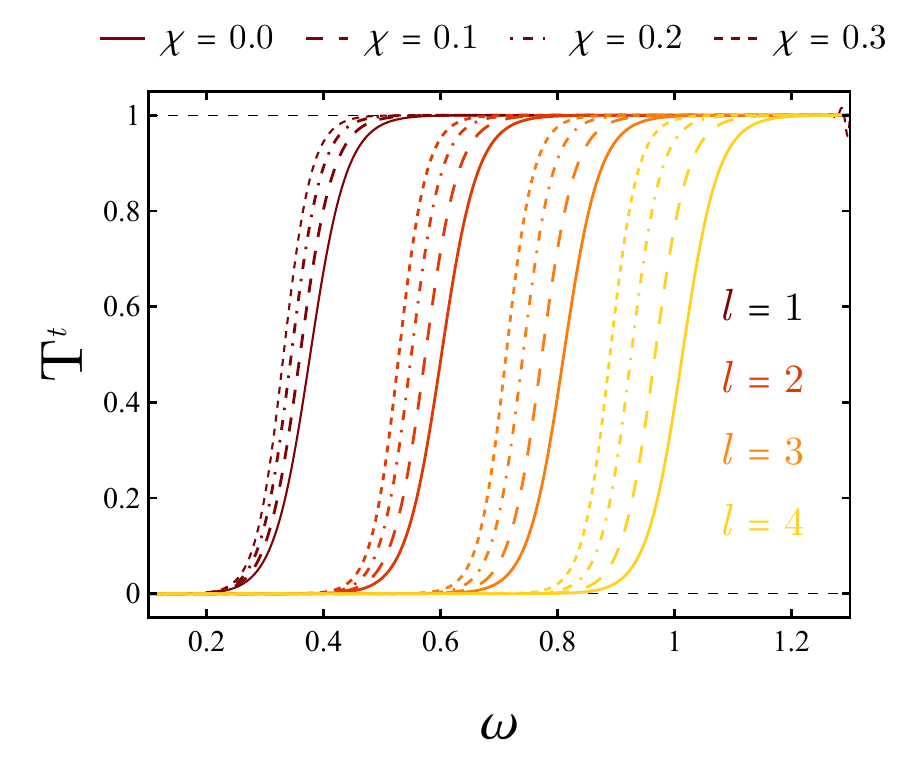}
    \caption{The greybody factors for tensor–type gravitational perturbations, $\text{T}^t$ are computed with the sixth--order WKB scheme for $M=1$ and various choices of the multipole number $l$ and the parameter $\chi$.}
    \label{fig:Tt}
\end{figure}

It shows that the tensor potential is also sensitive to the parameter $\chi$: as $\chi$ increases, the height of the barrier decreases. This trend indicates that waves of tensor type encounter a less restrictive barrier when the Lorentz--violating effects are stronger. The corresponding greybody factors, computed using the sixth--order WKB method, are plotted in Fig.~\ref{fig:Tt} for a selection of multipole numbers $l$. 

As depicted in Fig.~\ref{fig:Tt}, the transmission probability for a specific multipole number, at a fixed frequency, has larges values for higher value of $\chi$. This behavior can be seen across all multipole orders. Furthermore, for larger $l$, the curves become increasingly responsive to variations in $\chi$.


\subsection{Spin 1/2}
\begin{figure}
    \centering
     \includegraphics[width=90mm]{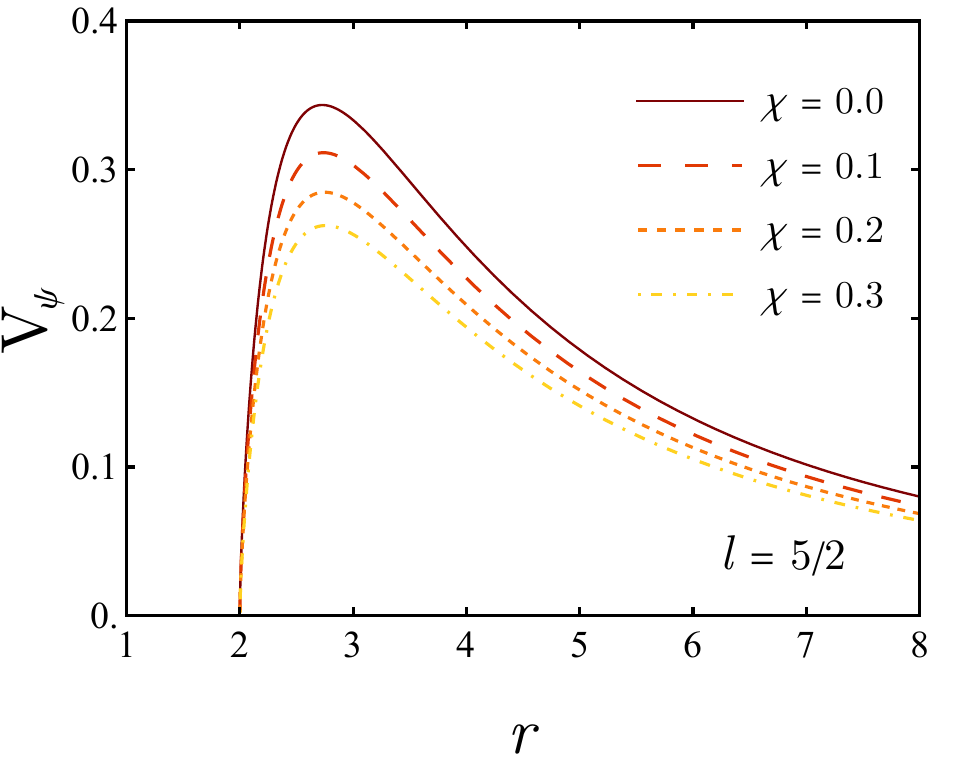}
    \caption{The effective potential for Dirac perturbations {$\mathrm{V}_{\psi}$} for $M=1$, $l=5/2$, and several values of the bumblebee parameter $\chi$.}
    \label{fig:veffd}
\end{figure}

\begin{figure}
    \centering
     \includegraphics[width=90mm]{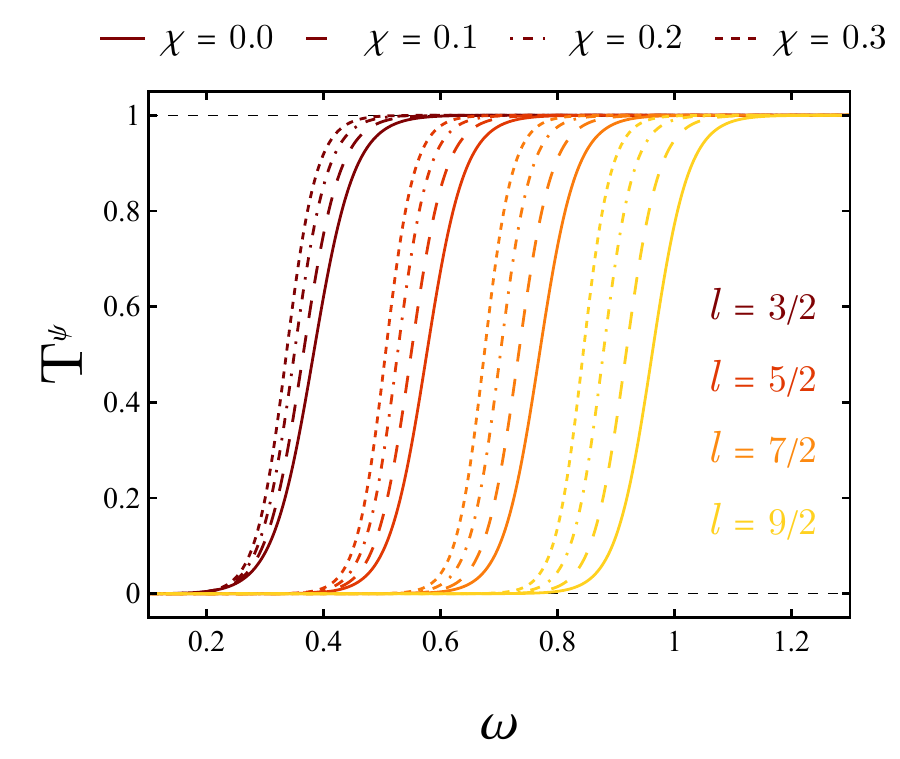}
    \caption{The greybody factors for the Dirac (spin–$1/2$) field are calculated using the sixth--order WKB formalism for $M=1$, multipole modes $l = 3/2$ - $9/2$, and several values of the bumblebee parameter $\chi$.}
    \label{fig:Td}
\end{figure}

For fermionic perturbations, we consider the Dirac field governed by the effective potential obtained from Eq.~\eqref{eq:Veffg} with {$s = 1/2$}. The resulting potential, denoted by $\text{V}_{\psi}$, is shown in Fig.~\ref{fig:veffd} for different values of the Lorentz--violating parameter $\chi$. 
As illustrated in Fig.~\ref{fig:veffd}, the fermionic potential exhibits a reduction in its peak height as $\chi$ increases. This tendency suggests that Lorentz--violating effects facilitate the penetration of the Dirac field through the potential barrier.
The corresponding greybody factors, calculated using the sixth--order WKB approximation, are presented in Fig.~\ref{fig:Td} for a range of half--integer multipole modes.

As shown in Fig.~\ref{fig:Td}, for any fixed multipole number and frequency, the transmission probability increases consistently with the value of $\chi$. This trend persists for all considered modes. In addition, the dependence on the bumblebee parameter becomes more pronounced at higher $l$, where the greybody spectra display a stronger response to variations in $\chi$.
\begin{figure}
    \centering
     \includegraphics[width=90mm]{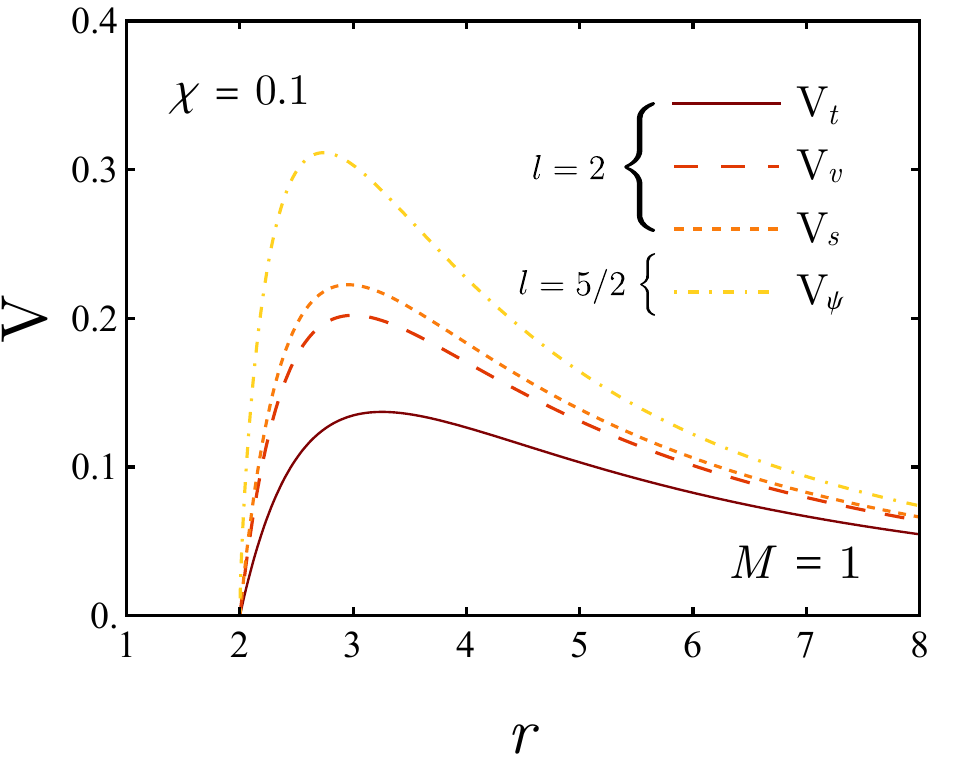}
    \caption{Comparison of effective potentials for different spin fields. Shown are $V_{t}$ (tensor), $V_{v}$ (vector), and $V_{s}$ (scalar) for $l=2$, together with the Dirac potential $V_{\psi}$ for $l=5/2$. The potentials are plotted for the same background parameters $M = 1$ and $\chi = 0.1$.}
    \label{fig:veffComp}
\end{figure}
A direct comparison of the effective potentials for the four perturbative sectors is presented in Fig.~\ref{fig:veffComp}. When the multipole numbers are fixed as $l=2$ for the bosonic fields and $l=5/2$ for the Dirac field, the relative peak heights exhibit a clear hierarchy as 
\begin{equation}
\mathrm{V}_{\psi} > \mathrm{V}_{s} > \mathrm{V}_{v} > \mathrm{V}_{t}.    
\end{equation}

\begin{figure}
    \centering
     \includegraphics[width=90mm]{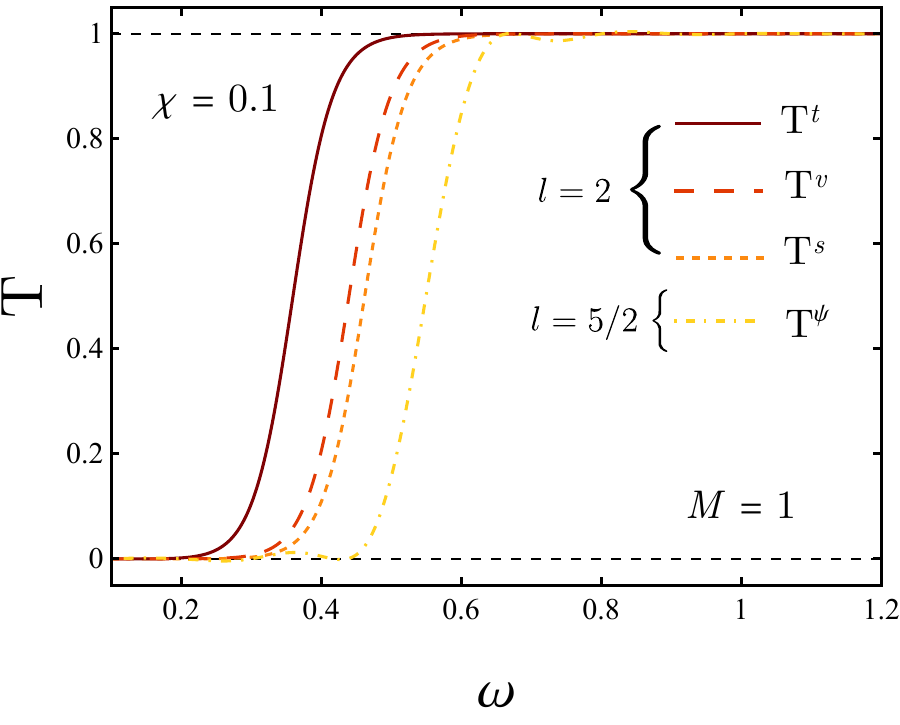}
    \caption{Comparison of the greybody factors for scalar, vector, and tensor fields with fixed multipole number $l=2$, together with the Dirac field for $l=5/2$, computed for $M=1$ and $\chi=0.1$.}
    \label{fig:GBFall}
\end{figure}

Since the height of the potential barrier determines the degree of suppression experienced by each mode, this hierarchy is directly reflected in the corresponding greybody factors in Fig.~\ref{fig:GBFall}, where
\begin{equation}\label{eq:compT}
\mathrm{T}_{t} > \mathrm{T}_{v} > \mathrm{T}_{s} > \mathrm{T}_{\psi}.    
\end{equation}

The greybody factors for all four perturbations increase monotonically with frequency and share a qualitatively similar profile. However, their transmission efficiencies differ systematically: at a fixed frequency, the tensor mode exhibits the largest transmission probability, followed by the vector and scalar modes, while the Dirac field maintains the smallest values throughout.  Additionally, the tensor mode reaches $\text{T}\simeq 1$ at the lowest frequency, indicating that it becomes fully transmitted more rapidly than the others. The vector and scalar fields approach unit transmission at moderately higher frequencies, whereas the Dirac field requires the largest $\omega$ to reach $\text{T}\simeq 1$. Hence, although the functional form of the spectra is similar, their relative magnitudes and transmission thresholds reveal clear spin-dependent distinctions in the propagation of perturbations across the black hole potential barrier. 
\section{Absorption cross section}

	The transmission coefficient, defined in Eq. \eqref{TransRef2} can be used to determine the partial absorption cross section \cite{crispino2009scattering,gogoi2024quasinormal,anacleto2020absorption}.

		\begin{equation}
			{\sigma^{i}_{\mathrm{abs}}} = \frac{{\pi (2l + 1)}}{{{{\omega}^2}}}\mathrm{T}^{i}
		\end{equation}
		where $\omega$ denotes the wave frequency, $l$ is the multipole number, and the index $i \in {s, v, t, \psi}$ labels the scalar, vector, tensor, and spinor perturbations, respectively.\\
		
\begin{figure}
    \centering
     \includegraphics[scale=0.5]{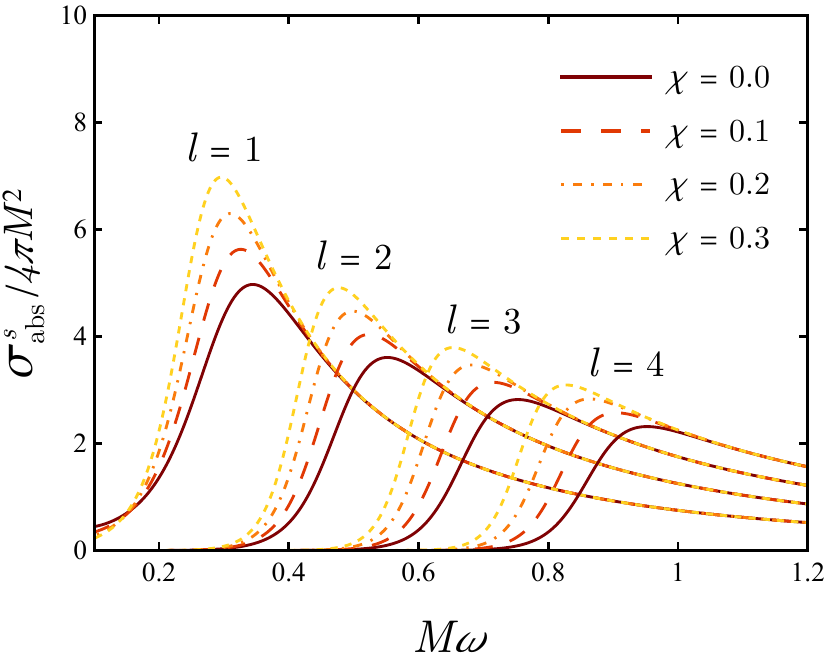}
     \includegraphics[scale=0.5]{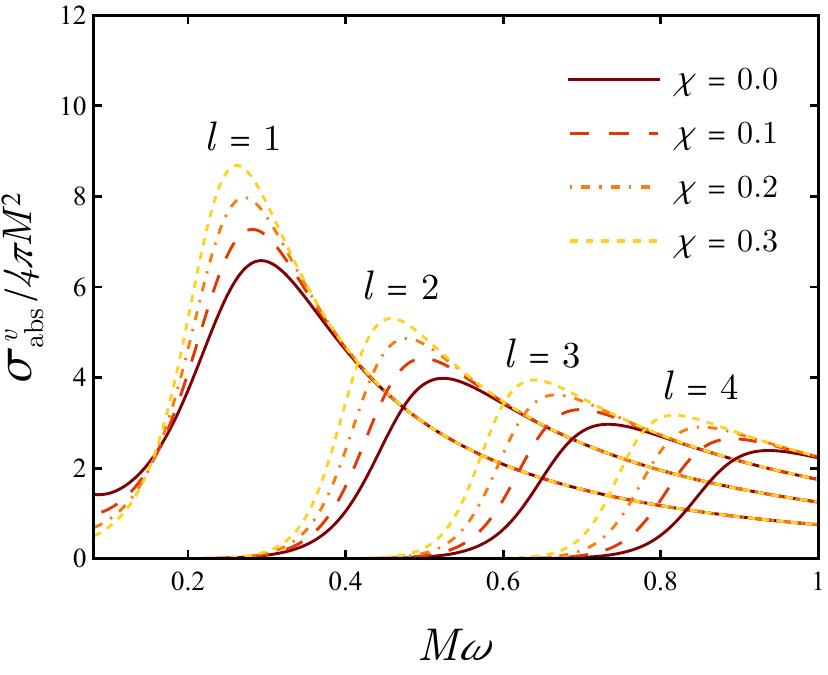}\\
     \includegraphics[scale=0.5]{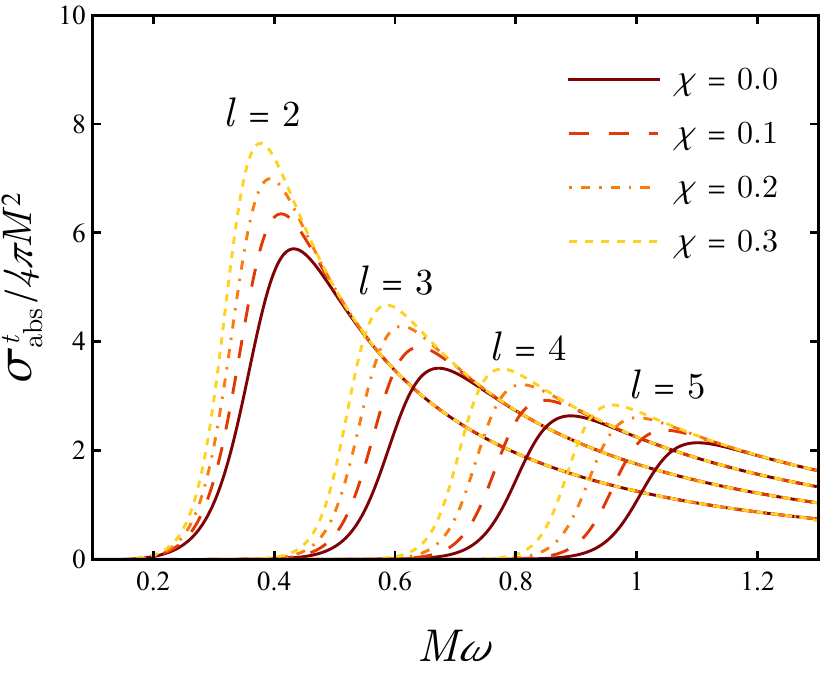}
     \includegraphics[scale=0.5]{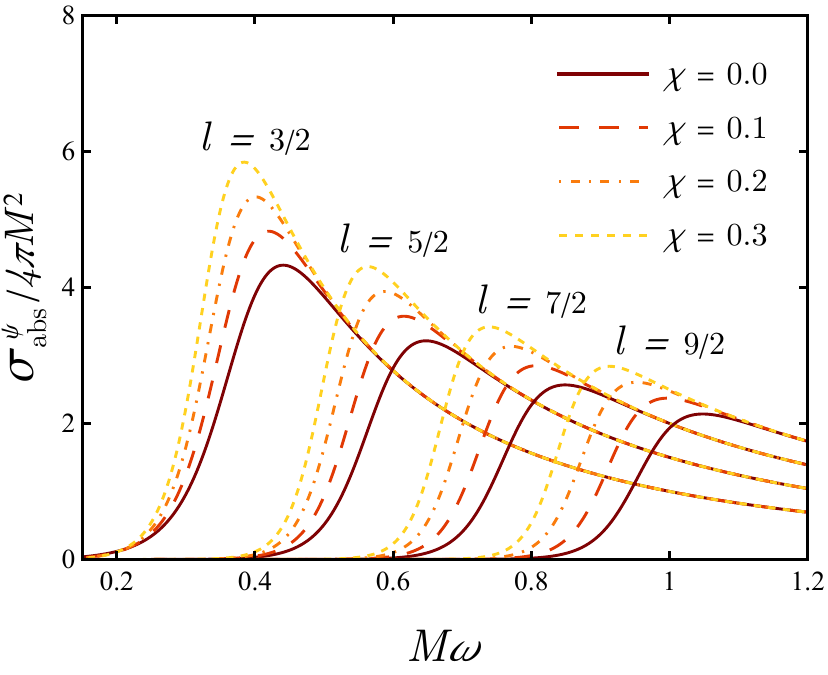}
    \caption{The partial absorption cross sections for scalar ($\sigma^s$), vector ($\sigma^v$), tensor ($\sigma^t$), and Dirac ($\sigma^\psi$) perturbations, in the mass unit, are demonstrated in the bumblebee framework with $M = 1$. For each field, corresponding multipole modes are shown. The Lorentz--violating parameter varies from $\chi = 0$ to $3$. }
    \label{Abschi}
\end{figure}

The absorption cross sections for scalar, vector, tensor, and Dirac perturbations are shown in Figs.~\ref{Abschi}, each plotted over the dimensionless frequency $M\omega$ for several multipole numbers. All fields exhibit a similar qualitative pattern. For every spin sector, the lowest multipole number yields the largest contribution, confirming that low-$l$ modes dominate the absorption spectrum.

A central feature common to all perturbations is the influence of the Lorentz--violating parameter $\chi$. For each spin and for every multipole value, increasing $\chi$ produces two systematic effects. First, the height of the peak in the absorption cross section increases, and second, the frequency at which this peak occurs shifts to lower values of $M\omega$.
These trends indicate that larger values of $\chi$ enhance the transmissivity of the black hole potential barrier, allowing incoming waves to be absorbed more efficiently and at earlier frequencies. This behavior is fully consistent with our previous findings for the effective potentials and greybody factors in Sec. \ref{sec:GBF}. A higher $\chi$ reduces the height of the corresponding effective potential barrier, which in turn leads to larger greybody factors and thus a higher absorption probability. 

\begin{figure}
    \centering
     \includegraphics[scale=0.7]{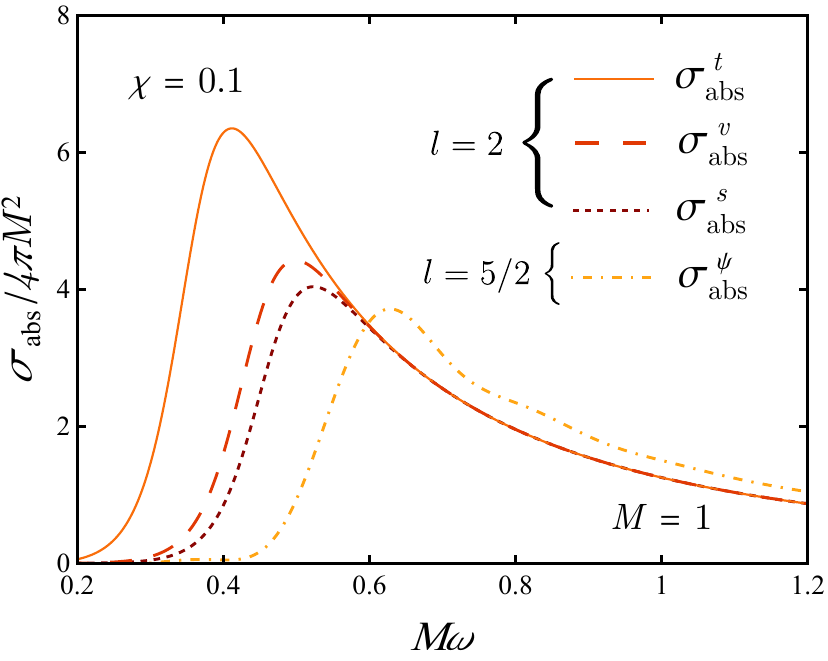}
    \caption{Partial absorption cross sections for scalar $(l = 2)$, vector $(l = 2)$, tensor $(l = 2)$, and Dirac $(l = 5/2)$ fields in the black hole background with $M = 1$ and bumblbee parameter $\chi = 0.1$. The curves show the dependence of the normalized absorption cross section $\sigma_{\text{abs}} / 4\pi M^{2}$ on the dimensionless frequency $M\omega$.}
    \label{fig:AbsCompare}
\end{figure}

A direct comparison of all four fields is provided in Fig.~\ref{fig:AbsCompare}, where a representative multipole mode from each spin sector is plotted. At fixed $\chi$, the peak amplitudes follow a clear hierarchy:
\begin{equation}
\sigma^{t} > \sigma^{v} > \sigma^{s} > \sigma^{\psi}.    
\end{equation}
This ordering matches the behavior of the effective potentials discussed earlier, where tensor modes encounter the lowest barrier and Dirac modes the highest in Fig. \ref{fig:veffComp}. The peak frequencies follow a similar ordering to greybody factor in Fig. \ref{fig:GBFall}. The tensor modes peak first, followed by vector and scalar modes, while Dirac modes require the largest $M\omega$ to reach their maximum. The combined behavior demonstrates how both the spin of the field and the Bumblebee framework influence the dominant absorption features.

\section{Greybody bounds }

As radiation emitted near the horizon propagates outward, the geometric structure of the surrounding spacetime reshapes the outgoing flux and prevents it from retaining a perfect thermal form. This distortion is quantified through greybody factors, which measure how the background modifies the transmission of different particle species. Furthermore, Ref.~\cite{Boonserm:2008zg} introduced analytic bounds that permit estimating the transmission probabilities without relying on numerical routines. These bounds are especially useful because many traditional techniques depend on approximations that lose accuracy in the intermediate--frequency regime or fail for certain spins, such as the electromagnetic case. The method developed in Ref.~\cite{Boonserm:2008zg} applies to arbitrary spin and angular momentum and does not require any assumption about the black hole interior. For this reason, the bounds provide an independent and powerful way to evaluate how the structure of the effective potential shapes the transmission process, complementing the direct computation of the greybody factors, which will be carried out in the next section.

With this point in mind, the next section examines these factors by treating the spin of the emitted particles as a central element of the discussion. Scalar, vector, tensor, and fermionic modes are considered separately. The construction of the corresponding effective potentials relies on developments recently presented in Ref.~\cite{AraujoFilho:2025zaj}, where the variable separation for each spin sector was laid out in detail.

According to Ref.~\cite{Boonserm:2008zg}, one may obtain a rigorous analytic bound that places a minimum value on the transmission probability $|T_b|$ as
\ie
\label{tgmetric}
|T_{b}| \ge {\mathop{\rm sech}\nolimits} ^2 \left(\int_\infty^ {+\infty} {\mathfrak{G} \,\rm{d}}r^{*}\right),
\fe
in which 
\ie
\mathfrak{G} = \frac{{\sqrt {{{(\Tilde{\varsigma}')}^2} + {{({\omega ^2} - \mathrm{V}_{s,v,t,\psi} - {\Tilde{\varsigma}^2})}^2}} }}{{2\Tilde{\varsigma}}}.
\fe

An important point in the derivation is that the auxiliary function $\Tilde{\varsigma}$ must remain positive everywhere and approach the frequency $\omega$ at both asymptotic ends and $\mathrm{V}_{s,v,t,\psi}$ is the effective potential for scalar, vector, tensor and Dirac field. Imposing $\Tilde{\varsigma}=\omega$ throughout the entire domain leads to a simplified version of Eq.~\eqref{tgmetric}, which then reduces to
\ie
\begin{split}
& |T_{b}^{s,v,t,\psi}|  \ge {\mathop{\rm sech}\nolimits} ^2 \left[\int_{-\infty}^ {+\infty} \frac{\mathrm{V}_{s,v,t,\psi}} {2\omega}\mathrm{d}r^{*}\right] \ge {\mathop{\rm sech}\nolimits} ^2 \left[\int_{r_{ h}}^ {+\infty} \frac{\mathrm{V}_{s,v,t,\psi}} {2\omega \sqrt{A(r,\chi) B(r,\chi)}}\mathrm{d} r\right].
\label{boundsgrey}
\end{split}
\fe

The analysis that follows is organized by spin sector, examining separately the transmission properties of scalar, vector, spinor, and tensor fields.


\subsection{Spin 0 }

To begin the analysis, the spin--0 sector is considered first. The scalar perturbations are governed by the corresponding effective potential, which can be obtained by considering $s = 0$ in Eq. \eqref{eq:Veffg} as
\ie
\label{scalarv}
\mathrm{V}_{s} = \frac{1}{\chi +1}\left(1-\frac{2 M}{r}\right) \left(\frac{l (l+1)}{r^2}+\frac{2 M}{r^3 (\chi +1)}\right).
\fe

By substituting the scalar potential from Eq.~(\ref{scalarv}) into the general expression (\ref{boundsgrey}), the resulting lower limits for the scalar greybody factors follow as
\ie
|T_{b}^{s}|  = {\mathop{\rm sech}\nolimits} ^2 \left[ \frac{2 l (l+1) (\chi +1)+1}{(2 \omega ) (4 M (\chi +1))}      \right].
\fe

\begin{figure}
    \centering
     \includegraphics[scale=0.55]{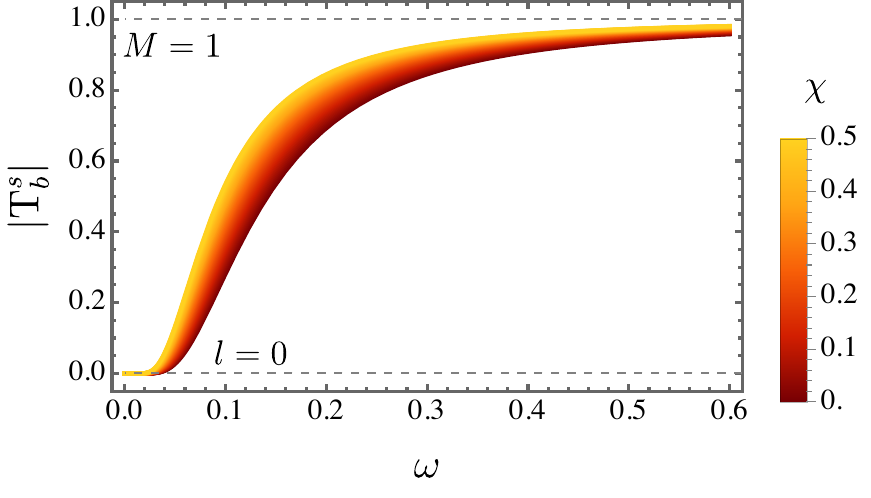}
     \includegraphics[scale=0.55]{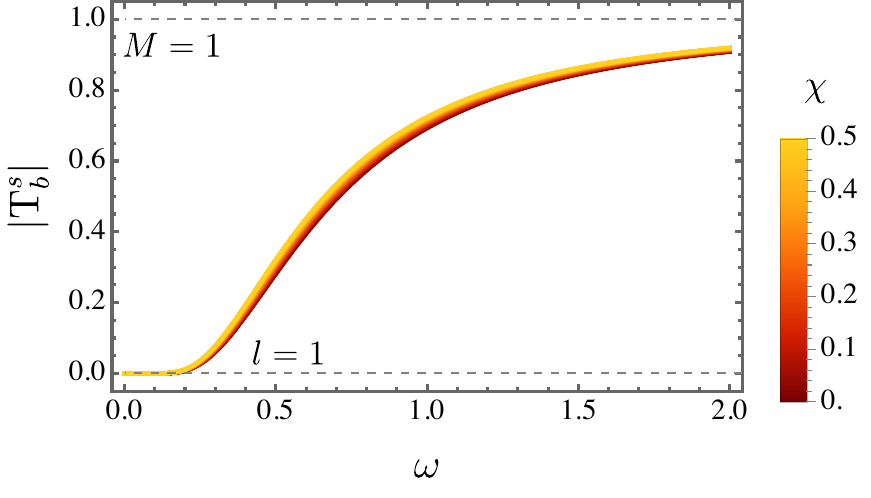}
     \includegraphics[scale=0.55]{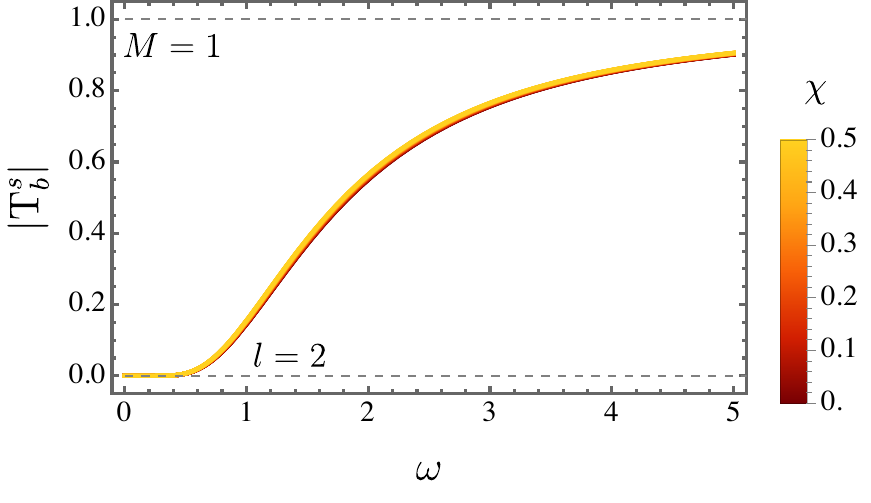}
    \caption{Scalar greybody bounds $|T_{b}^{s}|$ as functions of the frequency $\omega$ for several values of $\chi$. The cases $l=0$, $l=1$, and $l=2$ appear in the upper--left, upper--right, and lower panels, respectively.}
    \label{greybodyboundsscalar}
\end{figure}

Figure \ref{greybodyboundsscalar} presents the behavior of the scalar greybody bounds $|T_{b}^{s}|$ as functions of the frequency $\omega$. The panels show that larger values of $\chi$ enhance the transmission bound for spin--0 modes. The cases $l=0$, $l=1$, and $l=2$ are displayed in the upper--left, upper--right, and lower panels, respectively.

\subsection{Spin 1 }

Following the same procedure adopted for the scalar sector, the analysis of vector perturbations begins with the corresponding effective potential in Eq. \eqref{eq:Veffg}, which takes the following form  for $s = 1$
\ie
\label{vcalarv}
\mathrm{V}_{v} = \frac{1}{\chi +1}\left(1-\frac{2 M}{r}\right) \left(\frac{l (l+1)}{r^2}\right).
\fe

Unlike the Schwarzschild case or the standard bumblebee solution, the vector sector of this geometry reflects a direct influence of the Lorentz--violating parameter. Substituting the potential in Eq.~(\ref{vcalarv}) into the general expression (\ref{boundsgrey}) leads to the corresponding bounds for the vector greybody factors, which take the form
\ie
|T_{b}^{v}|  = {\mathop{\rm sech}\nolimits} ^2 \left[ \frac{l (l+1)}{(2 \omega ) 2 M}     \right].
\fe
Interestingly, even though the effective potential $\mathrm{V}_{v}$ carries an explicit dependence on the Lorentz--violating parameter $\chi$, the resulting greybody bounds for the vector sector do not inherit this dependence. The bounds remain unchanged, leading to the same outcome obtained for vector perturbations in the Schwarzschild black hole.


\subsection{Spin 2 }

In line with the procedure adopted for the lower--spin sectors, the tensor perturbations are governed by the corresponding effective potential, whose explicit form is derived by applying $s = 2$ in Eq. \eqref{eq:Veffg}
\ie
\label{tcalarv}
\mathrm{V}_{t} = \frac{1}{1+\chi} \left(1-\frac{2 M}{r}\right) \left(\frac{l (l+1)}{r^2}-\frac{6 M}{r^3 (\chi +1)}-\frac{2 \chi }{r^2 (\chi +1)}\right).
\fe

\begin{figure}
    \centering
     \includegraphics[scale=0.55]{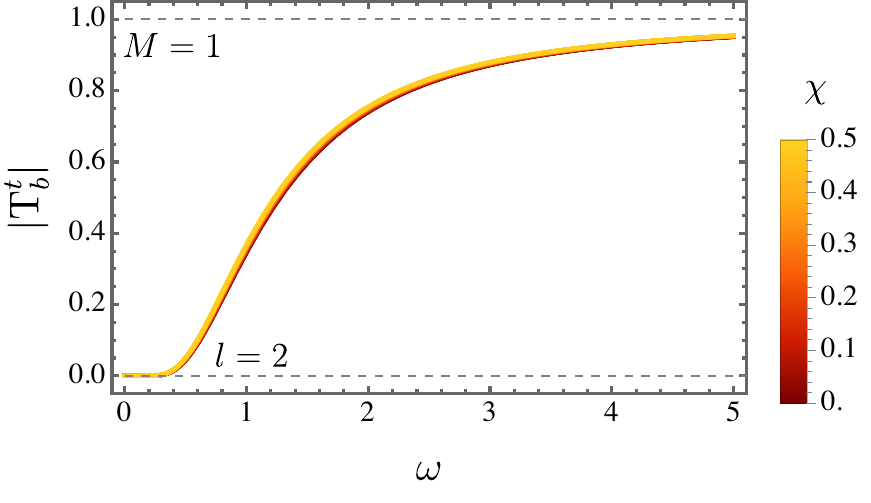}
     \includegraphics[scale=0.55]{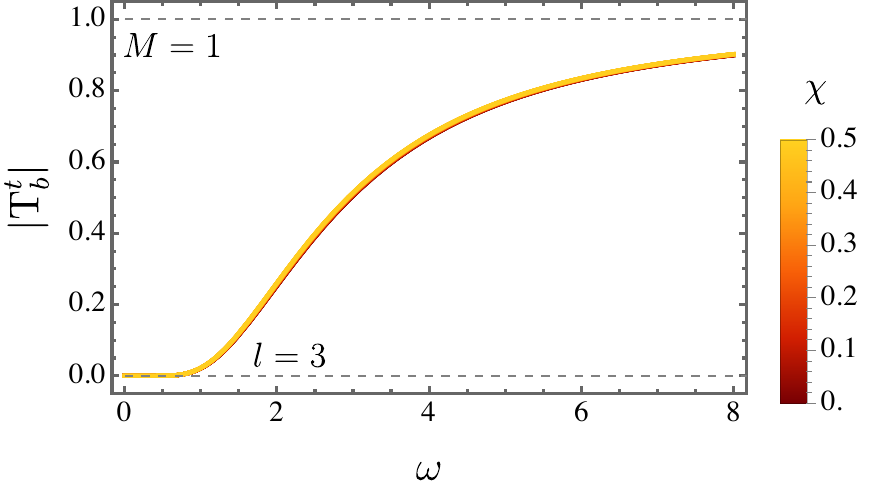}
     \includegraphics[scale=0.55]{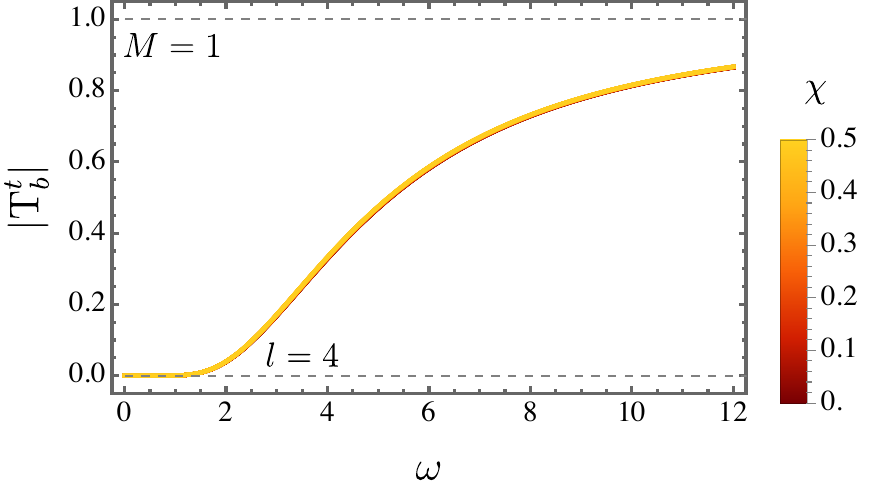}
    \caption{Greybody bounds $|T_{b}^{t}|$ as functions of the frequency $\omega$ for several values of $\chi$. The configurations {$l=2$}, {$l=3$}, and {$l=4$} are displayed in the upper--left, upper--right, and lower panels, respectively.}
    \label{greybodyboundstensor}
\end{figure}

By inserting the tensor potential from Eq.~(\ref{tcalarv}) into the general expression (\ref{boundsgrey}), the associated bounds for the tensor greybody factors follow as
\ie
|T_{b}^{t}|  = {\mathop{\rm sech}\nolimits} ^2 \left[  \frac{2 l (l+1) (\chi +1)-4 \chi -3}{(2 \omega ) (4 M (\chi +1))}   \right].
\fe
In contrast with the vector sector, the tensor greybody bounds do acquire a dependence on $\chi$, reflecting the influence of the Lorentz--violating contribution in this case. Figure \ref{greybodyboundstensor} illustrates the behavior of $|T_{b}^{t}|$ as a function of the frequency $\omega$ for the configurations {$l=2$} (upper--left panel), {$l=3$} (upper--right panel), and {$l=4$} (lower panel). In general lines, larger values of $\chi$ lead to an enhancement of the tensor bound $|T_{b}^{t}|$.

\subsection{Spin 1/2 }

Finally, the analysis of fermionic modes starts from the effective potential governing the spinor perturbations, whose expression is provided in Eq. \eqref{eq:Veffpsi} and has the following form
\ie
\label{spinpotential}
\mathrm{V}_{\psi} = \frac{\left(l+\frac{1}{2}\right)^2 \left(1-\frac{2 M}{r}\right)}{r^2 (\chi +1)}+\left(l+\frac{1}{2}\right) \left(\frac{M}{r^3 (\chi +1) \sqrt{\frac{1-\frac{2 M}{r}}{\chi +1}}}-\frac{\sqrt{\frac{1-\frac{2 M}{r}}{\chi +1}}}{r^2}\right) \sqrt{\frac{\left(1-\frac{2 M}{r}\right)^2}{(\chi +1)^2}}.
\fe

Substituting the fermionic potential from Eq.~(\ref{spinpotential}) into the general formula (\ref{boundsgrey}) yields the corresponding bounds for the spinor greybody factors, which take the form
\ie
|T_{b}^{\psi}|  = {\mathop{\rm sech}\nolimits} ^2 \left[  \frac{1}{2\omega} \left( \frac{(2 l+1)^2}{8 M} \right)   \right].
\fe

\begin{figure}
    \centering
     \includegraphics[scale=0.7]{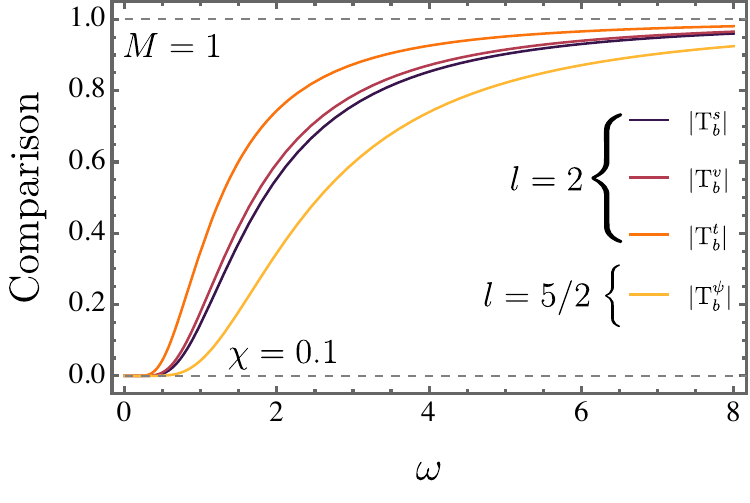}
    \caption{Comparison of all greybody bounds considered in this work: scalar, vector, tensor, and spinor sectors.}
    \label{boundscomparrr}
\end{figure}

As in the vector sector, the fermionic bound shows no dependence on the Lorentz–violating parameter $\chi$. Despite this feature, it is still instructive to contrast the behavior of all bounds obtained throughout this work. Figure \ref{boundscomparrr} presents such a comparison. For the choice $\chi = 0.1$ and angular momentum $l = 2$ for the bosonic modes and $l = 5/2$ for the spinor case, the resulting hierarchy is
$$
|T_{b}^{t}| > |T_{b}^{v}| > |T_{b}^{s}| >|T_{b}^{\psi}|.
$$
This result is completely consistent with the exploration of greybody factor in previos section (Eq. \eqref{eq:compT}.


\section{Evaporation lifetime }

This part of the paper turns to the qualitative behavior of the evaporation stage. Rather than beginning with the quantum description, the discussion is organized around the thermodynamic route, where the rate of energy loss is estimated by invoking the \textit{Stefan--Boltzmann} prescription. Within this approach, the luminosity associated with Hawking radiation is treated as the dominant mechanism driving the decrease of the black hole mass, allowing one to track the evaporation trend without committing to a specific particle spectrum \cite{ong2018effective}
\ie
\label{slawbotz}
\frac{\mathrm{d}M}{\mathrm{d}t}  =  - a  |T_{b}^{s,v,t,\psi}|  \sigma^{s,v,t,\psi}_{l \omega} \, T^{4},
\fe
with
\ie
\sigma^{s,v,t,\psi}_{l\,\omega} = \frac{\pi(2l + 1)}{\omega^{2}} |T_{b}^{s,v,t,\psi}|.
\fe

The symbols used in the emission rate deserve clarification before proceeding. The quantity $\sigma_{l\omega}$ corresponds to the partial absorption area for each mode, $a$ is the usual radiation constant, $T$ stands for the Hawking temperature, and the terms $|T_{b}^{s,v,t,\psi}|$ encode the transmission coefficients associated with the various perturbative sectors. To streamline the analytical treatment, the analysis shifts from the full greybody factors to their corresponding bounds, which provide manageable expressions (analytical) that take into account the relevant behavior.

As shown in the previous sections, the bounds associated with the scalar, vector, tensor, and spinor sectors—$|T_{b}^{s}|$, $|T_{b}^{v}|$, $|T_{b}^{t}|$, and $|T_{b}^{\psi}|$—have been obtained. All spin assignments ($0$, $1$, $2$, and $1/2$) will be used here to extract the corresponding evaporation lifetimes in analytical form. Each spin sector produces its own lifetime expression, allowing a direct comparison between them. In addition, the high--frequency regime will be included in the discussion.


\subsection{Spin 0 }

The evaluation of the evaporation time for the spin–$0$ sector begins by inserting the scalar transmission bound $|T_{b}^{s}|$ together with the partial absorption area $\sigma^{s}_{l\omega}$ into the radiative loss formula of Eq.~(\ref{slawbotz}). Notice that these inputs determine the mass–loss rate that governs the evolution of the black hole mass. In the scalar case, the resulting expression for $\mathrm{d}M/\mathrm{d}t$ becomes algebraically cumbersome, so it is not displayed. Once Eq.~(\ref{slawbotz}) is integrated with the scalar contributions, the corresponding evaporation time emerges as
\ie
\int \mathrm{d}t = \int_{M_{f}}^{M_{i}} \frac{\mathrm{d}M}{a  \Bar{\Gamma}_{l \omega}  \sigma_{l \omega} \, T^{4}},
\label{tgggg}
\fe
{where $\bar{\Gamma}_{l\omega}$ denotes the greybody factors as well. In addition, as will be discussed in the following sections, when addressing the high--frequency regime, we introduce, for simplicity, the constant $\xi \equiv a\,\bar{\Gamma}_{l\omega}$, which approaches unity within this approximation.}

The next step involves rewriting the integrand appearing in the lifetime expression. The combination $1/(a\, |T_{b}^{s}|\, \sigma_{l\omega}\, T^{4})$ is approximated to first order in $\chi$, which provides a workable expression for the scalar channel. Once this expansion is inserted into Eq.~(\ref{tgggg}) and the relevant substitutions are made, the evaporation time for this sector follows as
\ie
\begin{split}
t^{s}_{evap} = & \int_{M_{f}}^{M_{i}}  \frac{2048 \pi ^4 M^4 \omega ^2 \left(\cosh \left(\frac{2 l (l+1)+\frac{1}{\chi +1}}{4 M \omega }\right)+1\right)}{(2 l+1) (\chi -1)^4}\mathrm{d}M, \\
 \approx \, &  \int_{M_{f}}^{M_{i}}  \left\{ \frac{2048 \left[\pi ^4 M^4 \omega ^2 \left(\cosh \left(\frac{2 l (l+1)+1}{4 M \omega }\right)+1\right)\right]}{2 l+1}  \right. \\
 & \left. + \frac{512 \chi  \left(\pi ^4 M^3 \omega  \left(16 M \omega  \left(\cosh \left(\frac{2 l (l+1)+1}{4 M \omega }\right)+1\right)-\sinh \left(\frac{2 l (l+1)+1}{4 M \omega }\right)\right)\right)}{2 l+1} \right\} \mathrm{d}M.
\end{split}
\fe
Carrying out the integration described above yields an analytical expression for the evaporation time associated with the scalar sector:
\ie
\begin{split}
& t^{s}_{evap} =  \, \frac{\pi ^4}{120 (2 l+1) \omega ^3} \times \Bigg\{   -49152 (4 \chi +1) \omega ^5 \left(M_{f}^5-M_{i}^5\right) \\ 
& -8 M_{f} \omega \Bigg[  32 (2 l (l+1)+1) M_{f}^2 \omega ^2 (2 l (l+1) (4 \chi +1)-\chi +1) \\
& +(2 l (l+1)+1)^3 (2 l (l+1) (4 \chi +1)-\chi +1)+6144 M_{f}^4 (4 \chi +1) \omega ^4  \Bigg] \cosh \left(\frac{2 l (l+1)+1}{4 M_{f} \omega }\right)  \\
&  8 M_{i} \omega  \Bigg[ 32 (2 l (l+1)+1) M_{i}^2 \omega ^2 (2 l (l+1) (4 \chi +1)-\chi +1)\\
& +(2 l (l+1)+1)^3 (2 l (l+1) (4 \chi +1)-\chi +1)+6144 M_{i}^4 (4 \chi +1) \omega ^4   \Bigg] \cosh \left(\frac{2 l (l+1)+1}{4 M_{i} \omega }\right)  \\
&  -\ln \left(-\frac{1}{M_{f}}\right)-\ln (M_{f})-\ln \left(\frac{1}{M_{i}}\right)+\ln \left(-\frac{1}{M_{i}}\right)   \\
& - \Bigg[ (8 l (l+1)-1) (2 l (l+1)+1)^4 \chi \\
& +2 l (l+1) (4 l (l+1) (2 l (l+1) (l (l+1) (2 l (l+1)+5)+5)+5)+5)       \Bigg] \\
& \times \Bigg[ \ln \left(-\frac{1}{M_{f}}\right)+\ln (M_{f})+\ln \left(\frac{1}{M_{i}}\right)-\ln \left(-\frac{1}{M_{i}}\right)  \Bigg] + 32 \omega ^2 (2 l (l+1) (4 \chi +1)-\chi +1) \\
& \times \Bigg[ M_{i}^2 \left((2 l (l+1)+1)^2+96 M_{i}^2 \omega ^2\right) \sinh \left(\frac{2 l (l+1)+1}{4 M_{i} \omega }\right)\\
& -M_{f}^2 \left((2 l (l+1)+1)^2+96 M_{f}^2 \omega ^2\right) \sinh \left(\frac{2 l (l+1)+1}{4 M_{f} \omega }\right) \Bigg] \\
& + 2 (2 l (l+1)+1)^4 (2 l (l+1) (4 \chi +1)-\chi +1) \times \text{Shi}\left(\frac{2 l (l+1)+1}{4 M_{f} \omega }\right) -2 \text{Shi}\left(\frac{2 l (l+1)+1}{4 M_{i} \omega }\right)  \\
&  - 2 \Bigg[  (8 l (l+1)-1) (2 l (l+1)+1)^4 \chi \\
& +2 l (l+1) (4 l (l+1) (2 l (l+1) (l (l+1) (2 l (l+1)+5)+5)+5)+5)  \Bigg] \times \text{Shi}\left(\frac{2 l (l+1)+1}{4 M_{i} \omega }\right)    \Bigg\},
\end{split}
\fe
where $\text{Shi}(x)$ is the hyperbolic sine integral,
\ie
\operatorname{Shi}(x) = \int_{0}^{x} \frac{\sinh t}{t}\,\mathrm{d}t .
\fe

In Tab.~\ref{scalarevaporation}, the numerical values of $t^{s}_{\mathrm{evap}}$ are listed (for $l=2$). An increase in $\chi$ leads to a longer evaporation time, whereas fixing $\chi$ and raising $\omega$ shortens the lifetime. Therefore, in the scalar sector, the Lorentz--violating parameter $\chi$ acts to prolong the black hole’s evaporation process.

\begin{table}[!h]
\begin{center}
\begin{tabular}{c c c  || c c c } 
 \hline\hline \hline
$\chi$ & $\omega$ & $t^{s}_{evap}$ & $\chi$ & $\omega$ &  $t^{s}_{evap}$    \\ [0.2ex] 
 \hline
   0.01 & 0.90 & $3.39906\times 10^7$   & 0.1 & 0.10 & $2.41085\times 10^{66}$     \\
  
   0.1 & 0.90 & $4.22951\times 10^7$  & 0.1 & 0.20 & $5.14607\times 10^{32}$     \\
  
   0.2 & 0.90 & $5.15223\times 10^7$   & 0.1 & 0.30 & $3.96796\times 10^{21}$    \\
  
   0.3 & 0.90 & $6.07495\times 10^7$  & 0.1 & 0.40 & $1.43031\times 10^{16}$    \\
  
    0.4 & 0.90 & $6.99767\times 10^7$  & 0.1 & 0.50 & $9.17926\times 10^{12}$ \\
   
  0.5 & 0.90 & $7.92039\times 10^7$  & 0.1 & 0.60 & $7.69712\times 10^{10}$   \\ 

  0.6 & 0.90 & $8.84312\times 10^7$  & 0.1 & 0.70 & $2.77542\times 10^9$   \\
 
  0.7 & 0.90 & $9.76584\times 10^7$  & 0.1 & 0.80 & $2.49417\times 10^8$   \\
 
  0.8 & 0.90 & $1.06886\times 10^8$  & 0.1 & 0.90 & $4.22951\times 10^7$  \\
 
  0.9 & 0.90 & $1.16113\times 10^8$  & 0.1 & 0.99 & $1.33082\times 10^7$  \\ 
 [0.2ex] 
 \hline \hline \hline
\end{tabular}
\caption{\label{scalarevaporation} The evaporation time associated with the scalar sector, $t^{s}_{\mathrm{evap}}$, is displayed for various choices of $\chi$ and $\omega$. Here, it is considered $l=2$. }
\end{center}
\end{table}


\subsection{Spin 1 }

The analysis of the evaporation time for the spin--$1$ sector begins by inserting the vector bound $|T_{b}^{v}|$ and the corresponding mode--dependent cross section $\sigma^{v}_{l\omega}$ into the radiative loss relation of Eq.~(\ref{slawbotz}). These ingredients define the mass--loss rate for the vector channel, although the explicit form of $\mathrm{d}M/\mathrm{d}t$. To obtain the lifetime, the calculation proceeds through Eq.~(\ref{tgggg}). The integrand is reorganized by expanding up to first order in $\chi$, which provides a manageable approximation for the subsequent steps, i.e., as we did to the scalar case. Substituting this expansion into Eq.~(\ref{tgggg}) and inserting the appropriate expressions yields the evaporation time for the vector configuration
\ie
\begin{split}
t^{v}_{evap} = & \int_{M_{f}}^{M_{i}}  \frac{4096 \pi ^4 M^4 \omega ^2 \cosh ^2\left(\frac{l (l+1)}{4 M \omega }\right)}{(2 l+1) (\chi -1)^4} \mathrm{d}M, \\
 \approx \, &  \int_{M_{f}}^{M_{i}}  \left\{ \frac{8192 \chi  \left(\pi ^4 M^4 \omega ^2 \left(\cosh \left(\frac{l (l+1)}{2 M \omega }\right)+1\right)\right)}{2 l+1}+\frac{4096 \left(\pi ^4 M^4 \omega ^2 \cosh ^2\left(\frac{l (l+1)}{4 M \omega }\right)\right)}{2 l+1} \right\} \mathrm{d}M.
\end{split}
\fe

Once the integration is completed, an explicit closed form for the evaporation time associated with the vector perturbations:
\ie
\begin{split}
& t^{v}_{evap} =  \, \frac{4 \pi ^4 (4 \chi +1)}{15 (2 l+1) \omega ^3}  \Bigg\{ -4 M_{f} \omega  \left(l^4 (l+1)^4+8 l^2 (l+1)^2 M_{f}^2 \omega ^2+384 M_{f}^4 \omega ^4\right) \cosh \left(\frac{l (l+1)}{2 M_{f} \omega }\right) \\
& -1536 \omega ^5 \left(M_{f}^5-M_{i}^5\right)  +4 M_{i} \omega  \left(l^4 (l+1)^4+8 l^2 (l+1)^2 M_{i}^2 \omega ^2+384 M_{i}^4 \omega ^4\right) \cosh \left(\frac{l (l+1)}{2 M_{i} \omega }\right) \\
& +l^5 (l+1)^5 \left(\ln \left(\frac{1}{M_{f}}\right)-\ln \left(-\frac{1}{M_{f}}\right)-\ln \left(\frac{1}{M_{i}}\right)+\ln \left(-\frac{1}{M_{i}}\right)\right) \\
& +8 l (l+1) \omega ^2 \left(M_{i}^2 \left(l^2 (l+1)^2+24 M_{i}^2 \omega ^2\right) \sinh \left(\frac{l (l+1)}{2 M_{i} \omega }\right) \right. \\
& \left. -M_{f}^2 \left(l^2 (l+1)^2+24 M_{f}^2 \omega ^2\right) \sinh \left(\frac{l (l+1)}{2 M_{f} \omega }\right)\right) \\
& +2 l^5 (l+1)^5 \left(\text{Shi}\left(\frac{l (l+1)}{2 M_{f} \omega }\right)-\text{Shi}\left(\frac{l (l+1)}{2 M_{i} \omega }\right)\right)  \Bigg\}.
\end{split}
\fe

Table \ref{vectorevaporation} summarises the values obtained for $t^{v}_{\mathrm{evap}}$ across different choices of $\chi$ and $\omega$ (considering $l=2$). Larger values of $\chi$ push the evaporation time upward, while keeping $\chi$ fixed and increasing the frequency produces the opposite trend. In other words, within the spin--$1$ channel, the presence of the Lorentz--violating parameter effectively extends the lifetime of the black hole.

\begin{table}[!h]
\begin{center}
\begin{tabular}{c c c  || c c c } 
 \hline\hline \hline
$\chi$ & $\omega$ & $t^{v}_{evap}$ & $\chi$ & $\omega$ &  $t^{v}_{evap}$    \\ [0.2ex] 
 \hline
   0.01 & 0.90 & $1.07235\times 10^7$   & 0.1 & 0.10 & $8.65254\times 10^{61}$     \\
  
   0.1 & 0.90 & $1.44355\times 10^7$  & 0.1 & 0.20 & $1.93674\times 10^{30}$     \\
  
   0.2 & 0.90 & $1.85599\times 10^7$   & 0.1 & 0.30 & $9.50924\times 10^{19}$    \\
  
   0.3 & 0.90 & $2.26843\times 10^7$  & 0.1 & 0.40 & $8.82927\times 10^{14}$    \\
  
    0.4 & 0.90 & $2.68088\times 10^7$  & 0.1 & 0.50 & $1.00673\times 10^{12}$ \\
   
  0.5 & 0.90 & $3.09332\times 10^7$  & 0.1 & 0.60 & $1.24422\times 10^{10}$   \\ 

  0.6 & 0.90 & $3.50576\times 10^7$  & 0.1 & 0.70 & $5.95761\times 10^8$   \\
 
  0.7 & 0.90 & $3.9182\times 10^7$  & 0.1 & 0.80 & $6.74747\times 10^7$   \\
 
  0.8 & 0.90 & $4.33065\times 10^7$  & 0.1 & 0.90 & $1.44355\times 10^7$  \\
 
  0.9 & 0.90 & $4.74309\times 10^7$  & 0.1 & 0.99 & $5.80059\times 10^6$  \\ 
 [0.2ex] 
 \hline \hline \hline
\end{tabular}
\caption{\label{vectorevaporation} Evaporation time in the vector sector, $t^{v}_{\mathrm{evap}}$, for different values of $\chi$ and $\omega$. Here, it is considered $l=2$. }
\end{center}
\end{table}


\subsection{Spin 2 }

The treatment of the spin–$2$ sector follows a similar strategy to the previous analyses but is reorganized here for clarity. The tensor bound $|T_{b}^{t}|$ and its corresponding partial cross section $\sigma^{t}_{l\omega}$ are inserted into the radiative loss relation of Eq.~(\ref{slawbotz}), which determines the mass–loss rate governing this channel. The evaporation time is then extracted through Eq.~(\ref{tgggg}). As in the scalar and vector cases, the integrand is approximated by expanding it to first order in $\chi$, providing an expression that can be handled analytically. After introducing this expansion and substituting the relevant quantities into Eq.~(\ref{tgggg}), the resulting lifetime for the tensor configuration is obtained
\ie
\begin{split}
t^{t}_{evap} = & \int_{M_{f}}^{M_{i}}  \frac{4096 \pi ^4 M^4 \omega ^2 \cosh ^2\left(\frac{2 l (l+1) (\chi +1)-4 \chi -3}{8 M (\chi +1) \omega }\right)}{(2 l+1) (\chi -1)^4} \mathrm{d}M, \\
 \approx \, &  \int_{M_{f}}^{M_{i}}  \left\{ \frac{4096 \left(\pi ^4 M^4 \omega ^2 \cosh ^2\left(\frac{2 l (l+1)-3}{8 M \omega }\right)\right)}{2 l+1} \right. \\
 & \left. + \frac{512 \chi  \left(\pi ^4 M^3 \omega  \left(\sinh \left(\frac{3-2 l (l+1)}{4 M \omega }\right)+16 M \omega  \left(\cosh \left(\frac{3-2 l (l+1)}{4 M \omega }\right)+1\right)\right)\right)}{2 l+1} \right\} \mathrm{d}M.
\end{split}
\fe
Once the integration is carried out, the procedure yields a closed analytical form for the evaporation time associated with the tensor sector:
\ie
\begin{split}
& t^{t}_{evap} =  \, \frac{\pi ^4}{120 (2 l+1) \omega ^3} \times \Bigg\{ -(3-2 l (l+1))^4 (2 l (l+1) (4 \chi +1)-17 \chi -3) \text{Chi}\left(\frac{3-2 l (l+1)}{4 M_{f} \omega }\right) \\
& +(3-2 l (l+1))^4 (2 l (l+1) (4 \chi +1)-17 \chi -3) \text{Chi}\left(\frac{2 l (l+1)-3}{4 M_{f} \omega }\right)  \\
& +(3-2 l (l+1))^4 (2 l (l+1) (4 \chi +1)-17 \chi -3) \text{Chi}\left(\frac{3-2 l (l+1)}{4 M_{i} \omega }\right)  \\
& -(3-2 l (l+1))^4 (2 l (l+1) (4 \chi +1)-17 \chi -3) \text{Chi}\left(\frac{2 l (l+1)-3}{4 M_{i} \omega }\right)   \\
& + 8 M_{f} \omega  \Bigg[   -6144 M_{f}^4 (4 \chi +1) \omega ^4 - \left(  (2 l (l+1)-3)^3 (2 l (l+1) (4 \chi +1)-17 \chi -3) \right. \\
& \left. +32 (2 l (l+1)-3) M_{f}^2 \omega ^2 (2 l (l+1) (4 \chi +1)-17 \chi -3)+6144 M_{f}^4 (4 \chi +1) \omega ^4   \right) \cosh \left(\frac{3-2 l (l+1)}{4 M_{f} \omega }\right) \\ 
& +4 M_{f} \omega  (2 l (l+1) (4 \chi +1)-17 \chi -3) \left((3-2 l (l+1))^2+96 M_{f}^2 \omega ^2\right) \sinh \left(\frac{3-2 l (l+1)}{4 M_{f} \omega }\right) \Bigg]   \\
&  - 8 M_{i} \omega  \Bigg[   -6144 M_{i}^4 (4 \chi +1) \omega ^4 - \left(  (2 l (l+1)-3)^3 (2 l (l+1) (4 \chi +1)-17 \chi -3) \right. \\
& \left. +32 (2 l (l+1)-3) M_{i}^2 \omega ^2 (2 l (l+1) (4 \chi +1)-17 \chi -3)+6144 M_{i}^4 (4 \chi +1) \omega ^4   \right) \cosh \left(\frac{3-2 l (l+1)}{4 M_{i} \omega }\right) \\ 
& +4 M_{i} \omega  (2 l (l+1) (4 \chi +1)-17 \chi -3) \left((3-2 l (l+1))^2+96 M_{i}^2 \omega ^2\right) \sinh \left(\frac{3-2 l (l+1)}{4 M_{i} \omega }\right)    \\
& -(3-2 l (l+1))^4 (2 l (l+1) (4 \chi +1)-17 \chi -3) \text{Shi}\left(\frac{3-2 l (l+1)}{4 M_{f} \omega }\right) -243 \text{Shi}\left(\frac{2 l (l+1)-3}{4 M_{f} \omega }\right) \\
& \left((8 l (l+1)-17) (3-2 l (l+1))^4 \chi +2 l (l+1) (4 l (l+1) (2 l (l+1) (l (l+1) (2 l (l+1)-15)+45) \right. \\
& \left. -135)+405)\right) \text{Shi}\left(\frac{2 l (l+1)-3}{4 M_{f} \omega }\right) + (3-2 l (l+1))^4 (2 l (l+1) (4 \chi +1)-17 \chi -3) \text{Shi}\left(\frac{3-2 l (l+1)}{4 M_{i} \omega }\right)   \\
& +243 \text{Shi}\left(\frac{2 l (l+1)-3}{4 M_{i} \omega }\right)  \\
&  -\left((8 l (l+1)-17) (3-2 l (l+1))^4 \chi +2 l (l+1) (4 l (l+1) (2 l (l+1) (l (l+1) (2 l (l+1)-15) \right. \\
& \left. +45)-135)+405)\right) \text{Shi}\left(\frac{2 l (l+1)-3}{4 M_{i} \omega }\right)   \Bigg\}  ,
\end{split}
\fe
where the symbol $\text{Chi}(x)$ denotes the hyperbolic cosine integral, a standard special function. Its definition is
\[
\text{Chi}(x)
= \gamma + \ln|x|
  + \int_{0}^{x} \frac{\cosh t - 1}{t}\, \mathrm{d}t ,
\]
with $\gamma$ being the Euler\textendash Mascheroni constant.

Tab.~\ref{tensorevaporation} compiles the values of $t^{t}_{\mathrm{evap}}$ for several combinations of $\chi$ and $\omega$ (for $l=2$). Increasing $\chi$ lengthens the evaporation time, whereas fixing $\chi$ and raising $\omega$ reduces it. Thus, for the spin–$0$, $1$, and $2$ sectors alike, the Lorentz--violating parameter acts to prolong the black hole’s lifetime.

\begin{table}[!h]
\begin{center}
\begin{tabular}{c c c  || c c c } 
 \hline\hline \hline
$\chi$ & $\omega$ & $t^{t}_{evap}$ & $\chi$ & $\omega$ &  $t^{t}_{evap}$    \\ [0.2ex] 
 \hline
   0.01 & 0.90 & $1.06916\times 10^6$   & 0.1 & 0.10 & $6.99741\times 10^{44}$    \\
  
   0.1 & 0.90 & $1.40503\times 10^6$  & 0.1 & 0.20 & $1.07746\times 10^{22}$     \\
  
   0.2 & 0.90 & $1.77821\times 10^6$  & 0.1 & 0.30 & $3.53593\times 10^{14}$   \\
  
   0.3 & 0.90 & $2.15140\times 10^6$ & 0.1 & 0.40 & $8.45709\times 10^{10}$    \\
  
    0.4 & 0.90 & $2.52458\times 10^6$ & 0.1 & 0.50 & $6.86692\times 10^8$ \\
   
  0.5 & 0.90 & $2.89777\times 10^6$  & 0.1 & 0.60 & $6.86692\times 10^8$  \\ 

  0.6 & 0.90 & $3.27095\times 10^6$ & 0.1 & 0.70 & $4.98308\times 10^6$  \\
 
  0.7 & 0.90 & $3.64414\times 10^6$ & 0.1 & 0.80 & $1.93762\times 10^6$   \\
 
  0.8 & 0.90 & $4.01733\times 10^6$ & 0.1 & 0.90 & $1.40503\times 10^6$  \\
 
  0.9 & 0.90 & $4.39051\times 10^6$ & 0.1 & 0.99 & $1.33343\times 10^6$ \\ 
 [0.2ex] 
 \hline \hline \hline
\end{tabular}
\caption{\label{tensorevaporation} Evaporation time in the tensor sector, $t^{t}_{\mathrm{evap}}$, for different values of $\chi$ and $\omega$. Here, it is considered $l=2$. }
\end{center}
\end{table}


\subsection{Spin 1/2 }

Following the procedure adopted for the previous perturbative sectors, the evaporation analysis for the spinor case is developed in an analogous manner
\ie
\begin{split}
t^{\psi}_{evap} = & \int_{M_{f}}^{M_{i}}  \frac{2048 \pi ^4 M^4 \omega ^2 \left(\cosh \left(\frac{(2 l+1)^2}{8 M \omega }\right)+1\right)}{(2 l+1) (\chi -1)^4} \mathrm{d}M, \\
 \approx \, &  \int_{M_{f}}^{M_{i}}  \left\{ \frac{2048 \left(\pi ^4 M^4 \omega ^2 \left(\cosh \left(\frac{(2 l+1)^2}{8 M \omega }\right)+1\right)\right)}{2 l+1} \right. \\
 & \left. + \frac{8192 \chi  \left(\pi ^4 M^4 \omega ^2 \left(\cosh \left(\frac{(2 l+1)^2}{8 M \omega }\right)+1\right)\right)}{2 l+1} \right\} \mathrm{d}M.
\end{split}
\fe
Carrying out the integration leads to a closed analytical expression for the evaporation time, which reads
\ie
\begin{split}
& t^{\psi}_{evap} =  \,\frac{\pi ^4 (4 \chi +1)}{3840 (2 l+1) \omega ^3}\Bigg\{ -1572864 \omega ^5 \left(M_{f}^5-M_{i}^5\right)-\ln \left(-\frac{1}{M_{f}}\right)-\ln (M_{f})-\ln \left(\frac{1}{M_{i}}\right)+\ln \left(-\frac{1}{M_{i}}\right) \\
&  + 4\Bigg[  -4 M_{f} \omega  \left(128 (2 l+1)^4 M_{f}^2 \omega ^2+(2 l+1)^8+98304 M_{f}^4 \omega ^4\right) \cosh \left(\frac{(2 l+1)^2}{8 M_{f} \omega }\right)  \\
&  +4 M_{i} \omega  \left(128 (2 l+1)^4 M_{i}^2 \omega ^2+(2 l+1)^8+98304 M_{i}^4 \omega ^4\right) \cosh \left(\frac{(2 l+1)^2}{8 M_{i} \omega }\right) \\
& - l (l+1) (4 l (2 l+1) (2 l (l+1)+1)+1) (4 l (2 l (l (2 l+5)+5)+5)+5)\\
& \times \Bigg( \ln \left(-\frac{1}{M_{f}}\right)+\ln (M_{f})+\ln \left(\frac{1}{M_{i}}\right)-\ln \left(-\frac{1}{M_{i}}\right) \Bigg)  \\
&  +32 (2 l \omega +\omega )^2 \left(M_{i}^2 \left((2 l+1)^4+384 M_{i}^2 \omega ^2\right) \sinh \left(\frac{(2 l+1)^2}{8 M_{i} \omega }\right)-M_{f}^2 \left((2 l+1)^4+384 M_{f}^2 \omega ^2\right) \right. \\
& \left. \sinh \left(\frac{(2 l+1)^2}{8 M_{f} \omega }\right)\right)   \Bigg] + 2 (2 l+1)^{10} \text{Shi}\left(\frac{(2 l+1)^2}{8 M_{f} \omega }\right)-2 (2 l+1)^{10} \text{Shi}\left(\frac{(2 l+1)^2}{8 M_{i} \omega }\right)   \Bigg\}
\end{split}
\fe

Tab.~\ref{spinorevaporation} lists the values of $t^{\psi}_{\mathrm{evap}}$ for the different choices of $\chi$ and $\omega$ (maintaining $l=5/2$). As in the previous sectors, larger values of $\chi$ increase the evaporation time, while holding $\chi$ fixed and raising $\omega$ decreases it. Consequently, across all spin configurations considered—$0$, $1$, $2$, and $1/2$—the Lorentz--violating parameter consistently lengthens the black hole’s lifetime.

\begin{table}[!h]
\begin{center}
\begin{tabular}{c c c  || c c c } 
 \hline\hline \hline
$\chi$ & $\omega$ & $t^{\psi}_{evap}$ & $\chi$ & $\omega$ &  $t^{\psi}_{evap}$    \\ [0.2ex] 
 \hline
   0.01 & 0.90 & $1.73302\times 10^{10}$  & 0.1 & 0.10 & $1.76968\times 10^{94}$   \\
  
   0.1 & 0.90 & $2.33291\times 10^{10}$  & 0.1 & 0.20 & $2.01883\times 10^{46}$    \\
  
   0.2 & 0.90 & $2.99946\times 10^{10}$  & 0.1 & 0.30 & $3.63138\times 10^{30}$   \\
  
   0.3 & 0.90 & $3.66601\times 10^{10}$ & 0.1 & 0.40 & $6.38495\times 10^{22}$    \\
  
    0.4 & 0.90 & $4.33255\times 10^{10}$ & 0.1 & 0.50 & $1.67488\times 10^{18}$ \\
   
  0.5 & 0.90 & $4.99910\times 10^{10}$  & 0.1 & 0.60 & $1.65549\times 10^{15}$  \\ 

  0.6 & 0.90 & $5.66565\times 10^{10}$ & 0.1 & 0.70 & $1.28411\times 10^{13}$  \\
 
  0.7 & 0.90 & $6.33220\times 10^{10}$ & 0.1 & 0.80 & $3.58110\times 10^{11}$   \\
 
  0.8 & 0.90 & $6.99874\times 10^{10}$ & 0.1 & 0.90 & $2.33291\times 10^{10}$  \\
 
  0.9 & 0.90 & $7.66529\times 10^{10}$ & 0.1 & 0.99 & $3.33001\times 10^9$ \\ 
 [0.2ex] 
 \hline \hline \hline
\end{tabular}
\caption{\label{spinorevaporation} Evaporation time in the spinor sector, $t^{\psi}_{\mathrm{evap}}$, for different values of $\chi$ and $\omega$. Here, it is considered $l=5/2$. }
\end{center}
\end{table}

\begin{figure}
    \centering
     \includegraphics[scale=0.65]{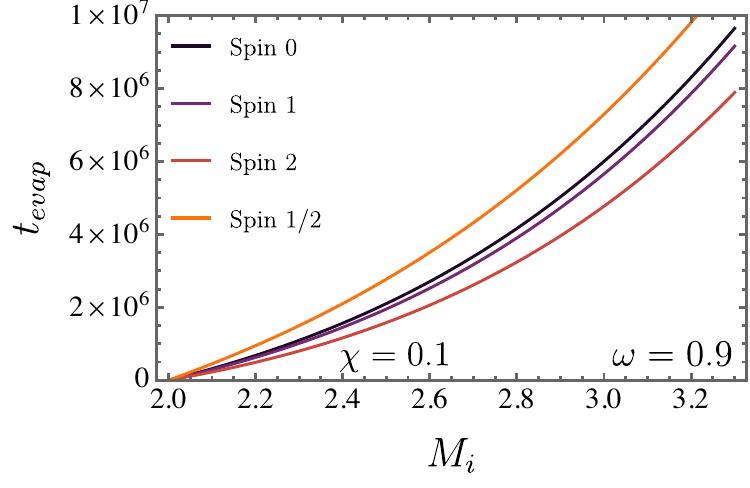}
      \caption{Evaporation times for all spin sectors, evaluated with $\chi=0.1$, $\omega=0.9$, $M_f=2$, and $M=1$ using $l=2$ (bosons) and $l=5/2$ (spinors). }
    \label{compevaporationallll}
\end{figure}

Figure~\ref{compevaporationallll} presents the evaporation times obtained for all perturbative sectors considered in this work. The comparison is performed for $\chi = 0.1$, $\omega = 0.9$, and $M_{f}=2$, adopting $l=2$ for the bosonic modes and $l=5/2$ for the spinor sector, with $M=1$ throughout. The resulting hierarchy is clear: tensor perturbations lead to the shortest evaporation time, while spinor modes yield the longest. As will be confirmed in the analysis of the energy–emission rates, this ordering persists across the different radiative channels.

\subsection{High-frequency regime }

In this subsection, we focus on a specific regime of the evaporation process: the high--frequency limit. In this case, two simplifications arise. First, the partial cross section approaches its limiting value, $\sigma_{l\omega} \to \sigma_{\mathrm{lim}} \simeq \pi \mathcal{R}^{2}$, where $\mathcal{R}$ is the shadow radius of the black hole. Second, the greybody factors tend to unity. The analysis is carried out for the new bumblebee black hole and, for comparison, the corresponding results for other Lorentz--violating configurations involving vector and tensor fields are presented both in a plot and in a table.


\subsubsection{New bumblebee black hole }

One should recall that the dominant contribution to the radiation spectrum comes from particles that behave as effectively massless, such as photons and neutrinos \cite{hiscock1990evolution,page1976particle}. In the high--frequency treatment adopted here, the relevant geometric scale entering the limiting cross section is the shadow radius characteristic and the other related quantities of the underlying black hole spacetime are \cite{AraujoFilho:2025zaj}
\ie
\begin{split}
& \mathcal{R} = 3 \sqrt{3} M, \quad \sigma_{lim} =  \, 27 \pi  M^2 , \quad T \approx \, \, \frac{1}{8 \pi  M} -\frac{\chi }{8 (\pi  M)}.
\nonumber
\end{split}
\fe
In this regime, the transmission probabilities turn out to $\bar{\Gamma}_{l\omega} \approx 1$ \cite{liang2025einstein}. With this simplification in place, Eq.~(\ref{slawbotz}) reduces to the form
\ie
\frac{\mathrm{d}M}{\mathrm{d}t} = -\frac{27 (\chi -1)^4}{4096 \pi ^3 M^2} \,.
\fe

The analysis then proceeds by computing the integral below
\ie
\begin{split}
& \int_{0}^{t_{\text{evap}}}  \mathrm{d}\tau = 
	- \int_{M_{i}}^{M_{f}} \mathrm{d}M
\left[ \frac{27 (\chi -1)^4}{4096 \pi ^3 M^2}  \right]^{-1} ,
\end{split}
\fe
in which $t_{\text{evap}}$ denotes the total duration of the evaporation process, which can therefore be expressed as
\ie
\begin{split}
t_{\text{evap}}  = -\frac{4096 \pi ^3 \left(M_{f}^3-M_{i}^3\right)}{81 (\chi -1)^4} \approx \,-\frac{1}{81} 4096 \left(\pi ^3 \left(M_{f}^3-M_{i}^3\right)\right) -\frac{16384}{81} \chi  \left(\pi ^3 \left(M_{f}^3-M_{i}^3\right)\right).
\end{split}
\fe

Imposing the condition that the temperature drops to zero, $T \to 0$, fixes the mass at the endpoint of the evolution and leads to a vanishing remnant, $M_{\mathrm{rem}} = 0$. This implies that the system evolves toward complete evaporation, with the final mass approaching this limiting value, $M_{f} \to M_{\mathrm{rem}}$. Under these circumstances, the expression for the total evaporation time takes the form
\ie
	t_{\text{evap-final}} =  \, \frac{4096 \pi ^3 M_{i}^3}{81} + \frac{16384}{81} \pi ^3 M_{i}^3 \chi.
\fe

\begin{figure}
    \centering
     \includegraphics[scale=0.7]{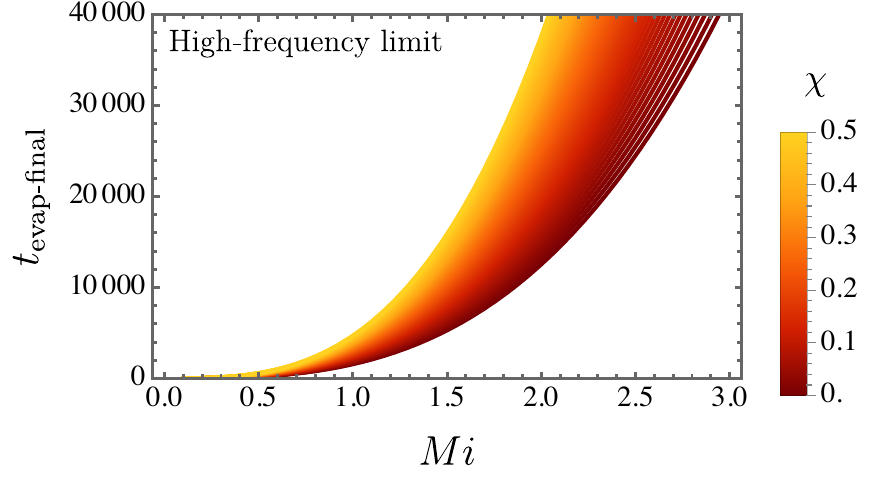}
    \caption{Evaporation time $t_{\text{evap-final}}$ for various initial masses $M_{i}$ and values of $\chi$. }
    \label{newbumblehighcase}
\end{figure}

The first contribution in the expression reproduces the Schwarzschild result, whereas the second term reflects the influence of the Lorentz--violating parameter $\chi$ introduced in this work.
To illustrate the physical implications, Fig.~\ref{newbumblehighcase} presents the behavior of the total evaporation time as $\chi$ varies. For all values of the parameter considered, the quantity $t_{\text{evap-final}}$ remained larger than its Schwarzschild counterpart ($\chi = 0$). This shows that the standard Schwarzschild solution evaporates more quickly, while increasing $\chi$ progressively delays the mass loss and extends the lifetime of the black hole.

\subsubsection{Bumblebee black hole }

“This subsubsection examines the black hole solution introduced in \cite{Casana:2017jkc}, namely the bumblebee black hole
\ie
\label{mmainin}
\mathrm{d}s^{2} =  -\left(1-\dfrac{2M}{r}\right)\mathrm{d}t^{2} + (1+\ell)\left(1-\dfrac{2M}{r}\right)^{-1}\mathrm{d}r^{2} + r^{2}\mathrm{d}\theta^{2} + r^{2} \sin^{2}\mathrm{d}\varphi^{2}.
\fe
The related quantities for this case are
\ie
\begin{split}
& \mathcal{R} = 3 \sqrt{3} M, \quad \sigma_{lim} =  \, 27 \pi  M^2 , \quad T \approx \, \, \frac{1}{8 \pi  M} -\frac{\ell }{16 (\pi  M)}.
\nonumber
\end{split}
\fe
In this manner, we have
\ie
\frac{\mathrm{d}M}{\mathrm{d}t} = -\frac{27 (\ell -2)^4}{65536 \pi ^3 M^2} \,.
\fe

The analysis then proceeds by evaluating the integral below
\ie
\begin{split}
& \int_{0}^{t_{\text{evap}}}  \mathrm{d}\tau = 
	- \int_{M_{i}}^{M_{f}} \mathrm{d}M
\left[ -\frac{27 (\ell -2)^4}{65536 \pi ^3 M^2}  \right]^{-1}.
\end{split}
\fe
Therefore, we have
\ie
\begin{split}
t_{\text{evap}}  = -\frac{65536 \pi ^3 \left(M_{f}^3-M_{i}^3\right)}{81 (\ell -2)^4} \approx \, -\frac{4096}{81}  \left(\pi ^3 \left(M_{f}^3-M_{i}^3\right)\right) -\frac{8192}{81}   \left(\pi ^3 \left(M_{f}^3-M_{i}^3\right)\right) \ell.
\end{split}
\fe

In this case, $M_{rem} = 0$, so that
\ie
\label{bumblebeemetricevaporation}
	t_{\text{evap-final}} =  \, \frac{4096 \pi ^3 M_{i}^3}{81} + \frac{8192}{81} \pi ^3 M_{i}^3 \ell ,
\fe
with the first contribution reproduces the Schwarzschild result, whereas the second term incorporates the Lorentz--violating parameter $\ell$ characteristic of the bumblebee geometry. At this stage, one observation is worth noting. In Ref.~\cite{araujo2025does}, the evaporation lifetime of the bumblebee black hole was given by
$t_{\text{evap-final}} = \frac{4096}{81}\,\pi^{3}\,(\ell+1)^{2}\,M_{i}^{3}$.
By expanding Eq.~(\ref{bumblebeemetricevaporation}) in the present work, one sees that the resulting expression matches exactly the form reported in Ref.~\cite{araujo2025does}.


\subsubsection{Bumblebee black hole (metric-afine) }

This subsection examines the black hole obtained by Araújo Filho et al. \cite{Filho:2022yrk}, namely the bumblebee solution formulated in the \textit{metric–affine} framework
\ie
  \mathrm{d}s^2= - \frac{\left(1-\frac{2M}{r}\right)\mathrm{d}t^2}{\sqrt{\left(1+\frac{3X}{4}\right)\left(1-\frac{X}{4}\right)}}+\frac{\mathrm{d}r^2}{\left(1-\frac{2M}{r}\right)}\sqrt{\frac{\left(1+\frac{3X}{4}\right)}{\left(1-\frac{X}{4}\right)^3}} +r^{2}\left(\mathrm{d}\theta^2 +\sin^{2}{\theta}\mathrm{d}\phi^2\right).
\fe
The fundamental quantities for our next calculations are
\ie
\begin{split}
 \mathcal{R} = 3 \sqrt{3} M, \quad \sigma_{lim} =  \, 27 \pi  M^2, \quad
T \approx \, \, \frac{1}{8 \pi  M} -\frac{X }{16 (\pi  M)},
\nonumber
\end{split}
\fe
In this manner,
\ie
\frac{\mathrm{d}M}{\mathrm{d}t} = -\frac{27 (X -2)^4}{65536 \pi ^3 M^2} \,.
\fe

Thereby, the integral becomes
\ie
\begin{split}
& \int_{0}^{t_{\text{evap}}}  \mathrm{d}\tau = 
	- \int_{M_{i}}^{M_{f}} \mathrm{d}M
\left[ -\frac{27 (X -2)^4}{65536 \pi ^3 M^2}  \right]^{-1} ,
\end{split}
\fe
so that
\ie
\begin{split}
t_{\text{evap}}  = -\frac{65536 \pi ^3 \left(M_{f}^3-M_{i}^3\right)}{81 (X -2)^4} \approx \, -\frac{4096}{81}  \left(\pi ^3 \left(M_{f}^3-M_{i}^3\right)\right) -\frac{8192}{81}   \left(\pi ^3 \left(M_{f}^3-M_{i}^3\right)\right) X.
\end{split}
\fe

In this case, $M_{rem} = 0$. Then,
\ie
	t_{\text{evap-final}} =  \, \frac{4096 \pi ^3 M_{i}^3}{81} + \frac{8192}{81} \pi ^3 M_{i}^3 X .
\fe
As it is straightforward to see, the first term reproduces the Schwarzschild result, while the second incorporates the Lorentz--violating parameter $X$ associated with the \textit{metric--affine} bumblebee black hole.

A brief remark is necessary at this point. As emphasized in Ref.~\cite{AraujoFilho:2025zaj}, the shadow radii originally obtained in Ref.~\cite{araujo2024gravitational} contained a typo, later corrected in Ref.~\cite{AraujoFilho:2025zaj}. Because this expression was used in Ref.~\cite{araujo2025does} to analyze the evaporation lifetimes of \textit{metric} bumblebee and \textit{metric--affine} bumblebee black holes, the discussion there ended up led to an incorrect conclusion regarding the \textit{metric--affine} case. The present work incorporated the corrected expression and updates the corresponding statements. It is also worth noting that an erratum has been submitted to the respective journal to implement the necessary correction.

By expanding the Hawking temperatures as done in Ref.~\cite{araujo2025does},
$T_{\text{metric}} \approx \frac{1}{8\pi M} - \frac{\ell}{16\pi M}$
and
$T_{\text{met-aff}} \approx \frac{1}{8\pi M} - \frac{X}{16\pi M}$,
and by noting that the shadow radii remain the same in both formalisms, it follows that the evaporation lifetimes of the bumblebee black holes in the \textit{metric} and \textit{metric--affine} approaches coincide (at least for the high--frequency regime).


\subsubsection{Kalb-Ramond (Model 1) }

This subsubsection examines the black hole solution introduced in \cite{Yang:2023wtu}, namely the Kalb--Ramond black hole (Model 1)
\ie
\label{model1}
\mathrm{d}s^{2}  =  - \left( \frac{1}{1-\ell} - \frac{2M}{r}   \right) \mathrm{d}t^{2} + \frac{\mathrm{d}r^{2}}{\frac{1}{1-\ell} - \frac{2M}{r} } r^{2}\mathrm{d}\theta^{2} + r^{2} \sin^{2}\mathrm{d}\varphi^{2},
\fe
and the essential quantities are 
\ie
\begin{split}
 \mathcal{R} = 3 \sqrt{3} (1-\ell) M, \quad \sigma_{lim} =  \, 27 \pi  (1-\ell)^2 M^2 ,
\quad T \approx \, \, \frac{1}{8 \pi  (\ell-1)^2 M} \approx \,  \frac{1}{8 \pi  M}+\frac{\ell}{4 \pi  M},
\nonumber
\end{split}
\fe
in a such way that
\ie
\frac{\mathrm{d}M}{\mathrm{d}t} = -\frac{27 (\ell-1)^2 (2 \ell+1)^4}{4096 \pi ^3 M^2} \,,
\fe
which leads to the integral below
\ie
\begin{split}
& \int_{0}^{t_{\text{evap}}} \mathrm{d}\tau = 
	- \int_{M_{i}}^{M_{f}} \mathrm{d}M
\left[ -\frac{27 (\ell-1)^2 (2 \ell+1)^4}{4096 \pi ^3 M^2} \right]^{-1} .
\end{split}
\fe
After its evaluation, we obtain
\ie
\begin{split}
t_{\text{evap}}  = -\frac{4096 \pi ^3 \left(M_{f}^3-M_{i}^3\right)}{81 (\ell-1)^2 (2 \ell+1)^4} \approx \, -\frac{4096}{81} \left(\pi ^3 \left(M_{f}^3-M_{i}^3\right)\right) + \frac{8192}{27} \pi ^3 \ell \left(M_{f}^3-M_{i}^3\right).
\end{split}
\fe

As the other cases, here, $M_{rem} = 0$ and, then,
\ie
\label{kalbtricevaporation111}
	t_{\text{evap-final}} = \frac{4096}{81}\,\pi^{3}\,(1-\ell)^{6}\,M_{i}^{3} \approx \, \frac{4096 \pi ^3 M_{i}^3}{81}-\frac{8192}{27} \pi ^3 \ell M_{i}^3,
\fe

The first term reproduces the Schwarzschild contribution, while the second incorporates the Lorentz--violating parameter $\ell$ characteristic of the Kalb--Ramond black hole (Model 1). It is important to highlight a minor typo in Ref.~\cite{araujo2025particleasdasd}: the evaporation time there was written as
$
t_{\text{evap-final}} = \frac{4096}{81}\,\pi^{3}\,(1-\ell)^{5}\,M_{i}^{3}$. Our results show that, without performing any expansion, the correct expression should read $t_{\text{evap-final}} = \frac{4096}{81}\,\pi^{3}\,(1-\ell)^{6}\,M_{i}^{3}$.
In addition, expanding Eq.~(\ref{kalbtricevaporation111}) obtained in the present work reproduces exactly the functional form appearing in Ref.~\cite{araujo2025particleasdasd}, aside from this minor exponent slip.


\subsubsection{Kalb-Ramond (Model 2) }

This subsubsection analyzes the black hole solution presented in \cite{Liu:2024oas}, referred to here as the Kalb--Ramond black hole (Model 2)
\ie
\label{mmainin}
\mathrm{d}s^{2} =  -\left(1-\dfrac{2M}{r}\right)\mathrm{d}t^{2} + (1-\ell)\left(1-\dfrac{2M}{r}\right)^{-1}\mathrm{d}r^{2} + r^{2}\mathrm{d}\theta^{2} + r^{2} \sin^{2}\mathrm{d}\varphi^{2}.
\fe
Also, the important quantities are
\ie
\begin{split}
& \mathcal{R} = 3 \sqrt{3} M, \quad \sigma_{lim} =  \, 27 \pi  M^2 , \quad T \approx \, \, \frac{1}{8 \pi  M} +\frac{\ell }{16 (\pi  M)},
\nonumber
\end{split}
\fe
so that, we have
\ie
\frac{\mathrm{d}M}{\mathrm{d}t} = -\frac{27 (\ell +2)^4}{65536 \pi ^3 M^2} \,.
\fe

This lead to the following integral
\ie
\begin{split}
& \int_{0}^{t_{\text{evap}}}  \mathrm{d}\tau = 
	- \int_{M_{i}}^{M_{f}} \mathrm{d}M
\left[ -\frac{27 (\ell +2)^4}{65536 \pi ^3 M^2}  \right]^{-1} ,
\end{split}
\fe
which results
\ie
\begin{split}
t_{\text{evap}}  = -\frac{65536 \pi ^3 \left(M_{f}^3-M_{i}^3\right)}{81 (\ell +2)^4} \approx \, -\frac{4096}{81}  \left(\pi ^3 \left(M_{f}^3-M_{i}^3\right)\right) +\frac{8192}{81}   \left(\pi ^3 \left(M_{f}^3-M_{i}^3\right)\right) \ell.
\end{split}
\fe

Again, since $M_{rem} = 0$, we obtain
\ie
\label{kalbtricevaporation1111}
	t_{\text{evap-final}} =  \, \frac{4096 \pi ^3 M_{i}^3}{81} - \frac{8192}{81} \pi ^3 M_{i}^3 \ell .
\fe

The first term reproduces the Schwarzschild contribution, while the second incorporates the Lorentz--violating parameter $\ell$ characteristic of the Kalb--Ramond black hole (Model 2). A brief observation is in order. Ref.~\cite{araujo2025particleasdasd} reported the evaporation time for this configuration as
$t_{\text{evap-final}} = \frac{4096}{81},\pi^{3},(1-\ell)^{2},M_{i}^{3}$.
However, once the expression is expanded, it coincides with the result obtained here in Eq.~(\ref{kalbtricevaporation1111}).


\subsubsection{Non-commutative Kalb-Ramond }

The corresponding line element can be written explicitly in the form \cite{AraujoFilho:2025jcu}:
\ie
\label{metrictensorss}
\mathrm{d}s^{2} = g^{(\Theta)}_{\mu\nu}\left(x,\Theta\right) \mathrm{d}x^{\mu} \mathrm{d}x^{\nu}   = - A^{(\Theta,\ell)} \mathrm{d}t^{2} +  B^{(\Theta,\ell)} \mathrm{d}r^{2} + C^{(\Theta,\ell)} \mathrm{d}\theta^{2} + D^{(\Theta,\ell)} \mathrm{d}\varphi^{2},
\fe
with the metric components being given by
\ie \label{gtt}
A^{(\Theta,\ell)} = \frac{1}{1-\ell} - \frac{2 M}{r} - \frac{\Theta ^2 M (11 (\ell-1) M+4 r)}{2 (\ell-1) r^4},
\fe
\ie \label{grr}
B^{(\Theta,\ell)} = \frac{1}{\frac{1}{1-\ell}-\frac{2 M}{r}} + \frac{\Theta ^2 (\ell-1) M (3 (\ell-1) M+2 r)}{2 r^2 (2 (\ell-1) M+r)^2},
\fe
\ie \label{gtheta}
C^{(\Theta,\ell)} = r^2 -\frac{\Theta ^2 \left(64 (\ell-1)^2 M^2+32 (\ell-1) M r+r^2\right)}{16 (\ell-1) r (2 (\ell-1) M+r)},
\fe
\ie \label{gphi}
D^{(\Theta,\ell)} = r^2 \sin ^2(\theta ) +\frac{1}{16} \Theta ^2 \left[5 \cos ^2(\theta )+\frac{4 \sin ^2(\theta ) \left(-2 (\ell-1) M^2+4 (\ell-1) M r+r^2\right)}{r (2 (\ell-1) M+r)}\right].
\fe

Here, we have
\ie
 \mathcal{R} = 3 \sqrt{3} M -3 \sqrt{3} \ell M -\frac{\Theta ^2 \ell}{8 \sqrt{3} M}-\frac{\Theta ^2}{8 \sqrt{3} M},
\fe
\ie
\sigma_{lim} =  \, 27 \pi  M^2 - 54 \pi  \ell M^2 -\frac{3 \pi  \Theta ^2}{4}  ,
\fe
and
\ie
T \approx \, \, \frac{3 \Theta ^2}{128 \pi  (1-\ell)^3 M^3}+\frac{\ell}{8 \pi  (1-\ell) M}+\frac{1}{8 \pi  (1-\ell) M},
\nonumber
\fe
so that
\ie
\frac{\mathrm{d}M}{\mathrm{d}t} = \frac{3 \left(\Theta ^2+36 (2 \ell-1) M^2\right) \left(3 \Theta ^2+16 (\ell-1)^2 (\ell+1) M^2\right)^4}{1073741824 \pi ^3 (\ell-1)^{12} M^{12}}\,.
\fe

The corresponding integral can be written as 
\ie
\begin{split}
& \int_{0}^{t_{\text{evap}}}  \mathrm{d}\tau = 
	- \int_{M_{i}}^{M_{f}} \mathrm{d}M
\left[\frac{3 \left(\Theta ^2+36 (2 \ell-1) M^2\right) \left(3 \Theta ^2+16 (\ell-1)^2 (\ell+1) M^2\right)^4}{1073741824 \pi ^3 (\ell-1)^{12} M^{12}} \right]^{-1} ,
\end{split}
\fe
which its result is given by
\ie
\begin{split}
t_{\text{evap}}  \approx &\, -\frac{4096}{81} \left(\pi ^3 \left(M_{f}^3-M_{i}^3\right)\right)\\
& + \frac{8192}{27} \pi ^3 \ell \left(M_{f}^3-M_{i}^3\right)+ \left(\frac{26624}{243} \pi ^3 (M_{f}-M_{i})-\frac{134144}{243} \ell \left(\pi ^3 (M_{f}-M_{i})\right)\right)\Theta ^2.
\end{split}
\fe

Since $M_{rem} = 0$, we get
\ie
\label{kalbtricevaporation}
	t_{\text{evap-final}} =  \, \frac{4096 \pi ^3 M_{i}^3}{81} -\frac{1}{27} 8192 \pi ^3 \ell M_{i}^3+\frac{134144}{243} \pi ^3 \Theta ^2 \ell M_{i}-\frac{26624}{243} \pi ^3 \Theta ^2 M_{i} .
\fe
As all other cases, the first term reproduces the Schwarzschild contribution, as we should expect; the second incorporates the Lorentz--violating parameter $\ell$ associated with the Kalb--Ramond black hole (Model 1), and the remaining terms arise from the non--commutative corrections. For comparison, all evaporation times obtained in this section are collected in Tab.~\ref{evaporationlifetimes}.

An additional remark is appropriate here. One might wonder why the non--commutative extension of the bumblebee black hole discussed in \cite{AraujoFilho:2025rvn} does not appear in this table. The reason is straightforward: for the specific Moyal twist adopted in that work ($\partial_{r} \wedge \partial_{\theta}$), the surface gravity cannot be consistently defined, which renders the Stefan–Boltzmann law inapplicable. Consequently, an evaporation time cannot be derived in that framework.

\begin{figure}
    \centering
     \includegraphics[scale=0.64]{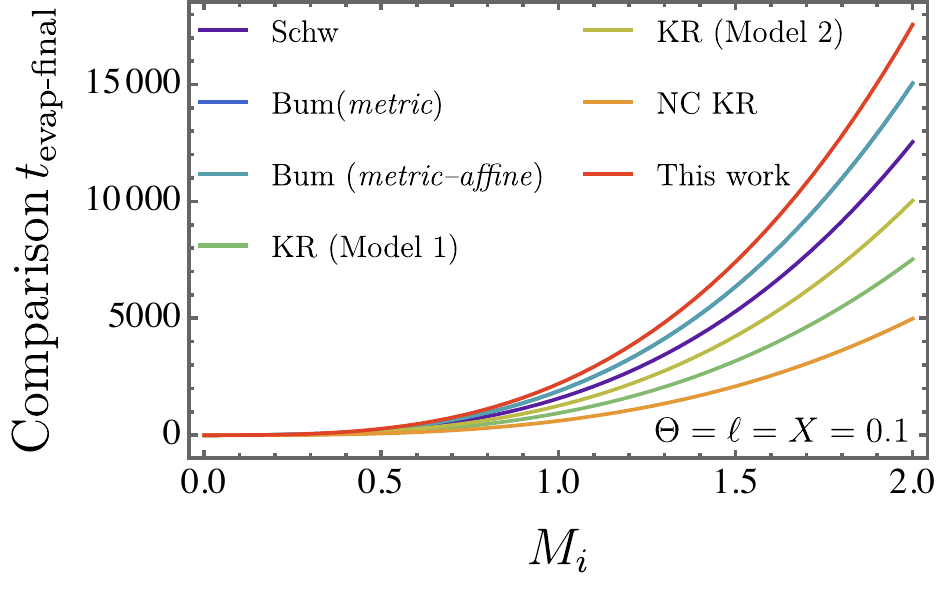}
      \caption{The comparison of final evaporation times for all black holes considered in this paper by taking into account the high--frequency limit. }
      \label{comparisonevaporationnn}
\end{figure}

A further question naturally arises: among the black holes compared in this paper, which one evaporates more rapidly when the high–frequency limit is considered? To address this point, Fig.~\ref{comparisonevaporationnn} presents the corresponding comparison. For simplicity, the parameters have been fixed to $\Theta = X = \ell = 0.1$. Under this choice, the following hierarchy becomes evident:
\[
t_{\text{evap-final}}^{\text{this work}}> t_{\text{evap-final}}^{\text{bum (\textit{metric})}} = t_{\text{evap-final}}^{\text{bum (\textit{met--aff})}} > t_{\text{evap-final}}^{\text{Schw}}  > t_{\text{evap-final}}^{\text{KR (Model 2)}} > t_{\text{evap-final}}^{\text{KR (Model 1)}} > t_{\text{evap-final}}^{\text{NC KR}} . 
\]
In other words, among the Lorentz--violating black holes examined here, the new bumblebee solution exhibits the slowest evaporation, while the non--commutative Kalb--Ramond black hole evaporates the quickest.

\begin{table}[!h]
\centering
\caption{\label{evaporationlifetimes}
Within the context of high--limit case, we comparison of the limiting $\sigma_{lim}$ and the evaporation lifetimes for the existing Lorentz--violating configurations associated with bumblebee and Kalb--Ramond black holes. In this context, $\ell$ and $X$ represent the Lorentz--violating parameters.}
\renewcommand{\arraystretch}{1.25}
\setlength{\tabcolsep}{6pt}
\hspace*{-2.4cm}\begin{tabular}{l|c|c}
\hline\hline\hline
\textbf{Black holes} & $\sigma_{\mathrm{lim}}$ & \textbf{Final evaporation lifetimes} \\
\hline
This work  &
$27 \pi  M^2$ &
$ \frac{4096 \pi ^3 M_{i}^3}{81} + \frac{16384}{81} \pi ^3 M_{i}^3 \chi  $ \\[4pt]
Bumblebee  (\textit{metric}) \cite{Casana:2017jkc} &
$27 \pi  M^2$ &
$ \frac{4096 \pi ^3 M_{i}^3}{81} + \frac{8192}{81} \pi ^3 M_{i}^3 \ell $ \\[4pt]
Bumblebee (\textit{metric--affine}) \cite{Filho:2022yrk} &
$27 \pi  M^2$ &
$\frac{4096 \pi ^3 M_{i}^3}{81} + \frac{8192}{81} \pi ^3 M_{i}^3 X $ \\[4pt]
Kalb--Ramond (Model~1) \cite{Yang:2023wtu} &
$27 \pi  (1-\ell)^2 M^2$ &
$\frac{4096 \pi ^3 M_{i}^3}{81}-\frac{8192}{27} \pi ^3 M_{i}^3 \ell$ \\[4pt]
Kalb--Ramond (Model~2) \cite{Liu:2024oas} &
$27 \pi  M^2$ &
$\frac{4096 \pi ^3 M_{i}^3}{81}-\frac{8192}{81} \pi ^3 M_{i}^3 \ell$ \\[4pt]
NC Kalb--Ramond \cite{AraujoFilho:2025jcu} &
$ 27 \pi  M^2 - 54 \pi  \ell M^2 -\frac{3 \pi  \Theta ^2}{4}$ &
$\frac{4096 \pi ^3 M_{i}^3}{81} -\frac{1}{27} 8192 \pi ^3 \ell M_{i}^3+\frac{134144}{243} \pi ^3 \Theta ^2 \ell M_{i}-\frac{26624}{243} \pi ^3 \Theta ^2 M_{i}$ \\[4pt]
\hline\hline\hline
\end{tabular}
\end{table}

\section{Radiative output: energy and particle fluxes}


\subsection{Spin 0 particle modes }

\begin{figure}
    \centering
     \includegraphics[scale=0.54]{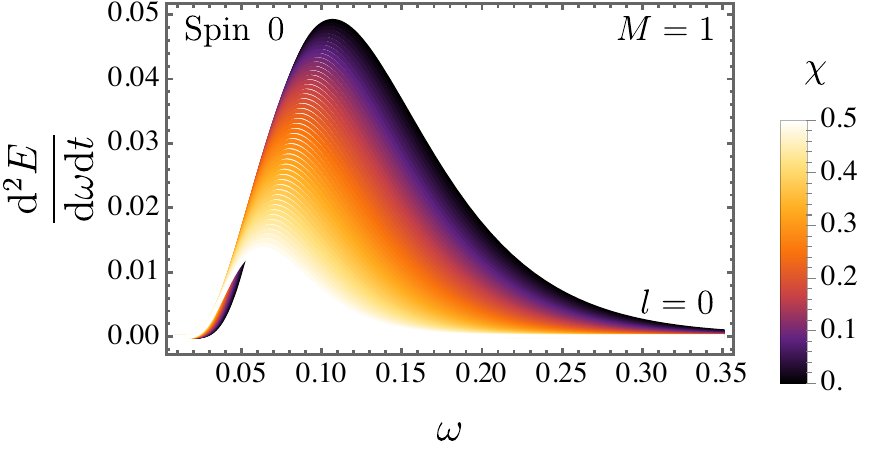}
     \includegraphics[scale=0.55]{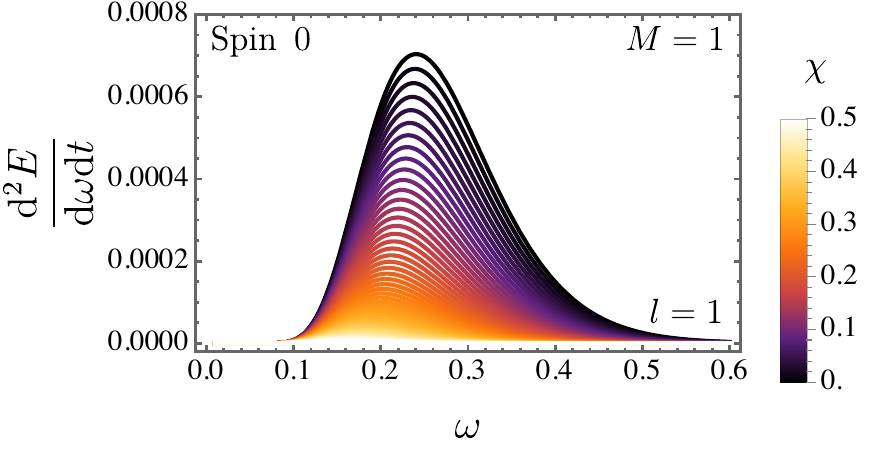}
      \includegraphics[scale=0.58]{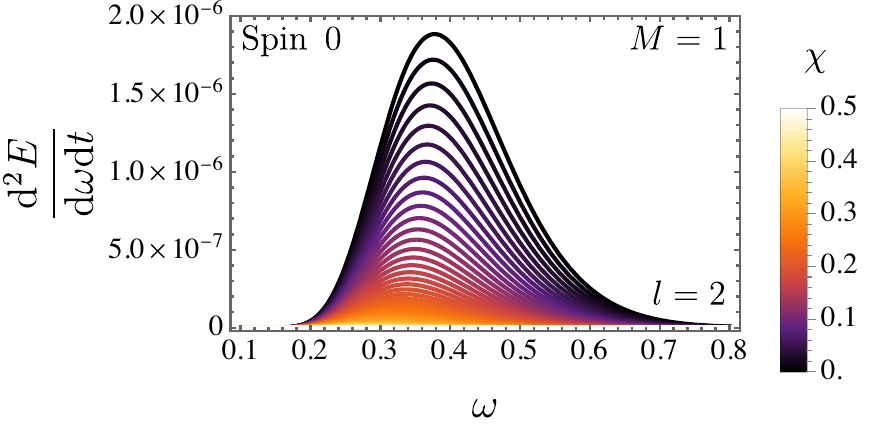}
      \caption{The energy--flux profiles are shown for the scalar perturbations for the choice $M=1$, with the cases $l=0$ (upper left), $l=1$ (upper right), and $l=2$ (lower panel) plotted separately.}
      \label{eeeemissionrate}
\end{figure}

The discussion now shifts to the behavior of the energy flux emitted by the black hole
\ie
\label{energyemission}
	\frac{{{\mathrm{d}^2}E}}{{\mathrm{d}\omega \mathrm{d}t}} = \frac{{2{\pi ^2}{\sigma}^{s}_{l \omega}}}{{{e^{\frac{\omega }{T}}} - 1}} {\omega ^3}.
\fe
Figure~\ref{eeeemissionrate} summarizes the behavior of the energy flux for the choices $M=1$ and angular momenta $l=0$ (upper left), $l=1$ (upper right), and $l=2$ (lower panel). As the parameter $\chi$ departs from the Schwarzschild limit, the corresponding curves reveal a gradual suppression of the emitted energy, indicating that the Lorentz–violating deformation diminishes the overall strength of the radiation.

The corresponding rate of particle production is given by
\ie
\frac{\mathrm{d}^{2}N}{\mathrm{d}\omega \mathrm{d}t}
= \frac{2\pi^{2}\,\sigma^{s}_{l \omega}\,\omega^{2}}
       {{{e^{\frac{\omega }{T}}} - 1}}.
\fe

\begin{figure}
    \centering
     \includegraphics[scale=0.54]{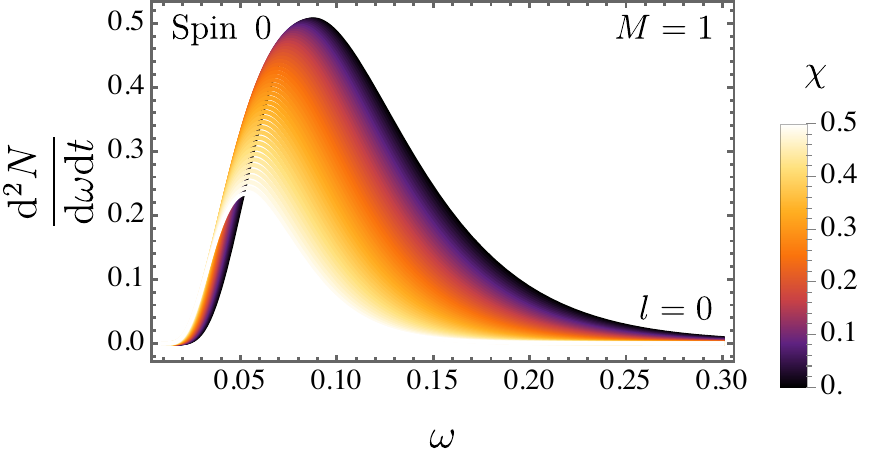}
     \includegraphics[scale=0.55]{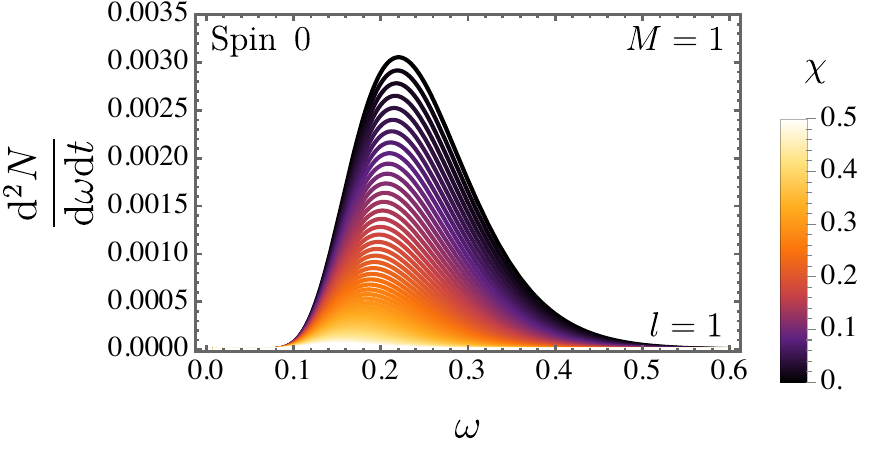}
      \includegraphics[scale=0.58]{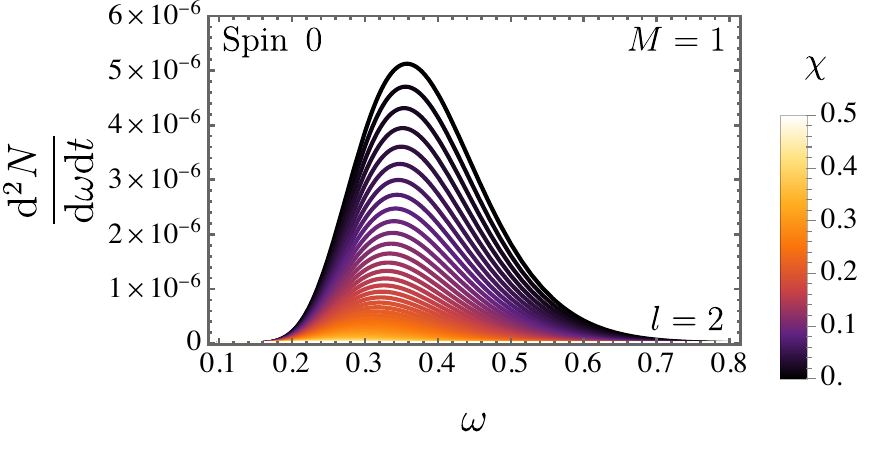}
      \caption{The particle--flux profiles are shown for the scalar perturbations for the choice $M=1$, with the cases $l=0$ (upper left), $l=1$ (upper right), and $l=2$ (lower panel) plotted separately.}
    \label{emissionrateparticle0}
\end{figure}

The emission rate curves are displayed in Fig.~\ref{emissionrateparticle0} for $M=1$ and for the angular momentum values $l=0$ (upper left), $l=1$ (upper right), and $l=2$ (lower panel). The trend mirrors what was previously identified in the energy--flux analysis: once the parameter $\chi$ departs from zero, the resulting profiles show a systematic reduction in amplitude. Thus, the Lorentz--violating contribution once again suppresses the overall strength of the emitted radiation, this time in the particle--production channel.

\subsection{Spin 1 particle modes }

\begin{figure}
    \centering
     \includegraphics[scale=0.55]{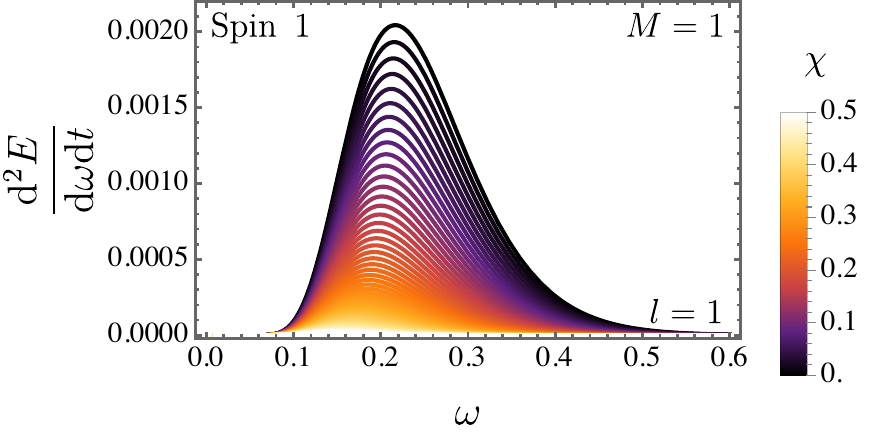}
     \includegraphics[scale=0.55]{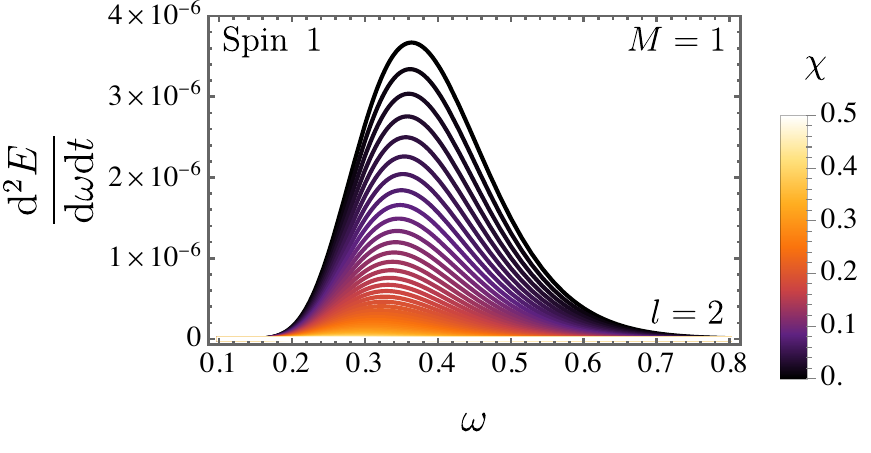}
      \includegraphics[scale=0.58]{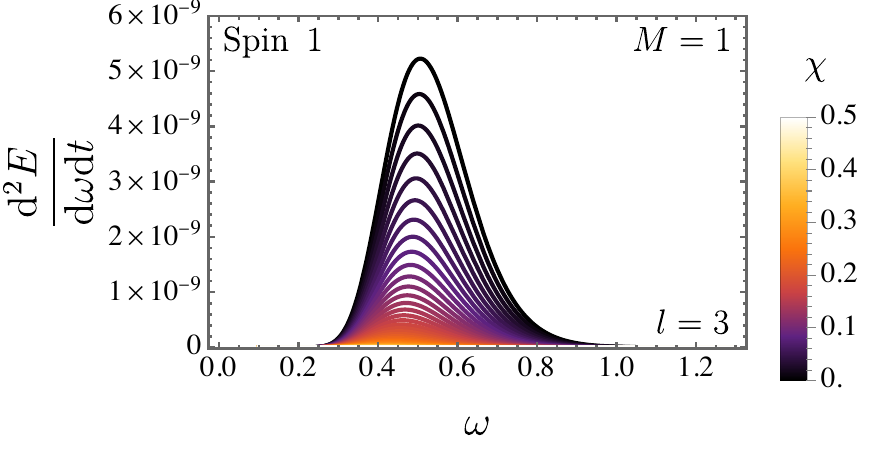}
      \caption{The energy--flux profiles are shown for the vector perturbations for the choice $M=1$, with the cases $l=1$ (upper left), $l=2$ (upper right), and $l=3$ (lower panel) plotted separately.}
      \label{vectoremissionrate}
\end{figure}

The discussion now shifts to the behavior of the emitted energy flux
\ie
\label{energyemission}
	\frac{{{\mathrm{d}^2}E}}{{\mathrm{d}\omega \mathrm{d}t}} = \frac{{2{\pi ^2}{\sigma}^{v}_{l \omega}}}{{{e^{\frac{\omega }{T}}} - 1}} {\omega ^3}.
\fe
Figure~\ref{vectoremissionrate} presents the energy--flux profiles for the vector sector with $M=1$ and the angular momentum values $l=1$ (upper left), $l=2$ (upper right), and $l=3$ (lower panel). Once the parameter $\chi$ is introduced, the corresponding curves exhibit a systematic reduction in amplitude. In other words, the Lorentz--violating deformation leads to a weaker energy output throughout the spectrum.

\begin{figure}
    \centering
     \includegraphics[scale=0.55]{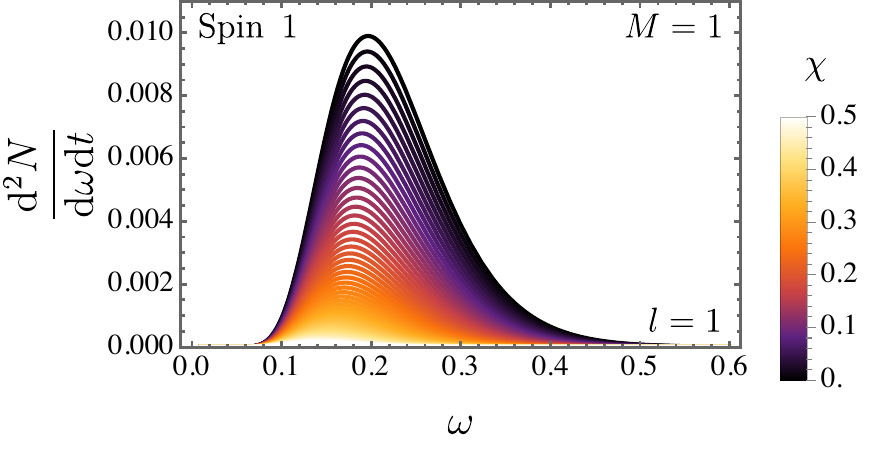}
     \includegraphics[scale=0.55]{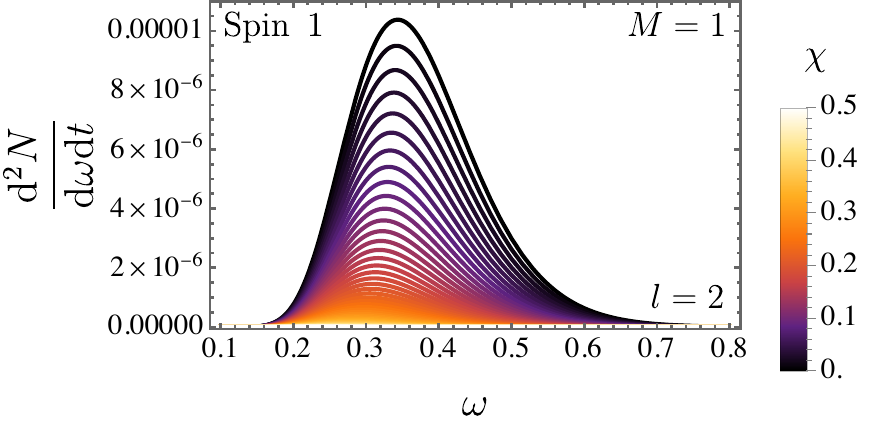}
      \includegraphics[scale=0.58]{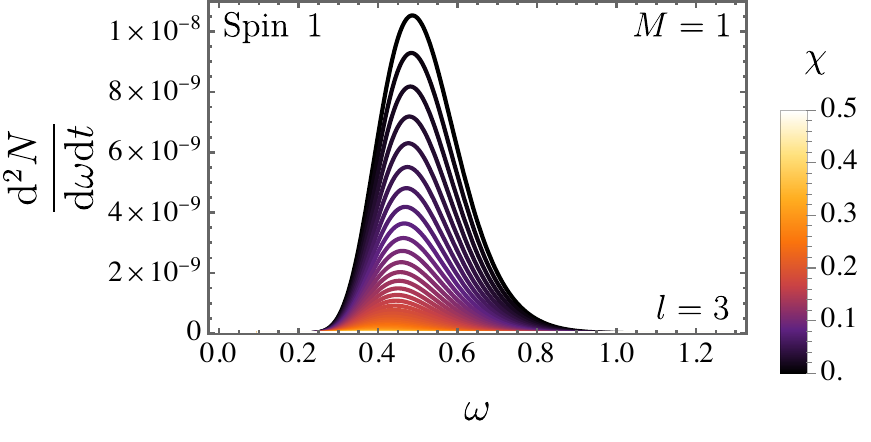}
      \caption{The particle--flux profiles are shown for the vector perturbations for the choice $M=1$, with the cases $l=1$ (upper left), $l=2$ (upper right), and $l=3$ (lower panel) plotted separately.}
    \label{vectorparticle}
\end{figure}

The corresponding particle emission rate takes the form
\ie
\frac{\mathrm{d}^{2}N}{\mathrm{d}\omega \mathrm{d}t}
= \frac{2\pi^{2}\,\sigma^{v}_{l \omega}\,\omega^{2}}
       {{{e^{\frac{\omega }{T}}} - 1}}.
\fe
The particle emission profiles for the vector sector appear in Fig.~\ref{vectorparticle} for $M=1$ and for the angular momentum values $l=1$ (upper left), $l=2$ (upper right), and $l=3$ (lower panel). Their overall behavior mirrors the trend found in the corresponding energy–flux curves: as the parameter $\chi$ departs from zero, the amplitudes of the particle emission diminish across the entire frequency range. Then, the Lorentz--violating contribution once again weakens the strength of the radiation, this time in the particle--generation channel.

\subsection{Spin 2 particle modes }

\begin{figure}
    \centering
     \includegraphics[scale=0.55]{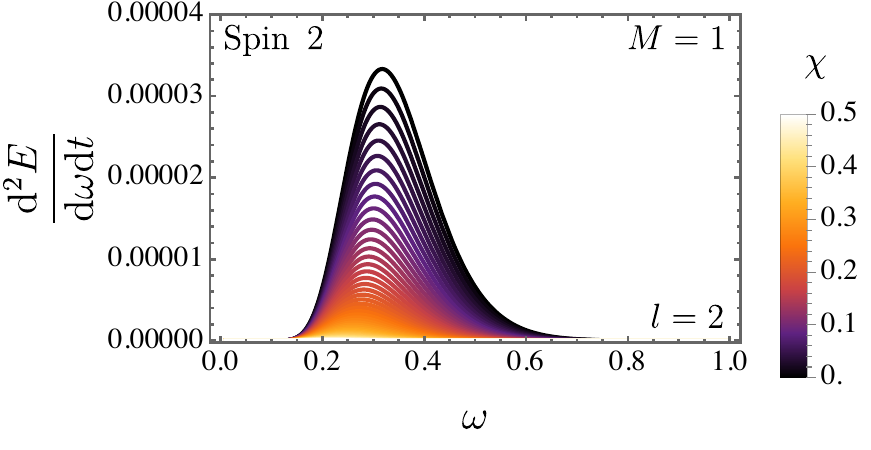}
     \includegraphics[scale=0.55]{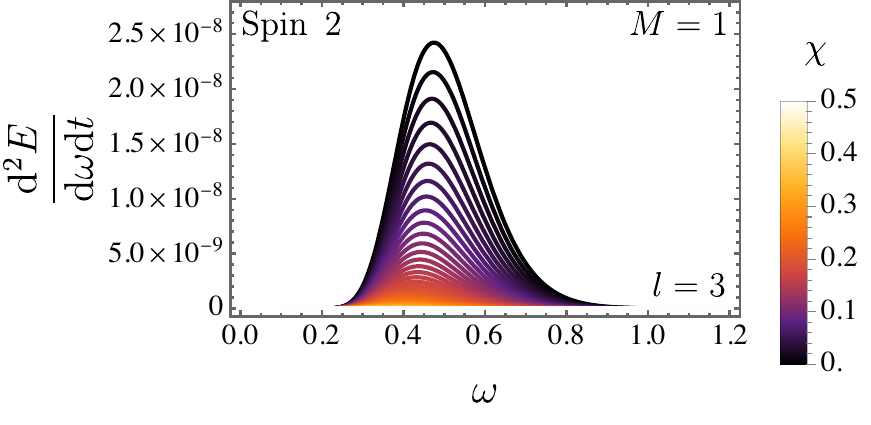}
      \includegraphics[scale=0.58]{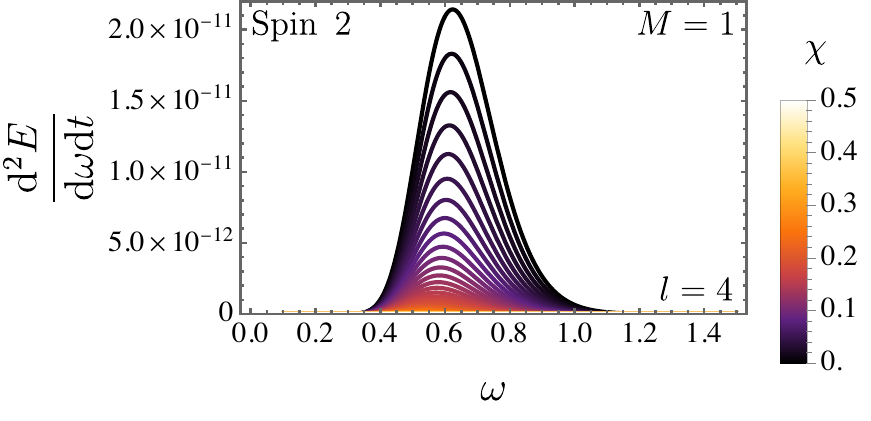}
      \caption{The energy--flux profiles are shown for the tensor perturbations for the choice $M=1$, with the cases $l=2$ (upper left), $l=3$ (upper right), and $l=4$ (lower panel) plotted separately.}
      \label{tensoremissionrate}
\end{figure}

The emission rate is 
\ie
\label{energyemission}
	\frac{{{\mathrm{d}^2}E}}{{\mathrm{d}\omega \mathrm{d}t}} = \frac{{2{\pi ^2}{\sigma}^{t}_{l \omega}}}{{{e^{\frac{\omega }{T}}} - 1}} {\omega ^3}.
\fe
Figure~\ref{tensoremissionrate} displays the energy--flux curves for the tensor sector with $M=1$ and angular momentum values $l=2$ (upper left), $l=3$ (upper right), and $l=4$ (lower panel). Once the parameter $\chi$ is introduced, the resulting profiles show a clear reduction in amplitude, indicating that the Lorentz--violating deformation consistently suppresses the emitted energy throughout the spectrum.

\begin{figure}
    \centering
     \includegraphics[scale=0.55]{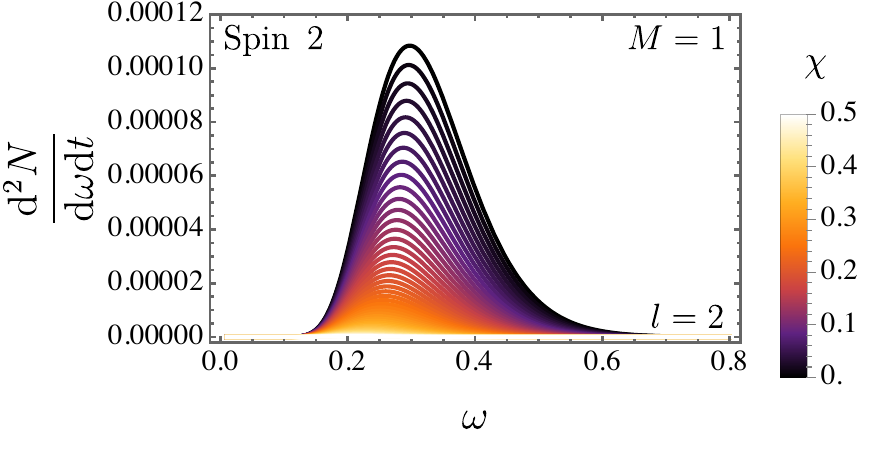}
     \includegraphics[scale=0.55]{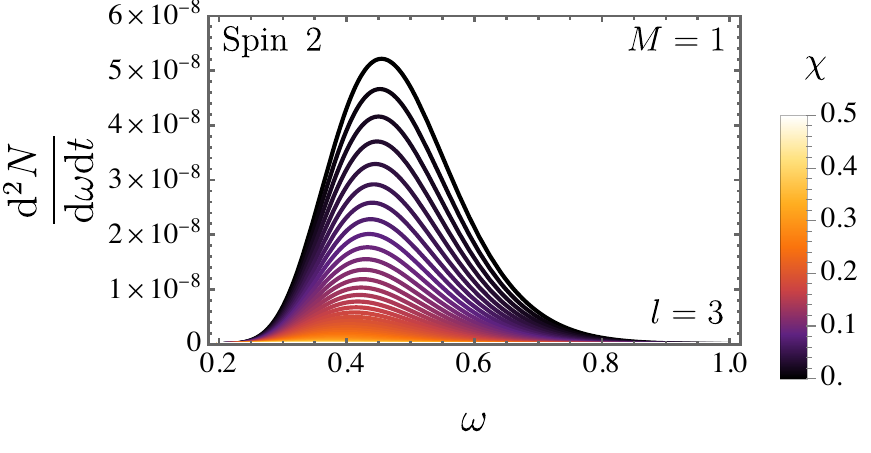}
      \includegraphics[scale=0.58]{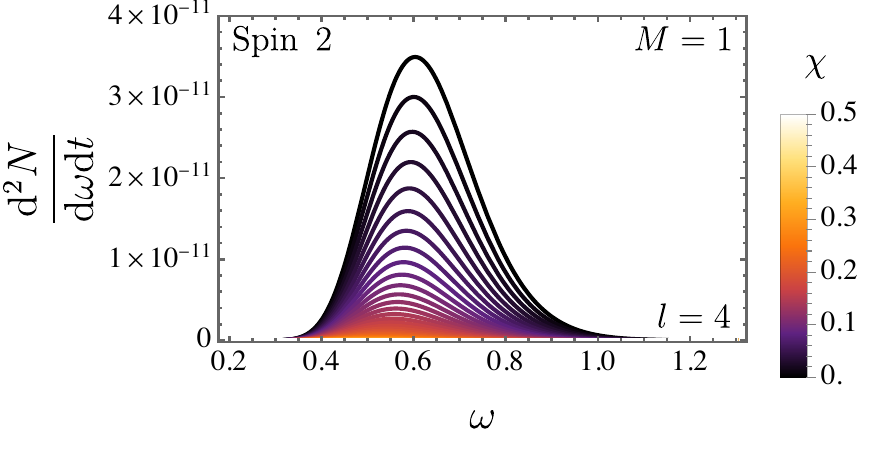}
      \caption{The particle--flux profiles are shown for the tensor perturbations for the choice $M=1$, with the cases $l=2$ (upper left), $l=3$ (upper right), and $l=4$ (lower panel) plotted separately.}
    \label{tensorparticle}
\end{figure}

Furthermore, the particle emission rate reads
\ie
\frac{\mathrm{d}^{2}N}{\mathrm{d}\omega \mathrm{d}t}
= \frac{2\pi^{2}\,\sigma^{t}_{l \omega}\,\omega^{2}}
       {{{e^{\frac{\omega }{T}}} - 1}}.
\fe
Figure~\ref{tensorparticle} displays the particle–emission profiles for the tensor sector with $M=1$ and angular momentum values $l=2$ (upper left), $l=3$ (upper right), and $l=4$ (lower panel). The pattern follows the same tendency identified in the energy--flux analysis: as soon as the parameter $\chi$ deviates from zero, the resulting curves exhibit a systematic reduction in amplitude. Thus, in this sector as well, the Lorentz--violating contribution acts to suppress the particle emission across the entire frequency range.

\subsection{Spin 1/2 particle modes }

\begin{figure}
    \centering
     \includegraphics[scale=0.55]{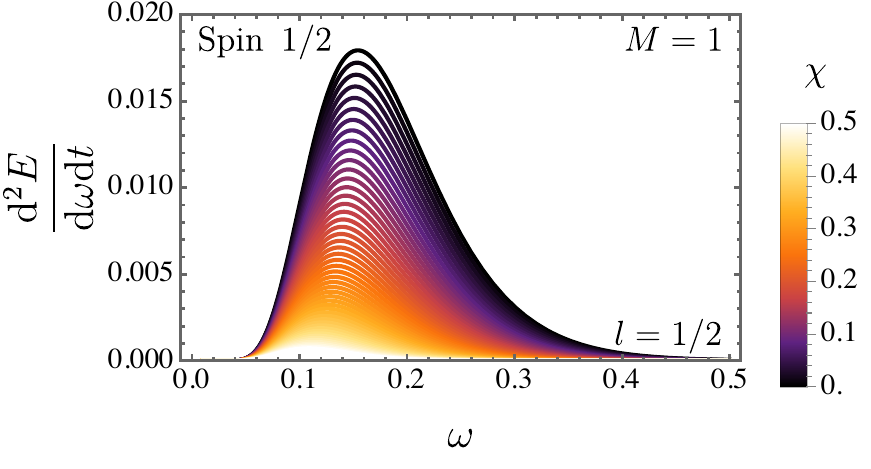}
     \includegraphics[scale=0.55]{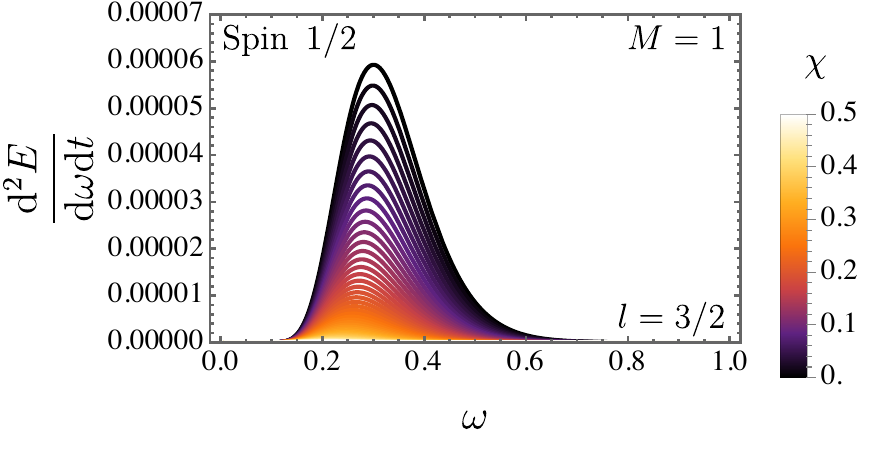}
      \includegraphics[scale=0.58]{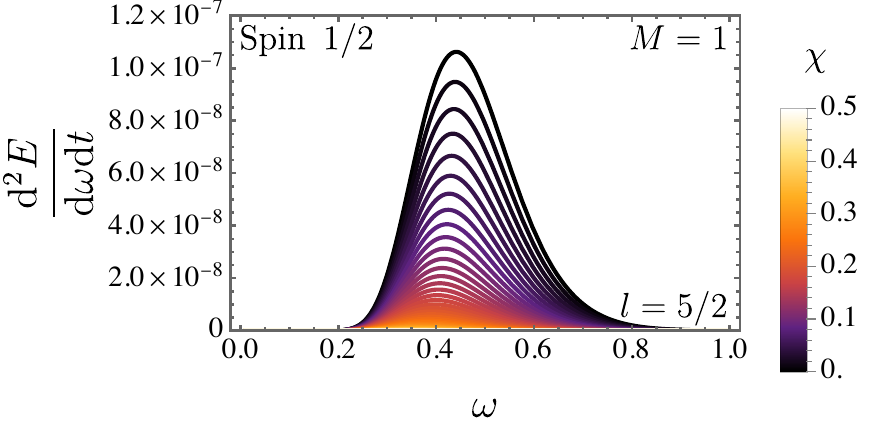}
      \caption{The energy--flux profiles are shown for the spinor perturbations for the choice $M=1$, with the cases $l=1/2$ (upper left), $l=3/2$ (upper right), and $l=5/2$ (lower panel) plotted separately}
      \label{spinoremissionrate}
\end{figure}

Finally, the energy emission rate for spinor perturbations is
\ie
	\frac{{{\mathrm{d}^2}E}}{{\mathrm{d}\omega \mathrm{d}t}} = \frac{{2{\pi ^2}{\sigma}^{\psi}_{l \omega}}}{{{e^{\frac{\omega }{T}}} - 1}} {\omega ^3}.
\fe
Figure~\ref{spinoremissionrate} presents the energy--flux profiles for the spinor sector with $M=1$ and angular momentum values $l=1/2$ (upper left), $l=3/2$ (upper right), and $l=5/2$ (lower panel). Once the parameter $\chi$ is introduced, as shown in other spin configurations, the resulting curves reveal a clear decrease in amplitude, indicating that the Lorentz--violating deformation systematically weakens the emitted energy throughout the spectrum.

\begin{figure}
    \centering
     \includegraphics[scale=0.55]{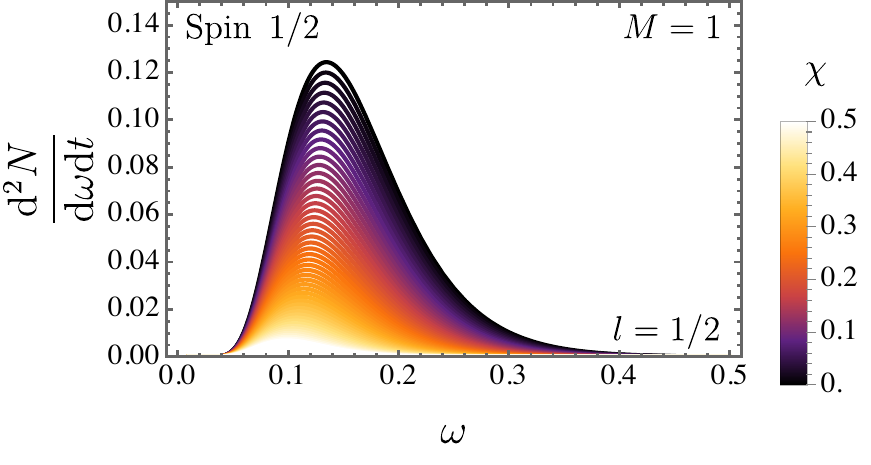}
     \includegraphics[scale=0.55]{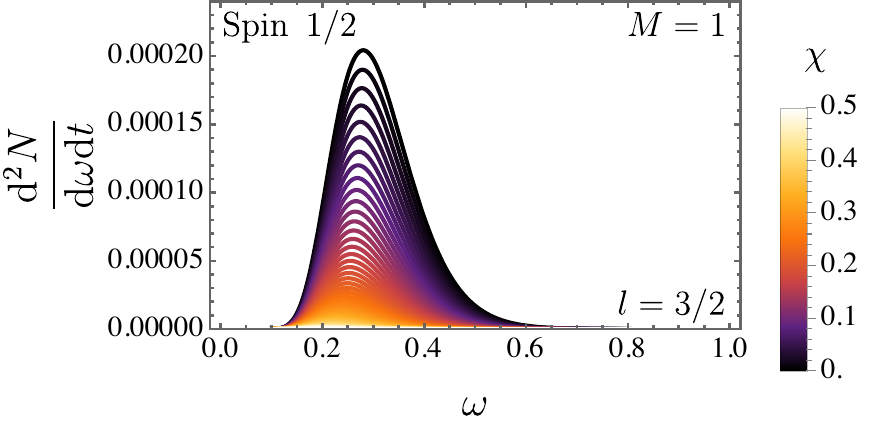}
      \includegraphics[scale=0.58]{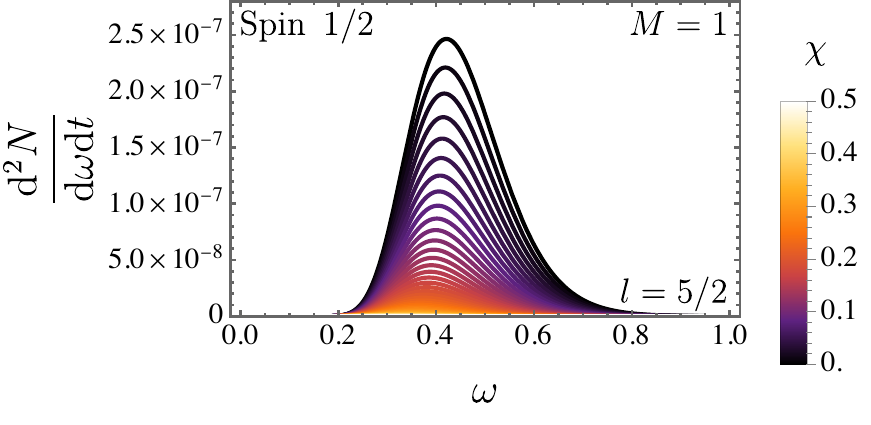}
      \caption{The particle--flux profiles are shown for the spinor perturbations for the choice $M=1$, with the cases $l=1/2$ (upper left), $l=3/2$ (upper right), and $l=5/2$ (lower panel) plotted separately. }
    \label{spinorparticle}
\end{figure}

On the other hand, the particle emission reads
\ie
\frac{\mathrm{d}^{2}N}{\mathrm{d}\omega \mathrm{d}t}
= \frac{2\pi^{2}\,\sigma^{\psi}_{l \omega}\,\omega^{2}}
       {{{e^{\frac{\omega }{T}}} - 1}}.
\fe
Figure~\ref{spinorparticle} shows the particle–production curves for the spinor sector with $M=1$ and the angular momentum values $l=1/2$ (upper left), $l=3/2$ (upper right), and $l=5/2$ (lower panel). The pattern follows the same tendency identified in the corresponding energy--flux analysis: once the parameter $\chi$ departs from the Schwarzschild limit, the amplitudes of the particle emission diminish across the entire frequency domain. Thereby, in this sector as well, the Lorentz--violating contribution acts to suppress the overall particle output.

\begin{figure}
    \centering
     \includegraphics[scale=0.65]{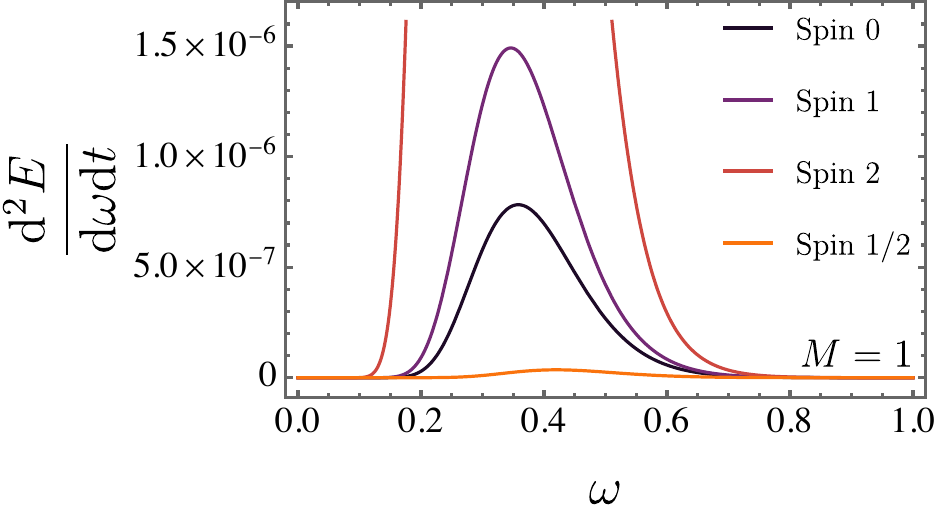}
      \caption{Energy–emission profiles for all spin sectors, evaluated at $l=2$ (bosons) and $l=5/2$ (spinors) with $M=1$. }
    \label{comparemission}
\end{figure}

Finally, Fig.~\ref{comparemission} contrasts the energy--emission curves obtained for all perturbative sectors. The comparison is carried out for $l=2$ in the bosonic cases and for $l=5/2$ in the spinor sector (for all cases we consider $M=1$). The hierarchy follows the same pattern observed in the evaporation–lifetime analysis: tensor perturbations produce the most pronounced energy output, whereas the spinor contribution remains the weakest across the spectrum.

\section{Linking quasinormal oscillations with greybody transmission }

The spectrum of quasinormal oscillations was obtained through a semi--analytical procedure rather than by solving the perturbation equations in closed form. Instead of working directly with the full lapse function—which complicates a purely numerical treatment—the analysis relied on the WKB framework, applied here in its third--order formulation. This approximation scheme, originally introduced and later refined in Refs.~\cite{iyer1987black,konoplya2011quasinormal,iyer1987black2,konoplya2019higher,cardoso2001quasinormal}, yields accurate estimates for the quasinormal frequencies by evaluating the effective potential near its peak
\begin{equation}
\label{vccssddd}
\begin{split}
    {\omega ^2}=&  \,\,{V_0} + \sqrt { - 2{V_0}^{\prime \prime }} \Lambda (n) - i\left(n + \frac{1}{2}\right)\sqrt { - 2{V_0}^{\prime \prime }} (1 + \Omega (n)),
\end{split}
 \end{equation}
with 
\begin{equation}
\begin{split}
    \Lambda (n) =   \frac{1}{{\sqrt { - 2{V_0}^{\prime \prime }} }}\left[\frac{1}{8}\left(\frac{{V_0^{\left(4\right)}}}{{{V_0}^{\prime \prime }}}\right)\left(\frac{1}{4} + {\alpha ^2}\right) - \frac{1}{{288}}{\left(\frac{{{V_0}^{\prime \prime \prime }}}{{{V_0}^{\prime \prime }}}\right)^2}\left(7 + 60{\alpha ^2}\right)\right],
\end{split}
\end{equation}
and 
\begin{eqnarray}
    \Omega (n) & = & \left( {\frac{1}{{ - 2{V_0}^{\prime \prime }}}} \right)
  \frac{5}{{6912}}{\left(\frac{{{V_0}^{\prime \prime \prime }}}{{{V_0}^{\prime \prime }}}\right)^4}\left( {77 + 188 \times {\alpha ^2}} \right)  - \frac{1}{{384}}  \left( {\frac{{{V_0}{{^{\prime \prime \prime }}^2}V_0^{(4)}}}{{{V_0}{{^{\prime \prime }}^3}}}} \right)\left( {51 + 100{\alpha ^2}} \right)  
  	\nonumber  \\ 
   && + \frac{1}{{2304}}{\left(\frac{{V_0^{(4)}}}{{{V_0}^{\prime \prime }}}\right)^2}\left( {67 + 68{\alpha ^2}} \right)  + \frac{1}{{288}}\left( {\frac{{{V_0}^{\prime \prime \prime }V_0^{(5)}}}{{{V_0}{{^{\prime \prime }}^2}}}} \right)\left( {19 + 28{\alpha ^2}} \right) 
   		\nonumber  \\ 
    && - \frac{1}{{288}}\left(\frac{{V_0^{(6)}}}{{{V_0}^{\prime \prime }}}\right)\left( {5 + 4{\alpha ^2}} \right). 
\end{eqnarray}
In the WKB prescription, the quantity $\alpha$ enters as $\alpha = n + \tfrac{1}{2}$, where $n$ denotes the overtone number, restricted by the usual requirement $n \leq l$.

The connection between quasinormal spectra and greybody behavior has recently been revisited from a different angle in Ref.~\cite{konoplya2024correspondence}. That work showed that, when the system approaches the eikonal domain (or equivalently the high–frequency limit), the greybody coefficients of any static and spherically symmetric geometry are essentially controlled by the lowest quasinormal frequency. Deviations from this pattern arise only when $l$ is small, since in that regime the higher overtones begin to influence the transmission probability. Within this approximation scheme, the transmission and reflection amplitudes follow from the standard WKB expression developed in Ref.~\cite{iyer1987black}
\begin{equation}
{\left| R \right|^2} = \frac{1}{{1 + {e^{ - 2\pi i{\mathcal{K}}}}}},
\end{equation}
\begin{equation}\label{Trans}
{\left| T \right|^2} = \frac{1}{{1 + {e^{  2\pi i{\mathcal{K}}}}}}.
\end{equation}

In the approach discussed in Ref.~\cite{konoplya2024correspondence}, the quantity $\mathcal{K}$ is not introduced directly; instead, it emerges from a specific combination of the first two quasinormal oscillations. These modes—labelled by $n=0$ and $n=1$—supply the pair of frequencies $(\omega_0,\omega_1)$ used to build the parameter. Each mode frequency is written as $\omega = \omega_R + i\,\omega_I$, where the real part encodes the oscillation rate, while the imaginary component determines the decay of the perturbation
\begin{equation}\label{Tfactor}
     - i{\mathcal{K}} =  - \frac{{{\omega ^2} - {\omega _{0R}}^2}}{{4{\omega _{0R}}{\omega _{0I}}}} + {\Delta _1} + {\Delta _2} + {\Delta _f},
\end{equation}
in which
\ie
{\Delta _1}  = \frac{{{\omega _{0R}} - {\omega _{1R}}}}{{16{\omega _{0I}}}},
\fe
\ie
\begin{split}
    \Delta_2 & =  - \frac{{{\omega ^2} - \omega_{0R}^2}}{{32{\omega _{0R}}{\omega _{0I}}}}\left[\frac{{{{({\omega _{0R}} - {\omega _{R1}})}^2}}}{{4{\omega _{0I}}^2}} - \frac{{3{\omega _{0I}} - {\omega _{1I}}}}{{3{\omega _{0I}}}}\right] + \frac{{{{({\omega ^2} - \omega _{0R}^2)}^2}}}{{16\omega _{0R}^3{\omega _{0I}}}}\left[1 + \frac{{{\omega _{0R}}({\omega _{0R}} - {\omega _{1R}})}}{{4\omega _{0I}^2}}\right], 
\end{split}
\fe
and
\ie
\begin{split}
\Delta_f &=  - \frac{{{{({\omega ^2} - \omega _{0R}^2)}^3}}}{{32\omega _{0R}^5{\omega _{0I}}}}\left\{1 + \frac{{{\omega _{0R}}({\omega _{0R}} - {\omega _{1R}})}}{{4{\omega _{0I}}^2}}  + \omega _{0R}^2\left[\frac{{{{({\omega _{0R}} - {\omega _{1R}})}^2}}}{{16\omega _{0I}^4}} - \frac{{3{\omega _{0I}} - {\omega _{1I}}}}{{12{\omega_{0I}}}}\right]\right\}.
\end{split}
\fe

The subsequent analysis applies the previously outlined scheme to the four perturbative sectors—scalar, vector, tensor, and spinor. The value of $\mathcal{K}$ is extracted from Eq.~(\ref{Trans}), whereas the quasinormal frequencies that enter this expression are determined through the third--order WKB prescription of Eq.~(\ref{vccssddd}). For consistency in the plots that follow, the resulting greybody quantities are represented by the notation $\Bar{\Gamma}(\omega,\chi)$.


\subsection{Spin--$0$ particle modes }

\begin{figure}
    \centering
      \includegraphics[scale=0.8]{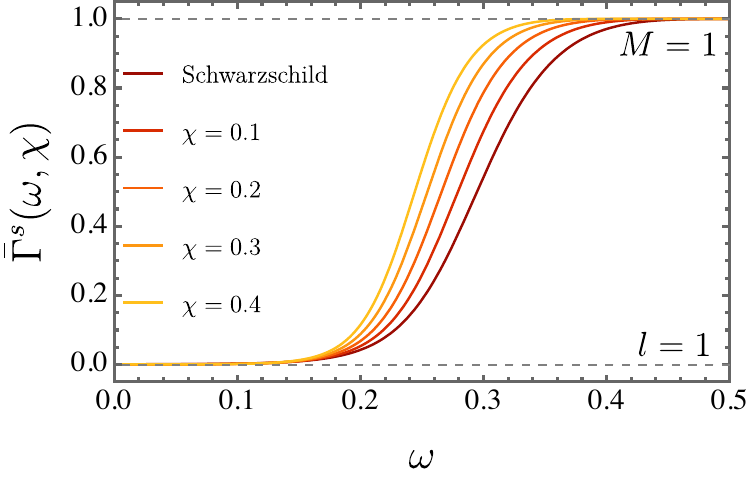}
    \caption{Relation between the scalar quasinormal spectrum and the corresponding greybody transmission for $l = 1$, displayed for multiple choices of the parameter $\chi$.}
    \label{qnmscorresscalar}
\end{figure}

The case of scalar perturbations with $l = 1$ is displayed in Fig.~\ref{qnmscorresscalar}, where the influence of the Lorentz--violating parameter becomes evident once the greybody behavior is contrasted with its Schwarzschild counterpart. The deformation governed by $\chi$ alters the quasinormal spectrum in such a way that both $\mathrm{Re}\,\omega$ and $\mathrm{Im}\,\omega$ decrease, a trend already identified in Ref.~\cite{AraujoFilho:2025zaj}. Because the real part sets the characteristic oscillation scale, its reduction displaces the principal absorption band toward lower frequencies. At the same time, a smaller imaginary component indicates a potential barrier that is less effective in reflecting the wave, thereby reducing damping.

When these modifications are translated into the greybody response, the resulting curve $\bar{\Gamma}^{s}(\omega,\chi)$ rises noticeably as $\chi$ becomes larger. The transmission becomes more efficient and the amplitude grows across the spectrum, signalling that the Lorentz--violating sector enhances the passage of scalar modes and shifts the dominant emission toward the infrared region.

\subsection{Spin--$1$ particle modes }

\begin{figure}
    \centering
      \includegraphics[scale=0.8]{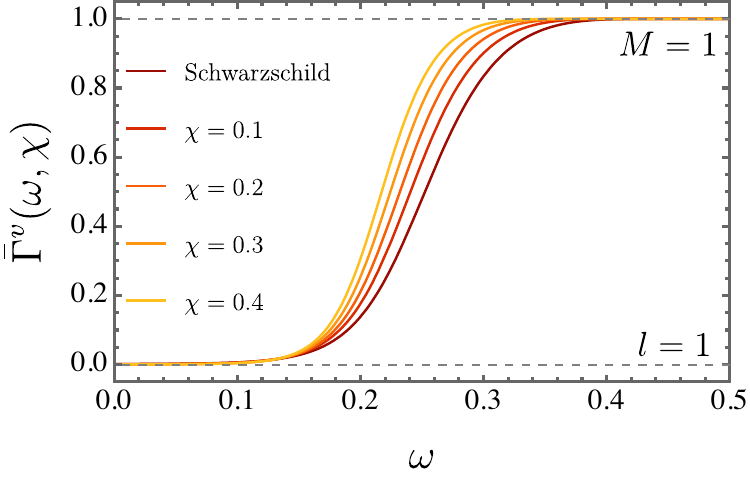}
    \caption{Relation between the vector quasinormal spectrum and the associated greybody transmission for $l = 1$.}
    \label{correlavectttt}
\end{figure}

Figure~\ref{correlavectttt} displays the behavior of vector perturbations for $l = 1$ once the quasinormal frequencies are compared with the corresponding greybody response. As soon as the parameter $\chi$ departs from the Schwarzschild limit, the transmission curves $\bar{\Gamma}^{v}(\omega,\chi)$ rise noticeably across the spectrum: for any fixed $\omega$, they stand above the undeformed case. This enhancement follows from the modifications introduced in the quasinormal structure, since increasing $\chi$ causes both the oscillation frequency and the damping rate to decrease. The lowering of $\mathrm{Re}\,\omega$ shifts the characteristic absorption window toward smaller values of $\omega$, while a reduced $\mathrm{Im}\,\omega$ reflects a potential barrier that dissipates the perturbations less efficiently.

These spectral adjustments ultimately translate into higher transmission probabilities and more prominent greybody profiles as $\chi$ grows. Such a pattern does not occur in the earlier bumblebee geometry of Ref.~\cite{Casana:2017jkc}. In that solution the temporal component of the metric coincides with the Schwarzschild one, so the effective potential governing vector modes remains unchanged. Without this deformation in $g_{tt}$, neither the quasinormal frequencies (as examined in Ref.~\cite{AraujoFilho:2025zaj}) nor the greybody factors experience the behavior found in the present analysis. Here, by contrast, the alteration in $g_{tt}$ reshapes the potential barrier, and it is precisely this modification that drives the observed evolution of the quasinormal spectrum and its greybody counterpart.

\subsection{Spin--$2$ particle modes }

\begin{figure}
    \centering
      \includegraphics[scale=0.8]{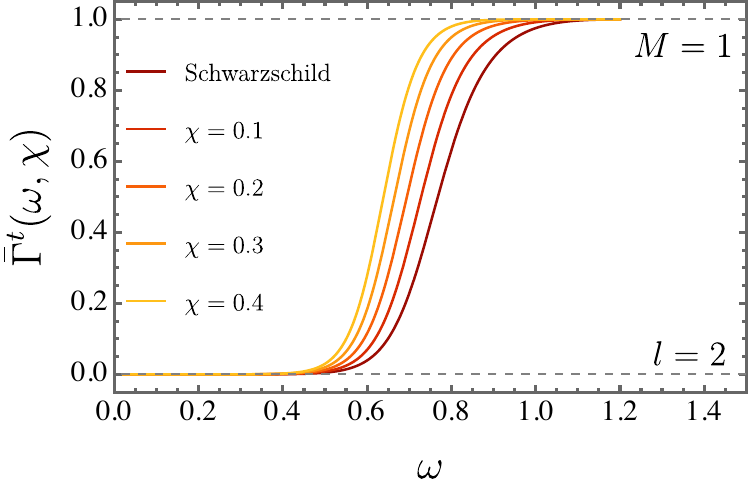}
    \caption{Relation between the tensor quasinormal spectrum and the corresponding greybody transmission for $l = 2$.}
    \label{correlatensor}
\end{figure}

Figure~\ref{correlatensor} displays the tensor sector for $l = 2$, highlighting how its quasinormal characteristics manifest in the corresponding greybody response. Once the parameter $\chi$ departs from the Schwarzschild limit, the curves $\bar{\Gamma}^{t}(\omega,\chi)$ rise systematically above their undeformed counterparts for every frequency considered. This behavior reflects the changes induced in the quasinormal spectrum: increasing $\chi$ pushes both the oscillation frequency and the damping rate to smaller values, a trend also identified in Ref.~\cite{AraujoFilho:2025zaj}.

A lowered real part of the frequency shifts the dominant absorption region toward the low--$\omega$ regime, while a smaller imaginary component signals a potential barrier that attenuates the perturbations less effectively. When both effects are combined, the transmission becomes more efficient and the greybody profiles develop more pronounced amplitudes as $\chi$ grows.

\subsection{Spin--$1/2$ particle modes }

\begin{figure}
    \centering
      \includegraphics[scale=0.8]{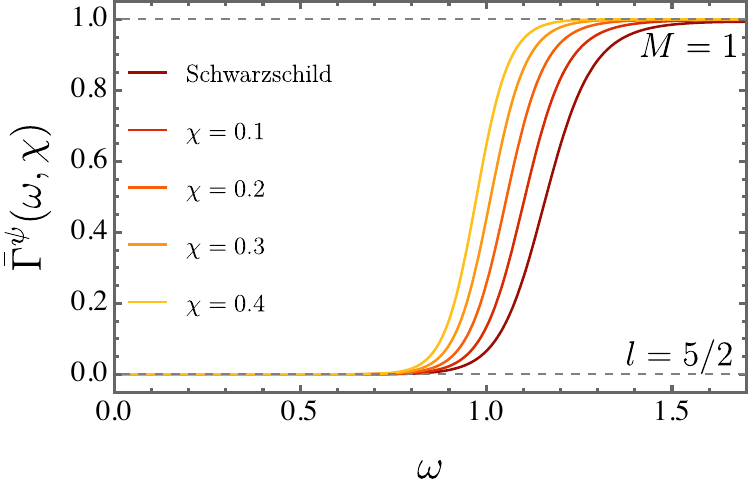}
    \caption{Relation between the spinor quasinormal spectrum and the associated greybody transmission for $l = 5/2$.}
    \label{correlaspinor}
\end{figure}

Figure~\ref{correlaspinor} depicts the spinor case with $l = 5/2$, revealing how its quasinormal behavior influences the corresponding greybody response. Once the parameter $\chi$ departs from the Schwarzschild limit, the transmission curves $\bar{\Gamma}^{t}(\omega,\chi)$ consistently rise above the undeformed profile across the entire frequency range. Because these spectral shifts relocate the characteristic absorption scale to lower $\omega$ and weaken the damping imposed by the potential barrier, the greybody response acquires larger amplitudes and more efficient transmission for growing $\chi$.

\section{Conclusion }\label{Sec:Conclusion}

This paper was aimed at examining quantum particle creation, radiative properties, and evaporation lifetimes for bosonic (spin–0, spin–1, spin–2) and fermionic (spin–1/2) fields in a recently proposed bumblebee black hole. In essence, we evaluated how the spin sector affected these phenomena.

We first presented the black hole solution and discussed its basic properties. The thermal quantities were then computed. The Hawking temperature was obtained from the surface–gravity prescription,
$T_{H} = \frac{1}{4\pi r_{h}(1+\chi)} \approx \frac{1}{4\pi r_{h}} - \frac{\chi}{4\pi r_{h}}$,
or, in terms of mass, $\frac{1}{8\pi M} - \frac{\chi}{8\pi M}$. In contrast, the entropy and heat capacity showed no dependence on the Lorentz–violating parameter $\chi$, matching the Schwarzschild case. The topological thermodynamic analysis was carried out as well.

Quantum particle creation for bosons was then derived. After quantizing the scalar field, the radiation spectrum was obtained from the Bogoliubov coefficients, yielding a blackbody--like distribution whose temperature coincided with the value obtained from the surface gravity. The tunneling method was subsequently applied to incorporate energy conservation. Using the Painlevé–Gullstrand form of the metric, the imaginary part of the action followed from the residue method:
$\mathrm{Im}\,\mathcal{S} = 4\pi (1+\chi)\,\omega\left( M - \frac{\omega}{2} \right)$,
so that $\Gamma \sim e^{-2,\mathrm{Im}\,\mathcal{S}} = e^{-8(1+\chi)\,\omega\left(M - \frac{\omega}{2}\right)}$.
The particle density therefore read
$n = \frac{1}{e^{8\pi(1+\chi),\omega\left(M-\frac{\omega}{2}\right)} - 1}$.
For the bosonic case, $\chi$ reduced the particle density. Comparison with other Lorentz–violating geometries showed the hierarchy $n^{\text{this work}} < n^{\text{bum (metric)}} \approx n^{\text{bum (met–aff)}} < n^{\text{Schw}} < n^{\text{KR (Model 2)}} < n^{\text{KR (Model 1)}} < n^{\text{NC KR}}$. In addition, for fermions, a near–horizon approximation allowed the particle density
$n_{\psi} = \frac{1}{e^{8\pi(1+\chi) M \omega} + 1}$.

Greybody bounds were examined for all spins. Scalar modes obeyed
$|T_{b}^{s}| = \mathrm{sech}^{2}\!\left[ \frac{2l(l+1)(1+\chi)+1}{(2\omega)(4M(1+\chi))} \right]$,
and tensor modes obeyed
$|T_{b}^{t}| = \mathrm{sech}^{2}\!\left[ \frac{2l(l+1)(1+\chi)-4\chi-3}{(2\omega)(4M(1+\chi))} \right]$,
both showing explicit dependence on $\chi$. No such dependence appeared for vector and spinorial modes. Overall, $\chi$ increased the intensities associated with the bounds.

The full greybody factors were then computed numerically with the sixth--order WKB method, followed by the partial absorption cross sections. Unlike the bounds, all spins—including vector and tensor sectors—became dependent on $\chi$. In each case, $\chi$ increased both the greybody intensities and the partial absorption cross section. The hierarchy $|T_{b}^{t}| > |T_{b}^{v}| > |T_{b}^{s}| > |T_{b}^{\psi}|$
was maintained for bounds, factors, and absorption.

Evaporation lifetimes were studied using the Stefan–Boltzmann law for all spins. Analytical estimates were obtained via the bounds. Spin–2 fields evaporated the fastest and spin–1/2 the slowest. The high--frequency regime was also explored, leading to the hierarchy
$t_{\text{evap-final}}^{\text{this work}}
> t_{\text{evap-final}}^{\text{bum (metric)}}
> = t_{\text{evap-final}}^{\text{bum (met–aff)}}
> t_{\text{evap-final}}^{\text{Schw}}
> t_{\text{evap-final}}^{\text{KR (Model 2)}}
> t_{\text{evap-final}}^{\text{KR (Model 1)}}
> t_{\text{evap-final}}^{\text{NC KR}}$ .

The emission rate for all spins followed the same pattern as the evaporation time: increasing $\chi$ reduced the emission of particle and energy modes; spin–2 exhibited the strongest emission, whereas spin–1/2 remained the weakest. Finally, the correlation between the quasinormal modes and the greybody factors was established.

As future work, scattering effects and the total absorption cross section appeared to be promising extensions of this study. Further topics included entanglement degradation, equivalence–principle tests, and HBAR {(Horizon Brightened Acceleration Radiation)} entropy { similar to \cite{Sen:2022tru}}. These analyses are under development and are expected to be released very soon on arXiv.


\section*{Acknowledgments}
\hspace{0.5cm} A.A.A.F. is supported by Conselho Nacional de Desenvolvimento Cient\'{\i}fico e Tecnol\'{o}gico (CNPq) and Fundação de Apoio à Pesquisa do Estado da Paraíba (FAPESQ), project numbers 150223/2025-0 and 1951/2025. 
N. Heidari is supported by the Conselho Nacional de Desenvolvimento Científico e Tecnológico (CNPq), grant No. 152891/2025-0. N. H. also acknowledges the networking support provided by COST Action CA22113 -- Fundamental challenges in theoretical physics (Theory and Challenges), CA21106 -- COSMIC WISPers in the Dark Universe: Theory, astrophysics and experiments (CosmicWISPers), CA21136 -- Addressing observational tensions in cosmology with systematics and fundamental physics (CosmoVerse), and CA23130 -- Bridging high and low energies in search of quantum gravity (BridgeQG).

\section*{Data Availability Statement}

Data Availability Statement: No Data associated with the manuscript

\bibliographystyle{ieeetr}
\bibliography{main}

\begin{thebibliography}{100}

\bibitem{kostelecky1989spontaneous}
V.~A. Kosteleck{\`y} and S.~Samuel, ``Spontaneous breaking of lorentz symmetry
  in string theory,'' {\em Physical Review D}, vol.~39, no.~2, p.~683, 1989.

\bibitem{colladay1997cpt}
D.~Colladay and V.~A. Kosteleck{\`y}, ``Cpt violation and the standard model,''
  {\em Physical Review D}, vol.~55, no.~11, p.~6760, 1997.

\bibitem{kostelecky2004gravity}
V.~A. Kosteleck{\`y}, ``Gravity, lorentz violation, and the standard model,''
  {\em Physical Review D}, vol.~69, no.~10, p.~105009, 2004.

\bibitem{kostelecky1999constraints}
V.~A. Kosteleck{\`y} and C.~D. Lane, ``Constraints on lorentz violation from
  clock-comparison experiments,'' {\em Physical Review D}, vol.~60, no.~11,
  p.~116010, 1999.

\bibitem{kostelecky2011data}
V.~A. Kosteleck{\`y} and N.~Russell, ``Data tables for lorentz and cpt
  violation,'' {\em Reviews of Modern Physics}, vol.~83, no.~1, pp.~11--31,
  2011.

\bibitem{Bluhm:2019ato}
R.~Bluhm, H.~Bossi, and Y.~Wen, ``{Gravity with explicit spacetime symmetry
  breaking and the Standard-Model Extension},'' {\em Phys. Rev. D}, vol.~100,
  no.~8, p.~084022, 2019.

\bibitem{Bluhm:2023kph}
R.~Bluhm and Y.~Zhi, ``{Spontaneous and Explicit Spacetime Symmetry Breaking in
  Einstein{\textendash}Cartan Theory with Background Fields},'' {\em Symmetry},
  vol.~16, no.~1, p.~25, 2024.

\bibitem{Maluf:2014dpa}
R.~V. Maluf, C.~A.~S. Almeida, R.~Casana, and M.~M. Ferreira, Jr.,
  ``{Einstein-Hilbert graviton modes modified by the Lorentz-violating
  bumblebee Field},'' {\em Phys. Rev. D}, vol.~90, no.~2, p.~025007, 2014.

\bibitem{Maluf:2013nva}
R.~V. Maluf, V.~Santos, W.~T. Cruz, and C.~A.~S. Almeida, ``{Matter-gravity
  scattering in the presence of spontaneous Lorentz violation},'' {\em Phys.
  Rev. D}, vol.~88, no.~2, p.~025005, 2013.

\bibitem{bluhm2008spontaneous}
R.~Bluhm, S.-H. Fung, and V.~A. Kosteleck{\`y}, ``Spontaneous lorentz and
  diffeomorphism violation, massive modes, and gravity,'' {\em Physical Review
  D—Particles, Fields, Gravitation, and Cosmology}, vol.~77, no.~6,
  p.~065020, 2008.

\bibitem{bluhm2005spontaneous}
R.~Bluhm and V.~A. Kosteleck{\`y}, ``Spontaneous lorentz violation,
  nambu-goldstone modes, and gravity,'' {\em Physical Review D—Particles,
  Fields, Gravitation, and Cosmology}, vol.~71, no.~6, p.~065008, 2005.

\bibitem{jacobson2004einstein}
T.~Jacobson and D.~Mattingly, ``Einstein-aether waves,'' {\em Physical Review
  D}, vol.~70, no.~2, p.~024003, 2004.

\bibitem{kostelecky1991photon}
V.~A. Kosteleck{\`y} and S.~Samuel, ``Photon and graviton masses in string
  theories,'' {\em Physical Review Letters}, vol.~66, no.~14, p.~1811, 1991.

\bibitem{Liu:2022dcn}
W.~Liu, X.~Fang, J.~Jing, and J.~Wang, ``{QNMs of slowly rotating
  Einstein{\textendash}Bumblebee black hole},'' {\em Eur. Phys. J. C}, vol.~83,
  no.~1, p.~83, 2023.

\bibitem{Bertolami:2005bh}
O.~Bertolami and J.~Paramos, ``{The Flight of the bumblebee: Vacuum solutions
  of a gravity model with vector-induced spontaneous Lorentz symmetry
  breaking},'' {\em Phys. Rev. D}, vol.~72, p.~044001, 2005.

\bibitem{Casana:2017jkc}
R.~Casana, A.~Cavalcante, F.~P. Poulis, and E.~B. Santos, ``{Exact
  Schwarzschild-like solution in a bumblebee gravity model},'' {\em Phys. Rev.
  D}, vol.~97, no.~10, p.~104001, 2018.

\bibitem{Liu:2024wpa}
W.~Liu, C.~Wen, and J.~Wang, ``{Lorentz violation alleviates gravitationally
  induced entanglement degradation},'' {\em JHEP}, vol.~01, p.~184, 2025.

\bibitem{AraujoFilho:2025hkm}
A.~A. Ara{\'u}jo~Filho, ``{How does non-metricity affect particle creation and
  evaporation in bumblebee gravity?},'' {\em JCAP}, vol.~06, p.~026, 2025.

\bibitem{AraujoFilho:2024ctw}
A.~A. Ara{\'u}jo~Filho, ``{Particle creation and evaporation in Kalb-Ramond
  gravity},'' {\em JCAP}, vol.~04, p.~076, 2025.

\bibitem{Neves:2022qyb}
J.~C.~S. Neves, ``{Kasner cosmology in bumblebee gravity},'' {\em Annals
  Phys.}, vol.~454, p.~169338, 2023.

\bibitem{Neves:2024ggn}
J.~C.~S. Neves and F.~G. Gardim, ``{Stars and quark stars in bumblebee
  gravity},'' {\em Annals Phys.}, vol.~475, p.~169950, 2025.

\bibitem{Liang:2022hxd}
D.~Liang, R.~Xu, X.~Lu, and L.~Shao, ``{Polarizations of gravitational waves in
  the bumblebee gravity model},'' {\em Phys. Rev. D}, vol.~106, no.~12,
  p.~124019, 2022.

\bibitem{amarilo2024gravitational}
K.~M. Amarilo, M.~B. Ferreira~Filho, A.~A. Ara{\'u}jo~Filho, and J.~A. A.~S.
  Reis, ``Gravitational waves effects in a lorentz--violating scenario,'' {\em
  Physics Letters B}, vol.~855, p.~138785, 2024.

\bibitem{Maluf:2020kgf}
R.~V. Maluf and J.~C.~S. Neves, ``{Black holes with a cosmological constant in
  bumblebee gravity},'' {\em Phys. Rev. D}, vol.~103, no.~4, p.~044002, 2021.

\bibitem{Uniyal:2022xnq}
A.~Uniyal, S.~Kanzi, and {\.I}.~Sakall{\i}, ``{Some observable physical
  properties of the higher dimensional dS/AdS black holes in Einstein-bumblebee
  gravity theory},'' {\em Eur. Phys. J. C}, vol.~83, no.~7, p.~668, 2023.

\bibitem{Filho:2022yrk}
A.~A.~A. Filho, J.~R. Nascimento, A.~Y. Petrov, and P.~J. Porf{\'\i}rio,
  ``{Vacuum solution within a metric-affine bumblebee gravity},'' {\em Phys.
  Rev. D}, vol.~108, no.~8, p.~085010, 2023.

\bibitem{AraujoFilho:2024ykw}
A.~A. Ara{\'u}jo~Filho, J.~R. Nascimento, A.~Y. Petrov, and P.~J.
  Porf{\'\i}rio, ``{An exact stationary axisymmetric vacuum solution within a
  metric-affine bumblebee gravity},'' {\em JCAP}, vol.~07, p.~004, 2024.

\bibitem{AraujoFilho:2025rvn}
A.~A. Ara{\'u}jo~Filho, N.~Heidari, I.~P. Lobo, Y.~Shi, and F.~S.~N. Lobo,
  ``{The Flight of the Bumblebee in a Non-Commutative Geometry: A New Black
  Hole Solution},'' 9 2025.

\bibitem{AraujoFilho:2025jcu}
A.~A. Ara{\'u}jo~Filho, N.~Heidari, and I.~P. Lobo, ``{A non-commutative
  Kalb-Ramond black hole},'' {\em JCAP}, vol.~09, p.~076, 2025.

\bibitem{Ovgun:2018xys}
A.~{\"O}vg{\"u}n, K.~Jusufi, and {\.I}.~Sakall{\i}, ``{Exact traversable
  wormhole solution in bumblebee gravity},'' {\em Phys. Rev. D}, vol.~99,
  no.~2, p.~024042, 2019.

\bibitem{AraujoFilho:2024iox}
A.~A. Ara{\'u}jo~Filho, J.~A. A.~S. Reis, and A.~{\"O}vg{\"u}n, ``{Modified
  particle dynamics and thermodynamics in a traversable wormhole in bumblebee
  gravity},'' {\em Eur. Phys. J. C}, vol.~85, no.~1, p.~83, 2025.

\bibitem{Magalhaes:2025lti}
R.~B. Magalh{\~a}es, L.~A. Lessa, and R.~Casana, ``{Lorentz-violating
  wormholes: The role of the matter coupled to Lorentz-violating fields},'' 7
  2025.

\bibitem{Magalhaes:2025nql}
R.~B. Magalh{\~a}es, L.~A. Lessa, and M.~M. Ferreira, ``{Wormholes in
  Lorentz-violating gravity},'' 5 2025.

\bibitem{Pereira:2025xnw}
C.~F.~S. Pereira, M.~V. d.~S. Silva, H.~Belich, D.~C.~Rodrigues, J.~C. Fabris,
  and M.~E. Rodrigues, ``{Black-bounce solutions in a k-essence theory under
  the effects of bumblebee gravity},'' {\em Phys. Rev. D}, vol.~111, no.~12,
  p.~124005, 2025.

\bibitem{Shi:2025plr}
Y.~Shi and A.~A. Ara{\'u}jo~Filho, ``Effects of bumblebee gravity on neutrino
  motion,'' {\em Journal of Cosmology and Astroparticle Physics}, vol.~2025,
  no.~11, p.~045, 2025.

\bibitem{Shi:2025ywa}
Y.~Shi and A.~A. Ara{\'u}jo~Filho, ``The role of non-metricity on neutrino
  behavior in bumblebee gravity,'' {\em arXiv preprint arXiv:2505.12551}, 2025.

\bibitem{Shi:2025rfq}
Y.~Shi and A.~A. Ara{\'u}jo~Filho, ``{Influence of a Kalb-Ramond black hole on
  neutrino behavior},'' {\em JHEP}, vol.~08, p.~028, 2025.

\bibitem{Khodadi:2023yiw}
M.~Khodadi, G.~Lambiase, and L.~Mastrototaro, ``{Spontaneous Lorentz symmetry
  breaking effects on GRBs jets arising from neutrino pair annihilation process
  near a black hole},'' {\em Eur. Phys. J. C}, vol.~83, no.~3, p.~239, 2023.

\bibitem{Khodadi:2022mzt}
M.~Khodadi, G.~Lambiase, and A.~Sheykhi, ``{Constraining the Lorentz-violating
  bumblebee vector field with big bang nucleosynthesis and gravitational
  baryogenesis},'' {\em Eur. Phys. J. C}, vol.~83, no.~5, p.~386, 2023.

\bibitem{Khodadi:2022dff}
M.~Khodadi, ``{Magnetic reconnection and energy extraction from a spinning
  black hole with broken Lorentz symmetry},'' {\em Phys. Rev. D}, vol.~105,
  no.~2, p.~023025, 2022.

\bibitem{Khodadi:2021owg}
M.~Khodadi, ``{Black Hole Superradiance in the Presence of Lorentz Symmetry
  Violation},'' {\em Phys. Rev. D}, vol.~103, no.~6, p.~064051, 2021.

\bibitem{Liu:2025oho}
J.-Z. Liu, S.-P. Wu, S.-W. Wei, and Y.-X. Liu, ``{Exact Black Hole Solutions in
  Bumblebee Gravity with Lightlike or Spacelike VEVS},'' 10 2025.
\newblock arXiv:2510.16731 [gr-qc].

\bibitem{Zhu:2025fiy}
J.~Zhu and H.~Li, ``{Full classification of static spherical vacuum solutions
  to bumblebee gravity with general VEVs},'' {\em Phys. Lett. B}, vol.~876,
  p.~140396, 2026.

\bibitem{AraujoFilho:2025zaj}
A.~A. Ara{\'u}jo~Filho, N.~Heidari, I.~P. Lobo, and V.~B. Bezerra,
  ``{Gravitational aspects of a new bumblebee black hole},'' {\em Annals
  Phys.}, vol.~489, p.~170469, 2026.

\bibitem{Shi:2025tvu}
Y.~Shi and A.~A. Ara{\'u}jo~Filho, ``{Neutrino oscillations induced by a new
  bumblebee black hole},'' 11 2025.
\newblock arXiv:2511.16181 [gr-qc].

\bibitem{Kumar:2025bim}
A.~Kumar, S.~U. Islam, and S.~G. Ghosh, ``{Probing Lorentz Symmetry Violation
  through Lensing Observables of Rotating Black Holes},'' 8 2025.
\newblock arXiv:2509.00127 [gr-qc].

\bibitem{Shi:2025hfe}
Y.~Shi and A.~A. Ara{\'u}jo~Filho, ``{Accretion of matter of a new bumblebee
  black hole},'' {\em Nucl. Phys. B}, vol.~1023, p.~117327, 2026.

\bibitem{Afrin:2024egb}
M.~Afrin, S.~G. Ghosh, and A.~Wang, ``Testing egb gravity coupled to bumblebee
  field and black hole parameter estimation with eht observations,'' {\em
  Physics of the Dark Universe}, vol.~46, p.~101642, 2024.

\bibitem{Islam:2024eht}
S.~U. Islam, S.~G. Ghosh, and S.~D. Maharaj, ``Investigating rotating black
  holes in bumblebee gravity: insights from eht observations,'' {\em Journal of
  Cosmology and Astroparticle Physics}, vol.~2024, no.~12, p.~047, 2024.

\bibitem{Singh:2024greybody}
Y.~P. Singh and T.~I. Singh, ``Greybody factor and quasinormal modes of scalar
  and dirac field perturbation in schwarzschild-de sitter-like black hole in
  bumblebee gravity model,'' {\em European Physical Journal C}, vol.~84,
  p.~1245, 2024.

\bibitem{Sarmah:2025aniso}
P.~Sarmah and U.~D. Goswami, ``Anisotropic cosmology in bumblebee gravity
  theory,'' {\em Physics of the Dark Universe}, vol.~49, p.~102057, 2025.

\bibitem{Neves:2023kasner}
J.~C.~S. Neves, ``Kasner cosmology in bumblebee gravity,'' {\em Annals of
  Physics}, vol.~454, p.~169338, 2023.

\bibitem{Zhu:2025cosmo}
X.~Zhu, R.~Xu, and D.~Xu, ``Bumblebee cosmology: Tests using distance- and
  time-redshift probes,'' {\em Physics of the Dark Universe}, vol.~50,
  p.~102127, 2025.

\bibitem{Sarmah:2026uhecr}
S.~P. Sarmah, P.~Sarmah, and U.~D. Goswami, ``Ultra-high energy cosmic rays
  with ufa-15 source model in bumblebee gravity theory,'' {\em Journal of High
  Energy Astrophysics}, vol.~49, p.~100451, 2026.

\bibitem{Parker:1968mv}
L.~Parker, ``{Particle creation in expanding universes},'' {\em Phys. Rev.
  Lett.}, vol.~21, pp.~562--564, 1968.

\bibitem{Parker:1969au}
L.~Parker, ``{Quantized fields and particle creation in expanding universes.
  I},'' {\em Phys. Rev.}, vol.~183, pp.~1057--1068, 1969.

\bibitem{dewitt1975quantum}
B.~S. DeWitt, ``Quantum field theory in curved spacetime,'' {\em Physics
  Reports}, vol.~19, no.~6, pp.~295--357, 1975.

\bibitem{wald1975particle}
R.~M. Wald, ``On particle creation by black holes,'' {\em Communications in
  Mathematical Physics}, vol.~45, no.~1, pp.~9--34, 1975.

\bibitem{fulling1989aspects}
S.~A. Fulling, {\em Aspects of quantum field theory in curved spacetime}.
\newblock No.~17, Cambridge university press, 1989.

\bibitem{lin2010quantum}
S.-Y. Lin, C.-H. Chou, and B.~L. Hu, ``Quantum entanglement and entropy in
  particle creation,'' {\em Physical Review D—Particles, Fields, Gravitation,
  and Cosmology}, vol.~81, no.~8, p.~084018, 2010.

\bibitem{calzetta1989dissipation}
E.~Calzetta and B.~Hu, ``Dissipation of quantum fields from particle
  creation,'' {\em Physical Review D}, vol.~40, no.~2, p.~656, 1989.

\bibitem{wald1994quantum}
R.~M. Wald, {\em Quantum field theory in curved spacetime and black hole
  thermodynamics}.
\newblock University of Chicago press, 1994.

\bibitem{Hawking:1974rv}
S.~W. Hawking, ``{Black hole explosions?},'' {\em Nature}, vol.~248,
  pp.~30--31, 1974.

\bibitem{Hawking:1975vcx}
S.~W. Hawking, ``{Particle Creation by Black Holes},'' {\em Commun. Math.
  Phys.}, vol.~43, pp.~199--220, 1975.
\newblock [Erratum: Commun.Math.Phys. 46, 206 (1976)].

\bibitem{Gibbons:1977mu}
G.~W. Gibbons and S.~W. Hawking, ``{Cosmological Event Horizons,
  Thermodynamics, and Particle Creation},'' {\em Phys. Rev. D}, vol.~15,
  pp.~2738--2751, 1977.

\bibitem{Birrell:1982ix}
N.~D. Birrell and P.~C.~W. Davies, ``{Quantum Fields in Curved Space},'' {\em
  Cambridge Monogr. Math. Phys.}, 1982.

\bibitem{Parker:2009uva}
L.~Parker and D.~J. Toms, {\em {Quantum Field Theory in Curved Spacetime:
  Quantized Fields and Gravity}}.
\newblock Cambridge University Press, 2009.

\bibitem{hawking1974black}
S.~W. Hawking, ``Black hole explosions?,'' {\em Nature}, vol.~248, no.~5443,
  pp.~30--31, 1974.

\bibitem{hawking1975particle}
S.~W. Hawking, ``Particle creation by black holes,'' in {\em Euclidean quantum
  gravity}, pp.~167--188, World Scientific, 1975.

\bibitem{Duane1984}
Y.~Duane, ``The structure of the topological current*,'' tech. rep., 1984.

\bibitem{wei2022black}
S.-W. Wei, Y.-X. Liu, and R.~B. Mann, ``Black hole solutions as topological
  thermodynamic defects,'' {\em Physical Review Letters}, vol.~129, no.~19,
  p.~191101, 2022.

\bibitem{yerra2022topology}
P.~K. Yerra and C.~Bhamidipati, ``Topology of black hole thermodynamics in
  gauss-bonnet gravity,'' {\em Physical Review D}, vol.~105, no.~10, p.~104053,
  2022.

\bibitem{wu2023topological}
D.~Wu, ``Topological classes of thermodynamics of the four-dimensional static
  accelerating black holes,'' {\em Physical Review D}, vol.~108, no.~8,
  p.~084041, 2023.

\bibitem{wu2023topological1}
D.~Wu, ``Topological classes of rotating black holes,'' {\em Physical Review
  D}, vol.~107, no.~2, p.~024024, 2023.

\bibitem{zhang2023bulk}
M.~Zhang and J.~Jiang, ``Bulk-boundary thermodynamic equivalence: a topology
  viewpoint,'' {\em Journal of High Energy Physics}, vol.~2023, no.~6,
  pp.~1--17, 2023.

\bibitem{gogoi2023thermodynamic}
N.~J. Gogoi and P.~Phukon, ``Thermodynamic topology of 4d dyonic ads black
  holes in different ensembles,'' {\em Physical Review D}, vol.~108, no.~6,
  p.~066016, 2023.

\bibitem{fan2023topological}
Z.-Y. Fan, ``Topological interpretation for phase transitions of black holes,''
  {\em Physical Review D}, vol.~107, no.~4, p.~044026, 2023.

\bibitem{calmet2023quantum}
X.~Calmet, S.~D. Hsu, and M.~Sebastianutti, ``Quantum gravitational corrections
  to particle creation by black holes,'' {\em Physics Letters B}, vol.~841,
  p.~137820, 2023.

\bibitem{vanzo2011tunnelling}
L.~Vanzo, G.~Acquaviva, and R.~Di~Criscienzo, ``Tunnelling methods and
  hawking's radiation: achievements and prospects,'' {\em Classical and Quantum
  Gravity}, vol.~28, no.~18, p.~183001, 2011.

\bibitem{araujo2025particleasdasd}
A.~A. Ara{\'u}jo~Filho, ``Particle creation and evaporation in kalb-ramond
  gravity,'' {\em Journal of Cosmology and Astroparticle Physics}, vol.~2025,
  no.~04, p.~076, 2025.

\bibitem{araujo2025does}
A.~A. Ara{\'u}jo~Filho, ``How does non-metricity affect particle creation and
  evaporation in bumblebee gravity?,'' {\em Journal of Cosmology and
  Astroparticle Physics}, vol.~2025, no.~06, p.~026, 2025.

\bibitem{AraujoFilho:2025rwr}
A.~A. Ara{\'u}jo~Filho, ``{Particle production induced by a Lorentzian
  non-commutative spacetime},'' {\em Annals Phys.}, vol.~481, p.~170167, 2025.

\bibitem{hollands2015quantum}
S.~Hollands and R.~M. Wald, ``Quantum fields in curved spacetime,'' {\em
  Physics Reports}, vol.~574, pp.~1--35, 2015.

\bibitem{parker2009quantum}
L.~Parker and D.~Toms, {\em Quantum field theory in curved spacetime: quantized
  fields and gravity}.
\newblock Cambridge university press, 2009.

\bibitem{o10}
P.~Kraus and F.~Wilczek, ``Self-interaction correction to black hole
  radiance,'' {\em Nuclear Physics B}, vol.~433, no.~2, pp.~403--420, 1995.

\bibitem{011}
M.~K. Parikh and F.~Wilczek, ``Hawking radiation as tunneling,'' {\em Physical
  review letters}, vol.~85, no.~24, p.~5042, 2000.

\bibitem{parikh2004energy}
M.~K. Parikh, ``Energy conservation and hawking radiation,'' {\em arXiv
  preprint hep-th/0402166}, 2004.

\bibitem{o69}
R.~Kerner and R.~B. Mann, ``Fermions tunnelling from black holes,'' {\em
  Classical and Quantum Gravity}, vol.~25, no.~9, p.~095014, 2008.

\bibitem{o71}
R.~Di~Criscienzo and L.~Vanzo, ``Fermion tunneling from dynamical horizons,''
  {\em Europhysics Letters}, vol.~82, no.~6, p.~60001, 2008.

\bibitem{o75}
M.~Rehman and K.~Saifullah, ``Charged fermions tunneling from accelerating and
  rotating black holes,'' {\em Journal of Cosmology and Astroparticle Physics},
  vol.~2011, no.~03, p.~001, 2011.

\bibitem{o70}
A.~Yale, ``Exact hawking radiation of scalars, fermions, and bosons using the
  tunneling method without back-reaction,'' {\em Physics Letters B}, vol.~697,
  no.~4, pp.~398--403, 2011.

\bibitem{o74}
A.~Yale and R.~B. Mann, ``Gravitinos tunneling from black holes,'' {\em Physics
  Letters B}, vol.~673, no.~2, pp.~168--172, 2009.

\bibitem{o72}
H.-L. Li, S.-Z. Yang, T.-J. Zhou, and R.~Lin, ``Fermion tunneling from a vaidya
  black hole,'' {\em Europhysics Letters}, vol.~84, no.~2, p.~20003, 2008.

\bibitem{o73}
R.~Kerner and R.~B. Mann, ``Charged fermions tunnelling from kerr--newman black
  holes,'' {\em Physics letters B}, vol.~665, no.~4, pp.~277--283, 2008.

\bibitem{o77}
A.~Yale, ``There are no quantum corrections to the hawking temperature via
  tunneling from a fixed background,'' {\em The European Physical Journal C},
  vol.~71, pp.~1--4, 2011.

\bibitem{o76}
B.~Chatterjee and P.~Mitra, ``Hawking temperature and higher order
  calculations,'' {\em Physics Letters B}, vol.~675, no.~2, pp.~240--242, 2009.

\bibitem{o83}
A.~Barducci, R.~Casalbuoni, and L.~Lusanna, ``Supersymmetries and the
  pseudoclassical relativistic electron,'' {\em Nuovo Cimento. A}, vol.~35,
  no.~3, pp.~377--399, 1976.

\bibitem{o84}
G.~Cognola, L.~Vanzo, S.~Zerbini, and R.~Soldati, ``On the lagrangian
  formulation of a charged spinning particle in an external electromagnetic
  field,'' {\em Physics Letters B}, vol.~104, no.~1, pp.~67--69, 1981.

\bibitem{konoplya2019higher}
R.~A. Konoplya, A.~Zhidenko, and A.~F. Zinhailo, ``{Higher order WKB formula
  for quasinormal modes and grey-body factors: recipes for quick and accurate
  calculations},'' {\em Class. Quant. Grav.}, vol.~36, p.~155002, 2019.

\bibitem{cardoso2001quasinormal}
V.~Cardoso and J.~P. Lemos, ``Quasinormal modes of schwarzschild--anti-de
  sitter black holes: Electromagnetic and gravitational perturbations,'' {\em
  Physical Review D}, vol.~64, no.~8, p.~084017, 2001.

\bibitem{konoplya2011quasinormal}
R.~Konoplya and A.~Zhidenko, ``Quasinormal modes of black holes: From
  astrophysics to string theory,'' {\em Reviews of Modern Physics}, vol.~83,
  no.~3, p.~793, 2011.

\bibitem{konoplya2003quasinormal}
R.~Konoplya, ``Quasinormal behavior of the d-dimensional schwarzschild black
  hole and the higher order wkb approach,'' {\em Physical Review D}, vol.~68,
  no.~2, p.~024018, 2003.

\bibitem{medved2004dirty}
A.~Medved, D.~Martin, and M.~Visser, ``Dirty black holes: quasinormal modes,''
  {\em Classical and Quantum Gravity}, vol.~21, no.~6, p.~1393, 2004.

\bibitem{nomura2005continuous}
H.~Nomura and T.~Tamaki, ``Continuous area spectrum of a regular black hole,''
  {\em Physical Review D—Particles, Fields, Gravitation, and Cosmology},
  vol.~71, no.~12, p.~124033, 2005.

\bibitem{iyer1987black2}
S.~Iyer, ``Black-hole normal modes: A wkb approach. ii. schwarzschild black
  holes,'' {\em Physical Review D}, vol.~35, no.~12, p.~3632, 1987.

\bibitem{crispino2009scattering}
L.~C. Crispino, S.~R. Dolan, and E.~S. Oliveira, ``Scattering of massless
  scalar waves by reissner-nordstr{\"o}m black holes,'' {\em Physical Review
  D—Particles, Fields, Gravitation, and Cosmology}, vol.~79, no.~6,
  p.~064022, 2009.

\bibitem{gogoi2024quasinormal}
D.~J. Gogoi, N.~Heidari, J.~Kriz, and H.~Hassanabadi, ``Quasinormal modes and
  greybody factors of de sitter black holes surrounded by quintessence in
  rastall gravity,'' {\em Fortschritte der Physik}, vol.~72, no.~3, p.~2300245,
  2024.

\bibitem{anacleto2020absorption}
M.~Anacleto, F.~Brito, J.~Campos, and E.~Passos, ``Absorption and scattering of
  a noncommutative black hole,'' {\em Physics Letters B}, vol.~803, p.~135334,
  2020.

\bibitem{Boonserm:2008zg}
P.~Boonserm and M.~Visser, ``{Bounding the greybody factors for Schwarzschild
  black holes},'' {\em Phys. Rev. D}, vol.~78, p.~101502, 2008.

\bibitem{ong2018effective}
Y.~C. Ong, ``An effective black hole remnant via infinite evaporation time due
  to generalized uncertainty principle,'' {\em Journal of High Energy Physics},
  vol.~2018, no.~10, pp.~1--11, 2018.

\bibitem{hiscock1990evolution}
W.~A. Hiscock and L.~D. Weems, ``Evolution of charged evaporating black
  holes,'' {\em Physical Review D}, vol.~41, no.~4, p.~1142, 1990.

\bibitem{page1976particle}
D.~N. Page, ``Particle emission rates from a black hole. ii. massless particles
  from a rotating hole,'' {\em Physical Review D}, vol.~14, no.~12, p.~3260,
  1976.

\bibitem{liang2025einstein}
X.~Liang, Y.-S. An, C.-H. Wu, and Y.-P. Hu, ``Einstein-horndeski gravity and
  the ultra slowly evaporating black hole,'' {\em Physics Letters B},
  p.~139303, 2025.

\bibitem{araujo2024gravitational}
A.~A. Ara{\'u}jo~Filho, H.~Hassanabadi, N.~Heidari, J.~Kr{\'\i}z, and S.~Zare,
  ``Gravitational traces of bumblebee gravity in metric--affine formalism,''
  {\em Classical and Quantum Gravity}, vol.~41, no.~5, p.~055003, 2024.

\bibitem{Yang:2023wtu}
K.~Yang, Y.-Z. Chen, Z.-Q. Duan, and J.-Y. Zhao, ``{Static and spherically
  symmetric black holes in gravity with a background Kalb-Ramond field},'' {\em
  Phys. Rev. D}, vol.~108, no.~12, p.~124004, 2023.

\bibitem{Liu:2024oas}
W.~Liu, D.~Wu, and J.~Wang, ``{Static neutral black holes in Kalb-Ramond
  gravity},'' {\em JCAP}, vol.~09, p.~017, 2024.

\bibitem{iyer1987black}
S.~Iyer and C.~M. Will, ``Black-hole normal modes: A wkb approach. i.
  foundations and application of a higher-order wkb analysis of
  potential-barrier scattering,'' {\em Physical Review D}, vol.~35, no.~12,
  p.~3621, 1987.

\bibitem{konoplya2024correspondence}
R.~A. Konoplya and A.~Zhidenko, ``{Correspondence between grey-body factors and
  quasinormal modes},'' {\em JCAP}, vol.~09, p.~068, 2024.

\bibitem{Sen:2022tru}
S.~Sen, R.~Mandal, and S.~Gangopadhyay, ``{Equivalence principle and HBAR
  entropy of an atom falling into a quantum corrected black hole},'' {\em Phys.
  Rev. D}, vol.~105, no.~8, p.~085007, 2022.

\end{thebibliography}

\end{document}